\documentclass[final,12p,times,twocolumn]{elsarticle}
\usepackage{lineno,hyperref}
\usepackage{graphicx}
\usepackage[linesnumbered,ruled,lined]{algorithm2e}
\usepackage{float}
\usepackage{tikz}
\usepackage{epstopdf}
\usepackage{amssymb}
\usepackage{amsmath}
\usepackage{subcaption}
\usepackage{footnote}
\makesavenoteenv{tabular*}
\makesavenoteenv{table*}
\renewcommand{\thefootnote}{\alph{footnote}}

\usetikzlibrary{arrows.meta}

\journal{Ocean Engineering}

\begin{document}
	\begin{frontmatter}
		\title{An efficient fully Lagrangian solver for modeling wave interaction with oscillating wave energy converter}
		\author[myfirstaddress]{Chi Zhang }
		\ead{c.zhang@tum.de}
		\author[mysecondaryaddress]{Yanji Wei}
		\ead{yanji.wei@aktishydraulics.com}
		\author[mythirdaddress]{Frederic Dias}
		\ead{frederic.dias@ucd.ie}
		\author[myfirstaddress]{Xiangyu Hu \corref{mycorrespondingauthor}}
		\ead{xiangyu.hu@tum.de}
		\address[myfirstaddress]{Department of Mechanical Engineering, 
			Technical University of Munich, 85748 Garching, Germany}
		\address[mysecondaryaddress]{Aktis Hydraulics BV, Zwolle 8017 JM,  Netherland}
		\address[mythirdaddress]{School of Mathematics and Statistics, University College Dublin, Belfield, Dublin 4, Ireland}
		\cortext[mycorrespondingauthor]{Corresponding author.}
		\begin{abstract}
			In this paper, 
			we present an efficient, accurate and fully Lagrangian numerical solver for modeling wave interaction with oscillating wave energy converter (OWSC). 
			The key idea is to couple SPHinXsys, 
			an open-source multi-physics library in unified smoothed particle hydrodynamic (SPH) framework, 
			with Simbody which presents an object-oriented Application Programming Interface (API) for multi-body dynamics. 
			More precisely, 
			the wave dynamics and its interaction with OWSC is resolved by Riemann-based weakly-compressible SPH method using SPHinXsys, 
			and the solid-body kinematics is computed by Simbody library. 
			Numerical experiments demonstrate that the proposed solver can accurately predict the wave elevations, 
			flap rotation and wave loading on the flap 
			in comparison with laboratory experiment. 
			In particularly, 
			the new solver shows optimized computational performance through CPU cost analysis and 
			comparison with commercial software package ANSYS FLUENT and other SPH-based solvers in literature. 
			Furthermore, 
			a linear damper is applied for imitating 
			the power take-off (PTO) system to study its effects on the hydrodynamics properties of OWSC and efficiency of energy harvesting. 
			In addition, 
			the present solver is used to model extreme wave condition 
			using the focused wave approach to investigate the extreme loads and motions of OWSC under such extreme wave conditions.
			It worth noting that though the model validation used herein is a bottom hinged oscillating Wave Energy Converter (WEC), 
			the obtained numerical results show	promising potential of the proposed solver to future applications in the design of high-performance WECs. 
		\end{abstract}
	\end{frontmatter}
%%%%%%%%%%%%%%%%%%%%%%%%%%%%%%%%%%%%%%%%%%%%%%%%%%%%%%%%%%%%%
%
% Section
%
%%%%%%%%%%%%%%%%%%%%%%%%%%%%%%%%%%%%%%%%%%%%%%%%%%%%%%%%%%%%%
\section{Introduction}\label{sec:introduction}
In the past decades, 
renewable ocean wave energy has received tremendous worldwide attention thanks to its abundant and dense energy form 
and in particular, low environmental impact nature. 
Consequently, 
various types of wave energy converters (WECs) have been developed to harvest electrical power from ocean waves \cite{day2015hydrodynamic}. 
In general, 
the working principle of the majority falls into four categories, 
namely, over-topping devices, oscillating water column devices, oscillating bodies and the others, 
and more details are referred to a comprehensive review \cite{antonio2010wave}.
As one of the most promising WECs, 
the oscillating wave surge converter (OWSC) has demonstrated its energy absorption capability and hydrodynamic performance 
\cite{whittaker2007development, renzi2013hydrodynamics, dias2017analytical}.  
The OWSC consists of a surface-piercing flap hinged near the seabed and oscillating back and forth under the interaction with the conforming incident waves. 
The flap's oscillating motion can be converted into electrical energy by pumping high pressure water ashore 
to drive a hydro-electric turbine \cite{whittaker2007development, wei2015wave}.
As experimental study of OWSC has the drawbacks of time consuming and economic expensive, 
numerical study is of great importance for understanding the hydrodynamics property, 
assessing the power production and optimizing control strategy \cite{renzi2013hydrodynamics, penalba2017mathematical}. 

Concerning the numerical study of OWSC, 
a set of mathematical models have been developed in literature \cite{penalba2017mathematical, dias2017analytical}. 
Folley et al. \cite{folley2007design} developed a linearized frequency domain model 
for small seabed-mounted bottom-hinged WECs and 
they applied a commercial package based on boundary element method (BEM), 
e.g., WAMIT,  
for predicting the wave force, radiation damping and added mass. 
Renzi and Dias \cite{renzi2012resonant, renzi2013hydrodynamics} proposed a semi-analytical model based on potential flow 
for three-dimensional modeling of OWSC in channel \cite{renzi2012resonant} and open ocean \cite{renzi2013hydrodynamics}. 
These models are simple and computationally efficient, and able to correctly predict the hydrodynamics properties of OWSC. 
However, 
they are unable to capture non-linear effects due to the notable assumptions of the potential flow theory. 
The non-linear effects, 
such as over-topping and slamming, 
are of great importance for correctly
predicting the flap motion as strong wave loads and large amplitude oscillations are expected during its operation. 
Therefore, 
Navier-Stokes (NS) CFD mesh-based solvers have been widely applied in modeling of wave interaction with OWSC. 
Wei et al. \cite{wei2015wave, wei2016wave} used the commercial package ANSYS FLUENT studied the viscous \cite{wei2015wave} and slamming \cite{wei2016wave, dias2018slamming} 
effects on OWSC. 
Schmitt et al . \cite{schmitt2015use} applied OpenFOAM toolbox to assess the applicability of Reynolds-averaged NS (RANS) 
solver for the simulation of OWSC. 
Mesh-based methods have demonstrated their accuracy in capturing the nonlinear effects, however, 
they are generally computational expensive as the complex mesh moving occurs during the flap's large oscillating motion. 

An alternative approach, 
meshless methods, such as smoothed particle hydrodynamics (SPH), 
have gained popularity in the simulation of nonlinear wave dynamics and wave-structure interaction 
(WSI) in the past decade \cite{ye2019smoothed, luo2019consistent, luo2017shared, khayyer2018development}. 
As a fully Lagrangian meshless method, 
SPH method was originally proposed by Lucy \cite{lucy1977numerical} 
and Gingold and Monaghan \cite{gingold1977smoothed} for astrophysical applications.
Since its inception,
SPH method has been successfully exploited in a broad variety of applications  
ranging from solid mechanics \cite{libersky1991smooth, monaghan2000sph} 
to fluid dynamics \cite{monaghan1994simulating, hu2006multi}
and fluid-structure interactions (FSI) \cite{antoci2007numerical, zhang2019multi, ye2019smoothed, gotoh2018state, khayyer2009wave}.
Thanks its Lagrangian feature, 
SPH method is particularly well suited for modeling problems involving significantly varying topology 
and free material surfaces \cite{zhang2017generalized, zhang2019weakly}. 
Recently, 
SPH method has been extended to simulate wave interaction with WECs. 
Dias et al. \cite{dias2017analytical, rafiee2013numerical, henry2013characteristics} developed an in-house UCD-SPH code 
based on OpenMP parallelization for modeling wave interaction with OWSC where the flap is considered as a rigid body and 
its kinematics is resolved directly in SPH framework.  
Crespo et al. \cite{crespo2018floating} conducted an SPH simulation for wave interaction with oscillating water column converter. 
Instead of directly computing Newton–Euler equation to capture the kinematics of rigid body in SPH framework,  
Brito \cite{brito2016coupling} presented a numerical approach by coupling DualSPHysics with Chrono project to study the wave interaction with OWSC. 
In their work, 
the mesh-free DualSPHysics implementation is considered for fluid descriptions and Chrono for mechanical systems.
Following Ref. \cite{brito2016coupling}, 
Wei et al. \cite{wei2019modeling} presented a similar approach by coupling Chrono with GPUSPH code. 
Despite of these developments, 
SPH method still suffers excessive computational efforts for three-dimensional large scale modeling of OWSC. 
Dias et al. \cite{dias2017analytical} reported that approximated $70 \text h$ CPU time is taken for $13 \text s$ physical simulation time 
with $3.2$ million particles on $72$ processors of  Intel(R) Xeon(R) CPU E5-2620 by using UCD-SPH code. 
Wei et al. \cite{wei2019modeling} shown that the computational efforts can be reduced to $2 \text h$ for $2 \text s$ physical simulation time 
with $7$ million particles 
with acceleration of graphics processing unit (GPU) on four NVIDIA Tesla K80 GPUs. 
More recently, 
Brito et al. \cite{brito2020numerical} reported $105 \text h$ computational time for $50 \text s$ physical simulation time 
with $11.4$ million particles with NVIDIA GTX 2080 alongside an Intel Xeon E5 CPU  by using DualSPHysics. 
With the advances in hardware, 
large-scale SPH modeling of WECs is becoming more and more possible, 
however, 
SPH-based solver implemented on conventional central processing units (CPUs) is still in its infancy. 

In this paper, 
we present an efficient, robust and fully Lagrangian numerical solver implemented on CPUs for modeling wave interaction with OWSC. 
The new solver is based on coupling two open-source libraries, 
SPHinXsys (\url{https://github.com/Xiangyu-Hu/SPHinXsys}) which is a multi-physics library based on SPH method
and Simbody (\url{https://simtk.org/projects/simbody}) 
which provides a high-performance multi-body physics object-oriented C++ Application Programming Interface (API). 
SPHinXsys \cite{zhang2020sphinxsys, zhang2021sphinxsys} has shown its robustness,
accuracy and versatility in modeling fluid dynamics \cite{zhang2020dual}, 
solid mechanics and fluid-structure interaction \cite{zhang2019multi} 
and multi-physics problems in cardiac function \cite{zhang2020integrative}. 
By coupling SPHinXsys with Simbody, 
the proposed solver provides an integrative interface for modeling fluid interaction with arbitrarily defined solid, flexible and the combined structures. 
The new solver is validated by modeling of regular wave interaction with OWSC  and comparing the results 
with experimental data \cite{wei2015wave} and those in literature \cite{dias2017analytical, brito2016coupling}. 
More importantly,  
the present solver shows great computational performance compared with the commercial software package ANSYS FLUENT \cite{wei2015wave}, 
UCD-SPH code \cite{dias2017analytical} and other open-source SPH library \cite{brito2020numerical}. 
Having the validation, 
a linear damper is applied to imitate the power take-off system 
to study its effects on the hydrodynamics properties of OWSC and efficiency in energy harvesting.  
Then, 
the extreme loads and motions of OWSC under extreme wave conditions 
are also investigated by modeling extreme wave condition using the focused wave approach.
 
The remainder of this paper is organized as follows.  
Section \ref{sec:methods} presents Riemann-based SPH method applied in SPHinXsys for modeling fluid dynamics,  
the principle characteristic for Simbody and the detailed coupling procedure. 
Numerical validations and applications for modeling wave interaction with OWSC are presented and discussed in Section \ref{sec:validation}. 
Concluding remarks are given in Section \ref{sec:conclusion} and all the codes and data-sets accompanying this work 
are available in repository of SPHinXsys \cite{zhang2020sphinxsys} on GitHub at \url{https://github.com/Xiangyu-Hu/SPHinXsys}.
%%%%%%%%%%%%%%%%%%%%%%%%%%%%%%%%%%%%%%%%%%%%%%%%%%%%%%%%%%%%%
%
% Section
%
%%%%%%%%%%%%%%%%%%%%%%%%%%%%%%%%%%%%%%%%%%%%%%%%%%%%%%%%%%%%%
\section{Methods}\label{sec:methods}
In this Section, 
we first briefly summarize the numerical principle characteristics of SPHinXsys 
whose detailed algorithms and rigorous validations 
are referred to Refs. \cite{zhang2020dual, zhang2020sphinxsys, zhang2020integrative, zhang2019multi, zhang2021sphinxsys}. 
Then, 
the main characteristics of Simbody library is introduced 
and the coupling procedure is presented in detail. 
%%%%%%%%%%%%%%%%%%%%%%%%%%%%%%%%%%%%%%%%%%%%%%%%%%%%%%%%%%%%%
% Section
%%%%%%%%%%%%%%%%%%%%%%%%%%%%%%%%%%%%%%%%%%%%%%%%%%%%%%%%%%%%%
\subsection{Governing equations}\label{sec:ns-equation}
The mass and momentum conservation equations for incompressbile fluid can be written in the Lagrangian frame as 
\begin{equation} 
\begin{cases}\label{governingeq}
\frac{\text d \rho}{\text d t}  =  - \rho \nabla \cdot \mathbf v \\
\rho \frac{\text d \mathbf v}{\text d t}  =   - \nabla p +  \mu \nabla^2 \mathbf v + \rho \mathbf g
\end{cases},
\end{equation}
where $\rho$ is the density, 
$\mathbf v$ the velocity, 
$p$ the pressure, 
$\mu$ the dynamic viscosity, 
$\mathbf g$ the acceleration due to gravity  
and $\frac{\text d}{\text d t}=\frac{\partial}{\partial t} + \mathbf v \cdot \nabla$ 
represents the material derivative. 
In weakly-compressible SPH (WCSPH) method, 
the weakly-compressible assumption \cite{monaghan1994simulating,morris1997modeling} 
is introduced for modeling incompressible flow 
where an artificial isothermal equation of state (EoS)
\begin{equation} \label{eqeos}
p = c^2(\rho - \rho^0).
\end{equation}
is used to close Eq. (\ref{governingeq}).  
With the weakly-compressible assumption, 
the density varies around $1 \% $ \cite{morris1997modeling} 
if an artificial sound speed of $ c = 10 U_{max}$ is employed, 
with $U_{max}$ being the maximum anticipated flow speed.
%%%%%%%%%%%%%%%%%%%%%%%%%%%%%%%%%%%%%%%%%%%%%%%%%%%%%%%%%%%%%
% Section
%%%%%%%%%%%%%%%%%%%%%%%%%%%%%%%%%%%%%%%%%%%%%%%%%%%%%%%%%%%%%
\subsection{Riemann-based WCSPH method}\label{sec:riemann-sph}
The SPHinXsys applies the Riemann-based WCSPH method for fluid dynamics where
the continuity and momentum equations are discretized as \cite{zhang2017weakly, zhang2020dual}
\begin{equation} 
\begin{cases}\label{riemannsph}
\frac{\text d \rho_i}{\text d t} = 2\rho_i \sum_j\frac{m_j}{\rho_j}(U^{\ast} - \mathbf v_{i} \mathbf e_{ij} ) \frac{\partial W_{ij}}{\partial r_{ij}} \\
\frac{\text d \mathbf v_i}{\text d t}  = - m_i \sum_j \frac{2P^{\ast}}{\rho_i \rho_j}  \nabla_i W_{ij} + m_i \sum_j \frac{2\mu}{\rho_i \rho_j} \frac{\mathbf v_{ij}}{r_{ij}} \frac{\partial W_{ij}}{\partial r_{ij}}
\end{cases}. 
\end{equation}
Here, $m$ is the mass of particle $i$, 
$\mathbf v_{ij} = \mathbf v_i - \mathbf v_j$ the relative velocity, 
$\nabla_i W_{ij}$ represents the gradient of the kernel function $W(|\mathbf v_{ij}|,h)$, 
where $\mathbf r_{ij} = \mathbf r_i - \mathbf r_j$ and $h$ is the smoothing length, 
with respect to particle $i$ and  $\mathbf e_{ij} = \mathbf r_{ij}/r_{ij}$. 
Also, 
$U^{\ast}$ and $P^{\ast}$ 
are the solutions of inter-particle one-dimensional Riemann problem constructed along the unit vector pointing from particle $i$ to $j$. 
Following the piece-wise constant reconstruction, 
the initial states of the Riemann problem are identical to those of particles $i$ and $j$, 
i.e., 
\begin{equation}\label{eq:rie-vector}
	\begin{cases}
		(\rho_L, U_L, P_L, c_L) = (\rho_i, -\mathbf v_i \cdot \mathbf e_{ij}, p_i, c_i)     \\
		(\rho_R, U_R, P_R, c_R) =  (\rho_j, -\mathbf v_j \cdot \mathbf e_{ij}, p_j, c_j) 
	\end{cases} .
\end{equation}

For solving the one-dimensional Riemann problem, 
the SPHinXsys applies the low-dissipation Riemann solver proposed by Zhang et al . \cite{zhang2017weakly} where
\begin{equation} \label{eqriesolverlinear}
	\begin{cases}
		U^{\ast} = \frac{\rho_L c_L U_L + \rho_R c_R U_R + P_L - P_R}{\rho_L c_L + \rho_R c_R} \\
		P^{\ast} = \frac{\rho_L c_L P_R  + \rho_R c_R P_L + \rho_L c_L\rho_R c_R \beta \left( U_L - U_R\right)}{\rho_L c_L + \rho_R c_R}
	\end{cases} ,
\end{equation}
with $\beta = \min \left(  3 \max \left( U_L - U_R, 0 \right) /\bar c, 1 \right) $ denotes
the low dissipation limiter \cite{zhang2017weakly} and $\bar c = \left( \rho_L c_L + \rho_R c_R \right) /\left( \rho_L + \rho_R \right)$.

For computational efficiency, 
the dual-criteria time-stepping method is applied for the time integration of fluid. 
Following Ref. \cite{zhang2020dual},  
two time-step criteria are defined as
the advection criterion $\Delta t_{ad}$
\begin{equation}\label{dt-advection}
\Delta t_{ad}   =  {CFL}_{ad} \min\left(\frac{h}{|\mathbf{v}|_{max}}, \frac{h^2}{\nu}\right),
\end{equation}
and the acoustic criterion $\Delta t_{ac}$
\begin{equation}\label{dt-relax}
\Delta t_{ac}   = {CFL}_{ac} \frac{h}{c + |\mathbf{v}|_{max}}. 
\end{equation}
Here, $CFL_{ad} = 0.25$, ${CFL}_{ac} = 0.6$,  
$|\mathbf{v}|_{max}$ the maximum particle advection velocity in the flow 
and $\nu$ the kinematic viscosity. 
Accordingly, 
the advection criterion controls the updating frequency of particle configuration and 
the acoustic criterion determines the frequency of the pressure relaxation process. 
More details and validations are referred to Ref. \cite{zhang2020dual}. 

Also, at the beginning of each advection time step, 
the density is reinitialized by  
\begin{equation} \label{eqrhosumnosurface}
	\rho_i =  \max \left( \rho^*, \rho^0 \frac{ \sum W_{ij}}{\sum W^0_{ij}}\right) , 
\end{equation}
where $\rho^*$ denotes the density before re-initialization and superscript $0$ represents the initial reference value.
%%%%%%%%%%%%%%%%%%%%%%%%%%%%%%%%%%%%%%%%%%%%%%%%%%%%%%%%%%%%%
% Section
%%%%%%%%%%%%%%%%%%%%%%%%%%%%%%%%%%%%%%%%%%%%%%%%%%%%%%%%%%%%%
\subsection{Simbody library}\label{sec:simbody}
As an open-source library licensed under Apache License 2.0, 
Simbody is distributed in binary form for multiple platforms 
and presents an object-oriented API to the application programmers 
who are of interest to handle the modeling and computational aspects of multi-body dynamics. 
The  Simbody 
can be applied for incorporating robust, high-performance and minimal-coordinate $O\left( n\right) $ 
multi-body dynamics into a variety range of domain-specific end-user applications, 
for example, 
it is used by biomechanists in OpenSim, 
by roboticists in Gazebo, 
and by biomolecular researcher in MacroMoleculeBuilder (MMB).

\begin{figure*}[htb!]
	\centering
	\includegraphics[trim = 5cm 6.5cm 1cm 4cm, clip,width=\textwidth]{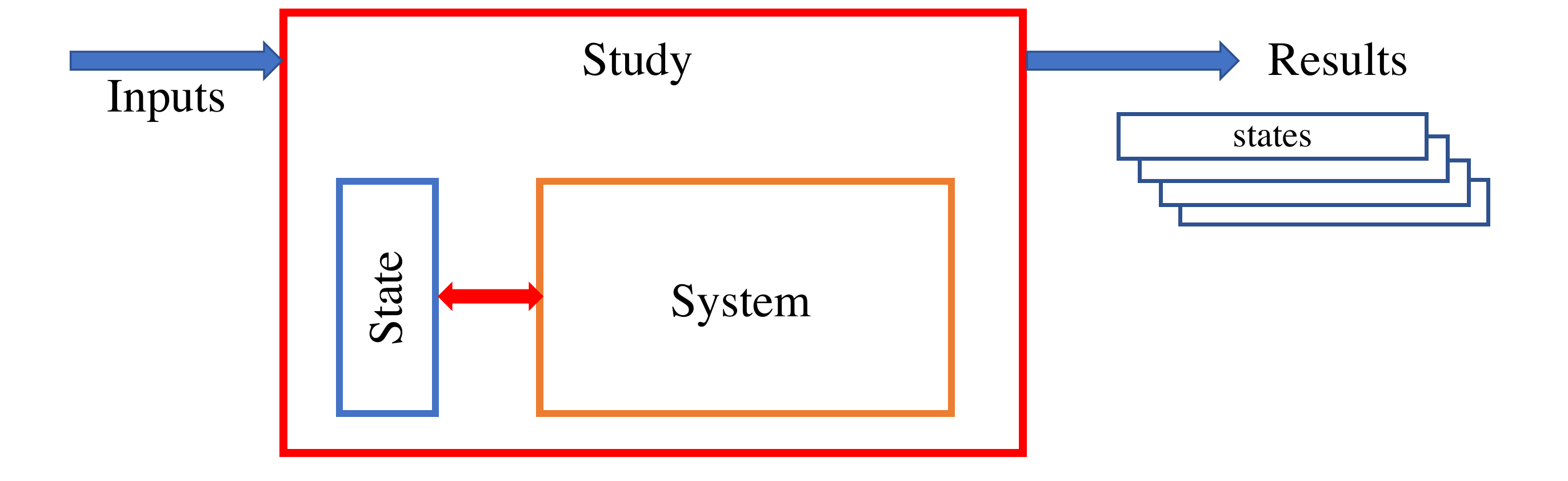}
	\caption{Architecture overview of Simbody.  
					A read-only System object contains the model components and defines the parameterizations.
					Those parameters' values are stored in State object. 
					The Study object generates a series of states which represent a specific solution. }
	\label{figs:simbody}
\end{figure*}

In the top-level architecture, 
Simbody consists of three primary objects, e.g., System, State and Study as shown in Figure \ref{figs:simbody}. 
The System object encapsulates of bodies, joints and forces of a model and defines its parameterization. 
A complete set of values for each of the System's parameters is called a "state" and 
the response of a System is determined by the state values. 
A System's compatible State object has the entries for the values of each "state", 
for example, time, position and velocity. 
A Study object couples a System and one or more States, 
and represents a computational experiment intended to reveal something about the System. 
For example, 
a simple evaluation Study merely asks the System to evaluate specific quantities, 
such as the position, using the values taken from a particular State. 
%
%%%%%%%%%%%%%%%%%%%%%%%%%%%%%%%%%%%%%%%%%%%%%%%%%%%%%%%%%%%%%
% Section
%%%%%%%%%%%%%%%%%%%%%%%%%%%%%%%%%%%%%%%%%%%%%%%%%%%%%%%%%%%
\subsection{SPHinXsys and Simbody coupling}\label{sec:coupling}
As mentioned in the previous Section \ref{sec:simbody}, 
an object-oriented C++ API is provided by Simbody 
and this feature makes its coupling with SPHinXsys straightforward.
In the present framework, 
the hydrodynamic force exerted on the rigid body is computed by SPHinXsys 
and passed to Simbody to predict the combined translational and rotational motion by solving the Newton–Euler equation. 
All the kinematic states, e.g., station location, velocity and acceleration, 
are stored in Simbody and passed back to SPHinXsys for 
updating the position, velocity and normal of an ensemble of particles belonging to the corresponding rigid body. 

\begin{figure*}[htb!]
	\centering
	\includegraphics[trim = 1cm 4cm 1cm 1.5cm, clip,width=0.95 \textwidth]{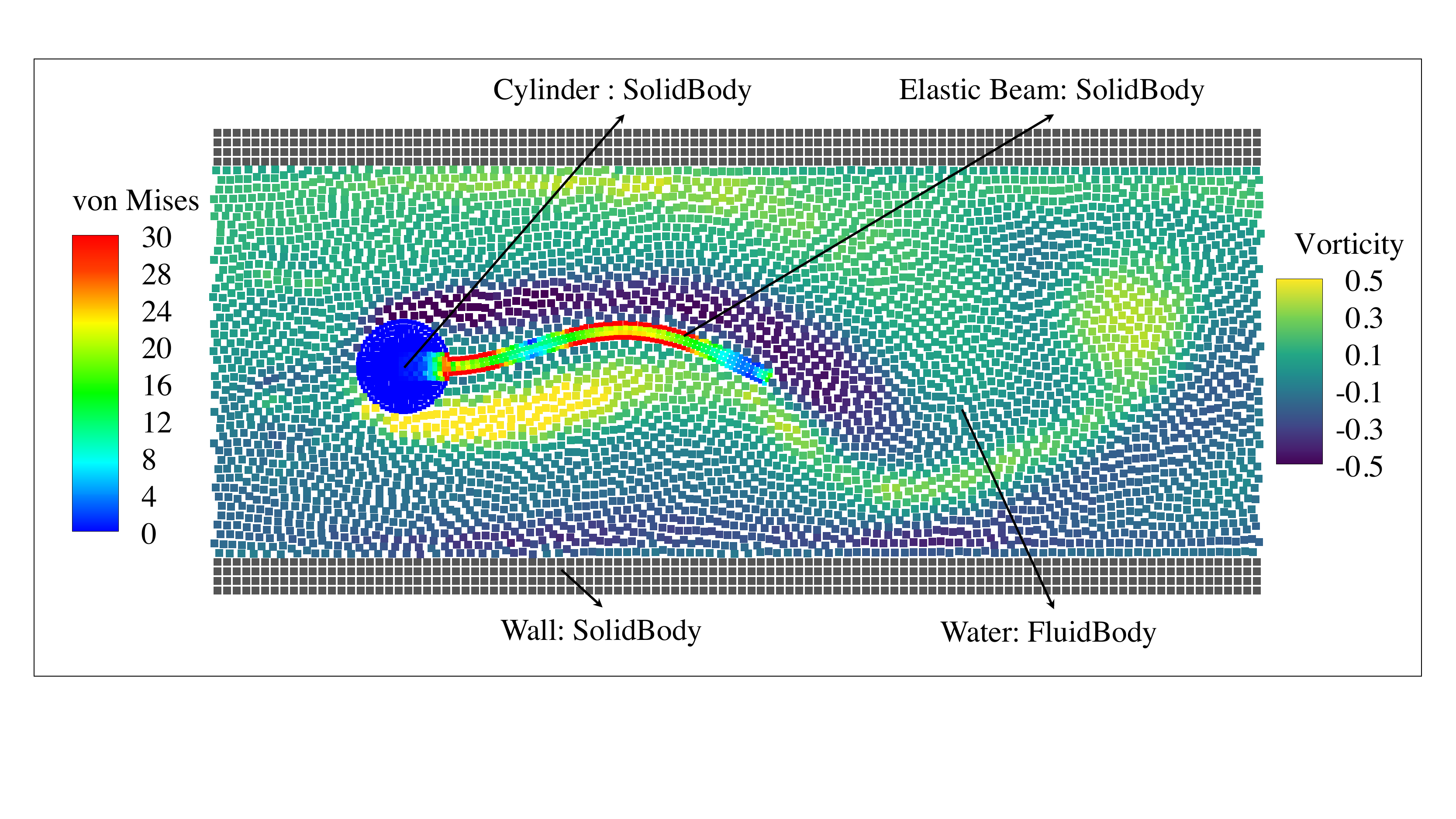}
	\caption{A typical fluid-structure interaction (FSI) involving a rigid solid (wall) body, 
		a composite solid (insert) body and a fluid body. The wall body has two (upper and lower) components. 
		The insert body is composed of a rigid (cylinder) and an elastic (beam) components (For color interpretation, the reader is referred to the web version of this paper).}
	\label{figs:fsi}
\end{figure*}

In SPHinXsys, 
all media are modeled as SPH bodies and each body is composed of an ensemble of SPH particles as shown in 
Figure \ref{figs:fsi} which represents a typical example of modeling flow induced vibration of a flexible beam attached to a rigid cylinder. 
In this framework, 
whole or parts of solid particles can be constrained to characterize the rigid-body dynamics. 
For example, 
the cylinder part can be fixed or moving accordingly by solving Newton-Euler equation 
and the beam part is constrained to the cylinder meanwhile deformed under the interaction with the surrounding flow. 

For modeling of fluid-structure interactions,  
the total force exerted on the structure by the surrounding fluid is evaluated through
\begin{equation}\label{eq:force}
\mathbf F = \sum_{a \in N} \mathbf f_a ,
\end{equation}
where $N$ is the total particle number for the solid structure and $\mathbf f_a$ is calculated through 
\begin{equation}\label{eq:fs-force}
\small
\mathbf f_a =  - 2  \sum_i V_i V_a \frac{p_i \rho_a^d + p^d_a \rho_i}{\rho_i + \rho^d_a} \nabla_a W_{ai} + 2\sum_i  \nu V_i V_a \frac{\mathbf v_i - \mathbf v^d_a}{r_{ai}} 
\frac{\partial W_{ai}}{\partial r_{ai}} , 
\end{equation}
where the subscript letter $a$ and $i$ represent particle belong to solid and fluid body, 
respectively. 
In Eq. \eqref{eq:fs-force}, 
the first and second term in the right-hand-side (RHS) denote the pressure and viscous force, 
respectively. 
The imaginary pressure $p_a^d$ and velocity $\mathbf{v}_a^d$ are defined by
\begin{equation}\label{eq:fs-coupling}
\begin{cases}
p_a^d = p_i + \rho_i \max \left( 0, \left( \mathbf g - {\frac{\text d \mathbf v_a}{\text d t}} \right)  \cdot \mathbf n \right)  \left( \mathbf r_{ai} \cdot \mathbf n \right)  \\
\mathbf v_a^d = 2\mathbf v_i - \mathbf v_a
\end{cases}, 
\end{equation}
where $\mathbf n$ denotes the norm point from solid to fluid. 
Then, 
the total torque acting about the center of mass of the solid body can be expressed as
\begin{equation} \label{eq:torque}
\mathbf \tau = \sum_{a \in N} \left( \mathbf r_a - \mathbf r_{com} \right) \times  \mathbf f_a,
\end{equation}
where $\mathbf r_{com}$ is the center of mass. 
At the end of each fluid time step, 
the total force and torque are obtained and passed to Simbody for solving the Newton-Euler equation
\begin{equation} \label{eq:newton-euler}
\left( 
\begin{array}{c}\mathbf F \\ \mathbf \tau\end{array} 
\right)  = 
\left( 
\begin{array}{cc}m \mathbb I & 0 \\0 & \mathbf I\end{array} 
\right)  
\left( 
\begin{array}{c} \frac{\text d \mathbf v}{\text d t} \\ \frac{\text d \mathbf \Omega}{\text d t}\end{array} 
\right)  
%+ 
%\left( 
%\begin{array}{c} 0 \\ \mathbf\Omega \times \mathbf I \mathbf\Omega\end{array} 
%\right) 
+
\left( 
\begin{array}{c} 0 \\ -k_d \mathbf\Omega \end{array} 
\right), 
\end{equation}
where $m$ is the mass of flap, 
$\mathbb I$ the identity matrix,
$\mathbf I$ the moment of inertia about the center of mass, 
$\mathbf\Omega$ is the angular velocity  and 
$k_d$ the damping coefficient. 
Note that Eq. \eqref{eq:newton-euler} introduces a linear damper for modeling the PTO damping of OWSC.  
The detailed coupling procedure are given in the following. 

\begin{figure*}[htb!]
	\centering
	\includegraphics[trim = 0mm 0mm 0mm 0mm, clip,width=0.99 \textwidth]{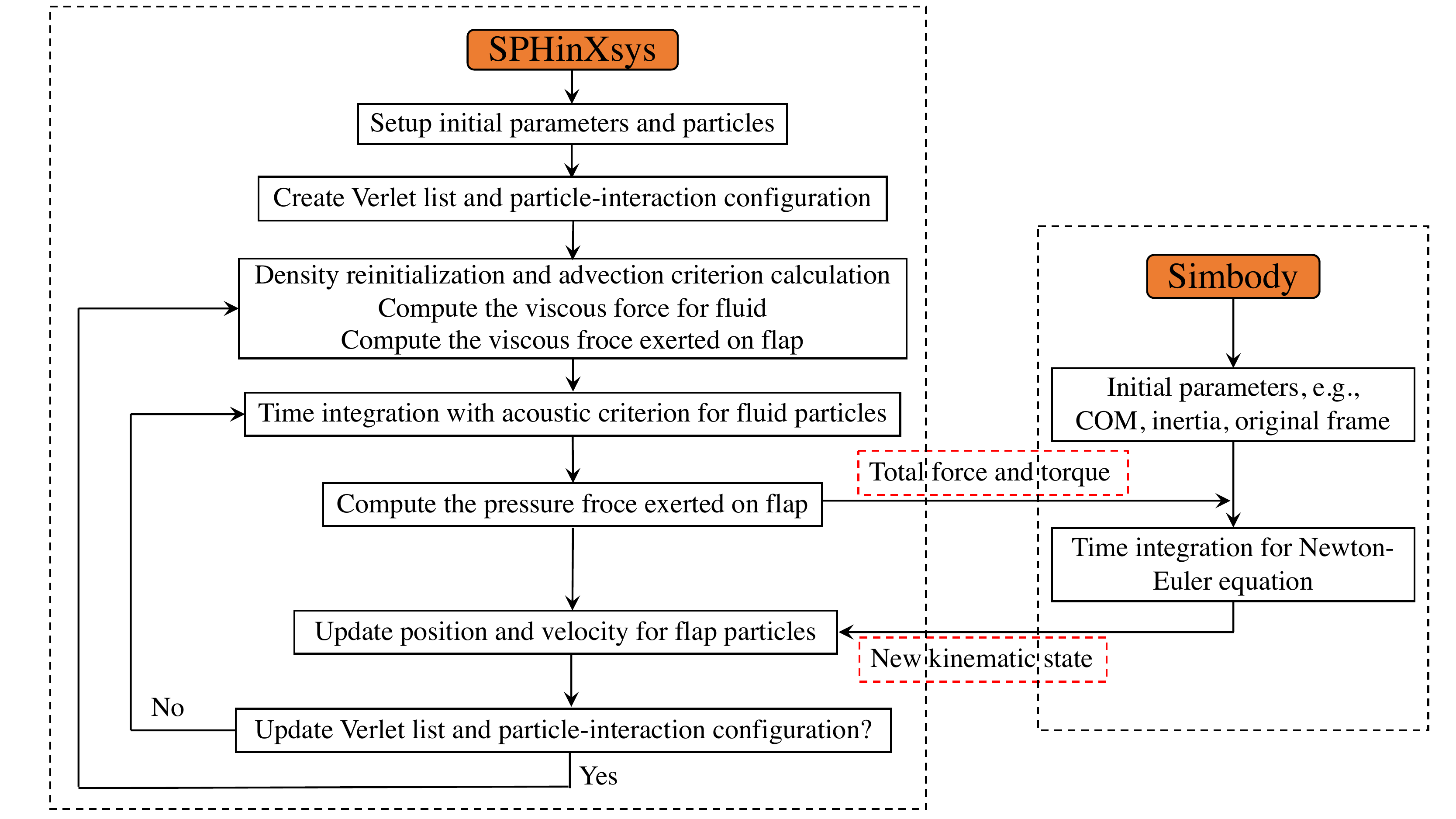}
	\caption{Flow chart of coupling SPHinXsys with Simbody.}
	\label{figs:sphinxsys-simbody}
\end{figure*}

At the beginning of the advection step, 
the fluid density is reinitialized by Eq. \eqref{eqrhosumnosurface} 
and the viscous force exerted on the solid body is also computed. 
Then the pressure relaxation process is repeated several times \cite{zhang2020dual} 
by using the position-based Verlet scheme proposed in Ref. \cite{zhang2019multi}. 
At first, the integration of the fluid is conducted as
\begin{equation}\label{verlet-first-half}
\begin{cases}
\rho_i^{n + \frac{1}{2}} = \rho_i^n + \frac{1}{2}\Delta t_{ac}  \frac{d \rho_i}{dt}\\
\mathbf r_i^{n + \frac{1}{2}} = \mathbf{r}_i^n + \frac{1}{2} \Delta t_{ac} \mathbf v_i^n
\end{cases}, 
\end{equation}
by updating the density and position fields into the mid-point. 
Then particle velocity is updated to the new time step
\begin{equation}\label{verlet-first-mediate}
\mathbf v_i^{n + 1} = \mathbf v_i^n +  \Delta t_{ac}  \frac{d \mathbf v_i} {dt}. 
\end{equation}
Finally, the position and density of fluid particles are updated to the new time step by 
\begin{equation}\label{verlet-first-final}
\begin{cases}
\mathbf r_i^{n + 1} = \mathbf r_i^ {n + \frac{1}{2}} +  \frac{1}{2} \Delta t_{ac} \mathbf v_i^{n + 1} \\
\rho_i^{n + 1} = \rho_i^{n + \frac{1}{2}} + \frac{1}{2} \Delta t_{ac} \frac{d \rho_i}{dt}
\end{cases}. 
\end{equation}
At this point, 
the total pressure force exerted on the solid body is also computed. 
Now, 
the total force and torque are obtained and passed to Simbody to solve the Newton–Euler equation 
and update the State with a Runge-Kutta-Merson integrator.
Having the new State, 
the position, velocity and norm of the whole or parts of solid particles are updated. 
An overview of the present coupling procedure is shown in Figure \ref{figs:sphinxsys-simbody}. 

It is worth noting that directly computing Newton–Euler equation to capture the kinematics of rigid body in SPH framework 
is also feasible and has been implemented in the work of Rafiee et al. \cite{rafiee2013numerical}, 
Henry et al. \cite{henry2013characteristics} and Dias et al. \cite{dias2017analytical}. 
However, 
coupling SPHinXsys with Simbody as presented in this work 
allows us to explore more challenging scenarios, 
e.g., passive and active flexible fish-like body swimming, wave interaction with elastic structures and flow driven multi-body collisions, 
which will be explored in the future work.
%
%%%%%%%%%%%%%%%%%%%%%%%%%%%%%%%%%%%%%%%%%%%%%%%%%%%%%%%%%%%%%
% Section
%%%%%%%%%%%%%%%%%%%%%%%%%%%%%%%%%%%%%%%%%%%%%%%%%%%%%%%%%%%%%
\subsection{Wave making and elimination}\label{sec:wave}
In Riemann-base WCSPH, 
the solid wall boundary can be treated as dummy particle whose interaction with fluid is determined 
by solving one-sided Riemann problem along the wall-normal direction \cite{zhang2017weakly}.
Subsequently, 
the regular wave can be generated by imposing a piston–type wave maker which consists of an ensemble of dummy particles 
whose displacement is determined by the linear wavemaker theory 
\begin{equation} \label{eq:wave-maker}
\mathbf r_a = S \sin \left( f t + \phi \right),
\end{equation}
where $S$ is the wave stroke, 
$f$ the wave frequency and $\phi$ the initial phase. 
Here, the wave stroke $S$ is determined by 
\begin{equation}
S = \frac{H \sinh\left(2kh_0\right) + 2kh_0 }{\sinh\left( 2kh_0 \right) \tanh\left(kh_0 \right) }, 
\end{equation}
where $H$ is the wave height, 
$h_0$ the water depth and $k$ is the wave number. 

For wave elimination, 
a damping zone is implemented as passive wave elimination system,
where the velocity of the fluid particles reduces at each time step according to their locations in the zone with quadratic decay. 
Subsequently, 
the velocity is modified by
\begin{equation} \label{eq:wave_elimination}
\mathbf v = \mathbf v_0 \left( 1.0  - \Delta t \alpha \left( \frac{\mathbf r - \mathbf r_0}{\mathbf r_1 - \mathbf r_0}\right)  \right), 
\end{equation}
where $\mathbf v_0$ denotes the initial velocity of the fluid particle at the entry of the damping zone,
$\mathbf v$ is the velocity after damping, 
$\Delta t$ the time step,  
$\mathbf r_0$ and $\mathbf r_1$ are the initial and final positions of the damping zone, respectively.
Also, 
the reduction coefficient $\alpha$ controls the modifications on the velocity at each time step and herein we adopt $\alpha =  5.0$.
%%%%%%%%%%%%%%%%%%%%%%%%%%%%%%%%%%%%%%%%%%%%%%%%%%%%%%%%%%%%%
%
% Section
%
%%%%%%%%%%%%%%%%%%%%%%%%%%%%%%%%%%%%%%%%%%%%%%%%%%%%%%%%%%%%%
\section{Modeling of wave interaction with OWSC}\label{sec:validation}
In this Section, 
we present the modeling of wave interaction with OWSC with the proposed numerical solver. 
We first conduct the validation by comparing the numerical results 
with experimental data \cite{wei2015wave} and those in literature \cite{dias2017analytical, brito2016coupling}. 
Then, 
we analyze the computational efficiency by comparing the normalized CPU time 
with that of ANSYS FLUENT \cite{ansys2011ansys}, 
UCD-SPH code \cite{dias2017analytical} and DualSPHysics \cite{brito2020numerical}. 
Subsequently, 
the new solver is applied for studying the effects and energy harvesting efficiency of PTO system 
and extended to investigate the extreme loads on OWSC under extreme wave condition. 
%%%%%%%%%%%%%%%%%%%%%%%%%%%%%%%%%%%%%%%%%%%%%%%%%%%%%%%%%%%%%
% Section
%%%%%%%%%%%%%%%%%%%%%%%%%%%%%%%%%%%%%%%%%%%%%%%%%%%%%%%%%%%%%
\subsection{Numerical setup}\label{sec:experiment} 
\begin{figure*}[htb!]
	\centering
	\includegraphics[trim = 1cm 1cm 1cm 1cm, clip,width=\textwidth]{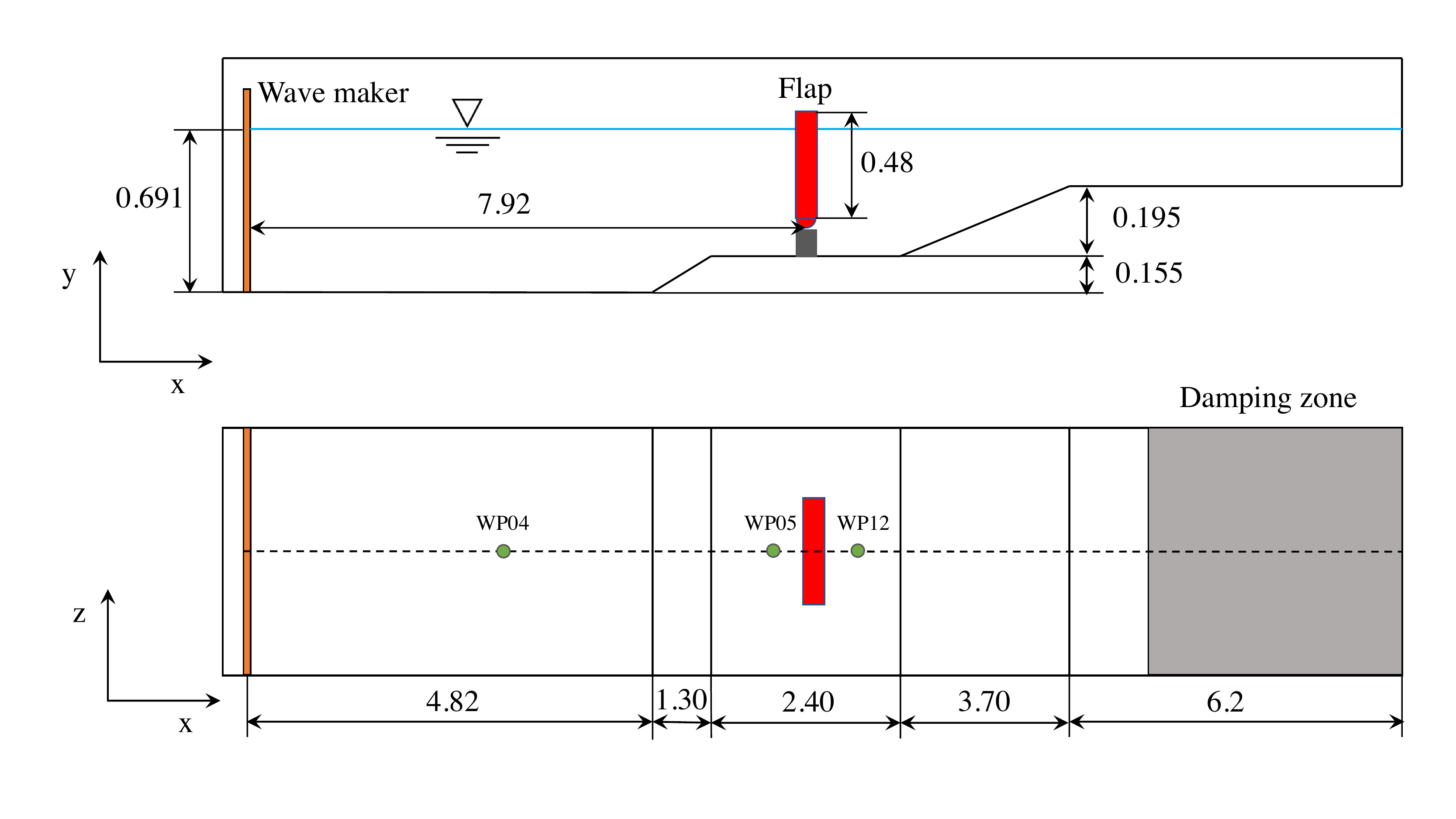}
	\caption{Schematic of the wave tank and the OWSC model. Dimensions are in meters.}
	\label{figs:owsc-setup}
\end{figure*}

The present numerical setup is identical to the experiment conducted at the Marine Research Group's hydraulics laboratory at Queen's University Belfast \cite{wei2015wave}. 
Following Ref. \cite{wei2015wave}, 
the numerical wave tank (NWT) is $18.4 \text m$ long, 
$4.58 \text m$ wide and $1.0 \text m$ high as shown in Figure \ref{figs:owsc-setup}. 
The OWSC shape is simplified as a $1.04 \times 0.48 \times 0.12 \text m$ box-type flap which is located $7.92 \text m$ far from the wave maker in $x$-axis and in the center of the
NWT in $z$-axis, and hinged to a $0.16 \text m$ high base. 
The mass of the flap is $33 \text {kg}$ and the inertia of the flap is $1.84 \text{kg}\text m^2$. 
In order to investigate the time variation of the pressure loads, 
an array of 6 pressure sensors located one side of the front face of the flap (toward the wave maker) is used.   
The positions of the sensors are given in Table \ref{tab:ps} and sensors $\text{PS}03$ and $\text{PS}11$ pierce the initial water free surface. 
Note that there are 13 pressure sensors in the experiment \cite{wei2015wave} and herein we choose $6$ of them for comparison following the work of Wei et al. \cite{wei2015wave}.
Similarly, 
three wave probes (there are 14 wave probes in the experiment) were placed in the NWT to elevate the free surface and their locations are given in Table \ref{tab:wp}. 
For clarity, 
the pressure sensors and wave probes are termed identical to those of the experiment. 
Note that the wave propagates along the $x$-axis, 
the $y$-axis is along the vertical direction (height of the NWT), 
the flap rotates along the $z$-axis (width of the NWT) and the rotation angle is positive when the flap pitches landward. 
Also,
the present model is $1:25$ scale and all the results presented in this paper have been converted to full scale as in Ref. \cite{wei2015wave}. 
\begin{table*}[htb!]
	\centering
	\caption{Positions of the pressure sensors on the front flap face. The position along the z-axis is measured from the center of the device, and $y = 0$ denotes the mean water level.}
	\begin{tabular}{cccccc}
		\hline
		No.    & $y$-axis (m)  & $z$-axis (m)     &  No.    & $y$-axis (m)  & $z$-axis (m)  \\ 
		\hline
		PS01  &$-0.046$       & $0.468$   & PS09  &$-0.117$       & $0.156$ \\ 
		%\hline
		PS03  &$ 0.050$       & $0.364$   & PS11  &$ 0.025$       & $0.052$ \\ 
		%\hline
		PS05  &$-0.300$       & $0.364$   & PS13  &$-0.239$       & $0.052$ \\ 
		\hline
	\end{tabular}
	\label{tab:ps}
\end{table*}
\begin{table}[htb!]
	\centering
	\caption{Positions of the wave probes. The position along the z-axis is measured from the center of the device, and $y = 0$ denotes the mean water level. }
	\begin{tabular}{ccc}
		\hline
		No.    & $x$-axis (m)  & $z$-axis (m)       \\ 
		\hline
		WP04  &$3.99$       & $0$   \\ 
		%\hline
		WP05  &$7.02$       & $0$   \\ 
		%\hline
		WP12  &$8,82$       & $0$   \\ 
		\hline
	\end{tabular}
	\label{tab:wp}
\end{table}

To discretize the system, 
the initial particle space is set as $dp = 0.03 \text m$ resulting in a number of $1.542$ million fluid particles 
and $0.628$ million solid particles (including tank, wave maker and flap). 
The $5th$-order Wendland kernel \cite{wendland1995piecewise} 
with a smoothing length of $h = 1.3dp$ and a cut-off radius of $2.6dp$ 
is employed in all the following simulations.
%%%%%%%%%%%%%%%%%%%%%%%%%%%%%%%%%%%%%%%%%%%%%%%%%%%%%%%%%%%%%
% Section
%%%%%%%%%%%%%%%%%%%%%%%%%%%%%%%%%%%%%%%%%%%%%%%%%%%%%%%%%%%%%
\subsection{Model validation}\label{sec:model-validation}
In this Section, 
we consider the regular wave interaction with OWSC in condition of wave height $\text H = 5.0 \text m$ and wave period $\text T = 10 \text s$ in full scale. 
For rigorous and comprehensive validation, 
comparisons of the main principle aspects of wave-flap interaction, 
e.g., 
wave elevation, wave loads on the flap and the flap's rotation, 
will be conducted same as Ref. \cite{wei2015wave}. 
%%%%%%%%%%%%%%%%%%%%%%%%%%%%%%%%%%%%%%%%%%%%%%%%%%%%%%%%%%%%%
% Section
%%%%%%%%%%%%%%%%%%%%%%%%%%%%%%%%%%%%%%%%%%%%%%%%%%%%%%%%%%%%%
\subsubsection{Wave propagation and its interaction with OWSC}\label{sec:wavedynmaics}
\begin{figure*}
	\centering
	%add desired spacing between images, e. g. ~, \quad, \qquad, \hfill etc. 
	\begin{subfigure}[b]{0.49\textwidth}
		\centering
		\includegraphics[trim = 2mm 2mm 2mm  2mm, clip, width=\textwidth]{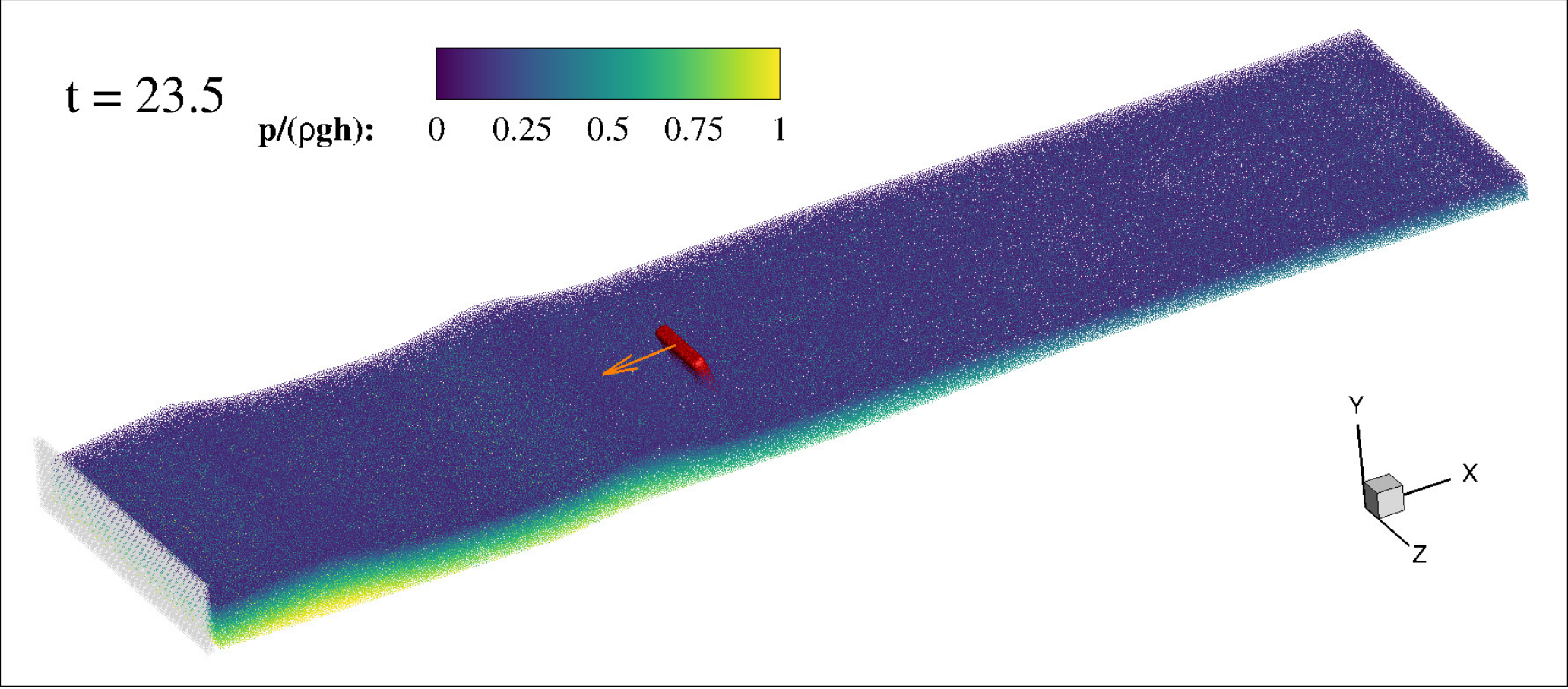}\\
		\includegraphics[trim = 2mm 2mm 2mm  2mm, clip, width=\textwidth]{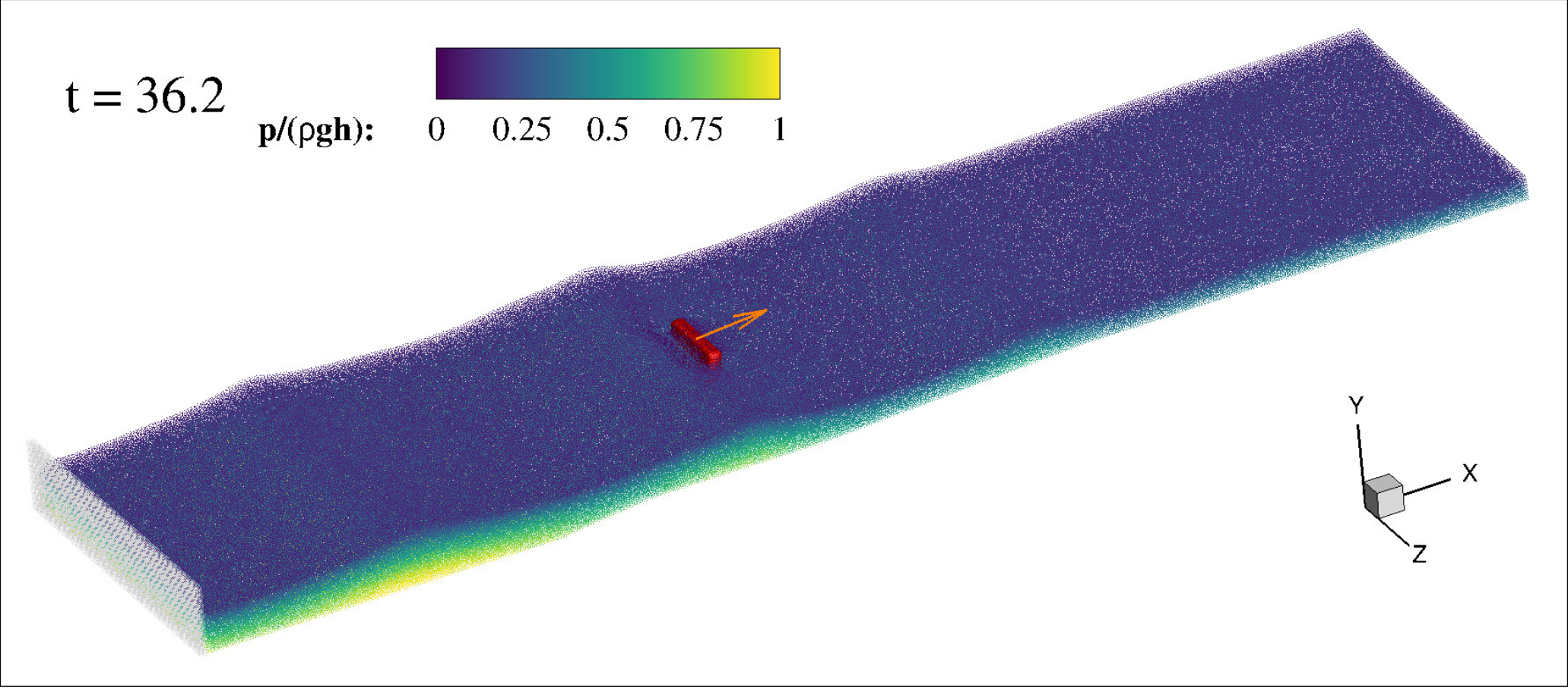}\\
		\includegraphics[trim = 2mm 2mm 2mm  2mm, clip, width=\textwidth]{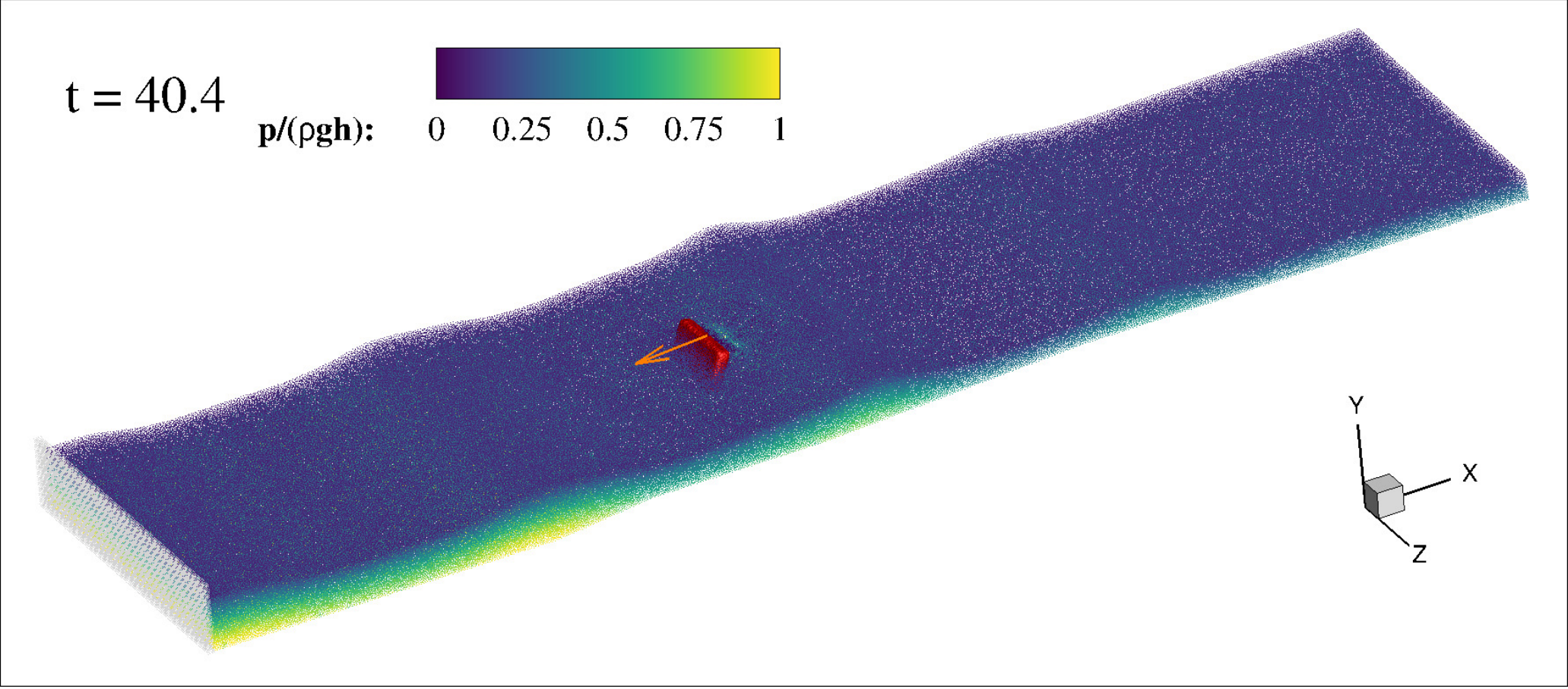}\\
		\includegraphics[trim = 2mm 2mm 2mm  2mm, clip, width=\textwidth]{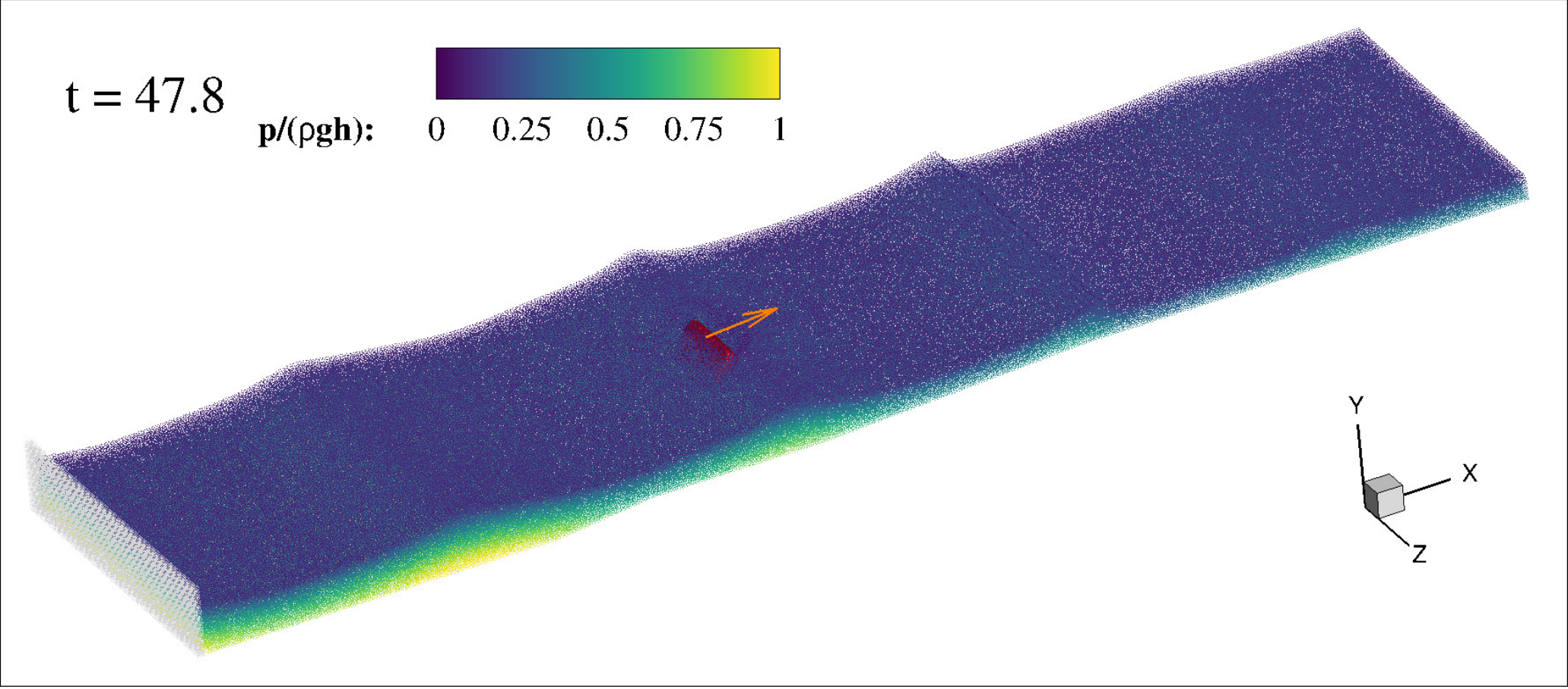}\\
		\includegraphics[trim = 2mm 2mm 2mm  2mm, clip, width=\textwidth]{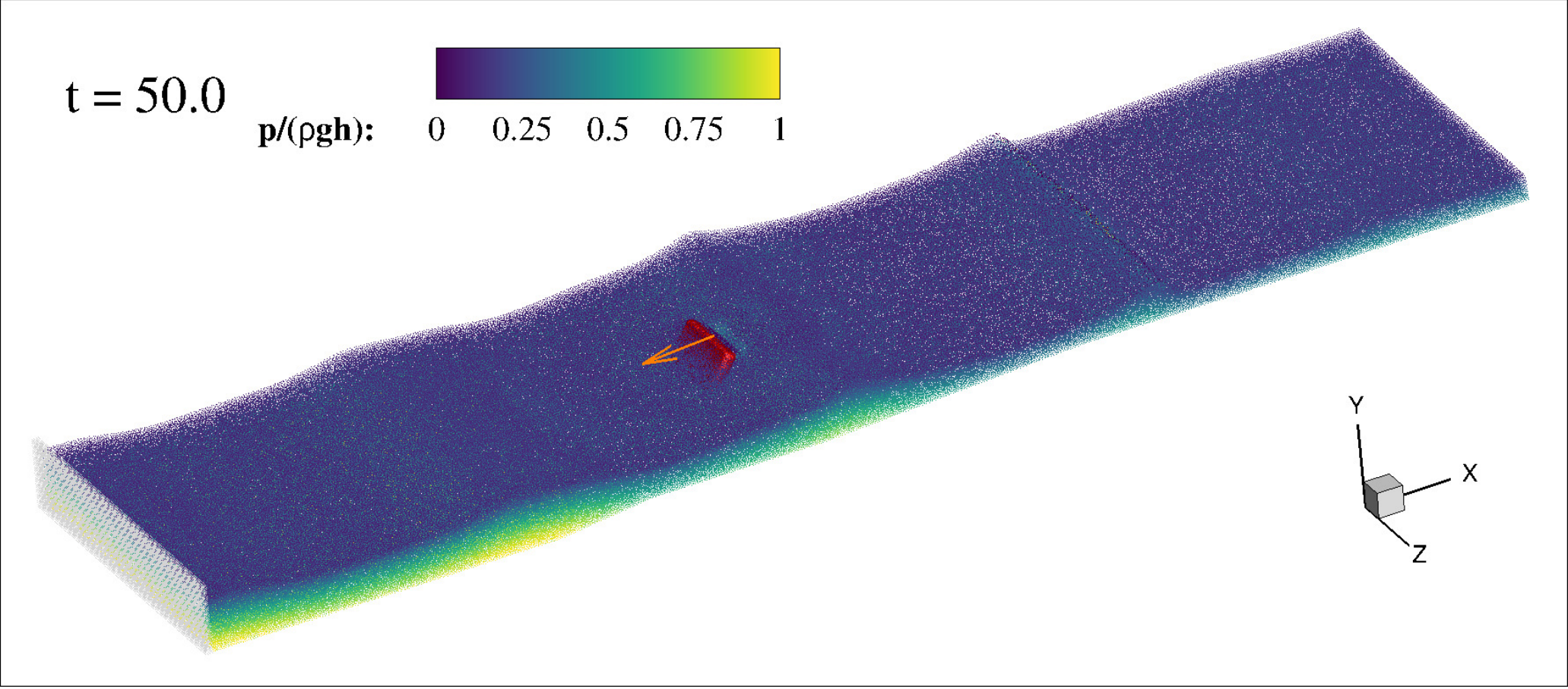}\\
		\includegraphics[trim = 2mm 2mm 2mm  2mm, clip, width=\textwidth]{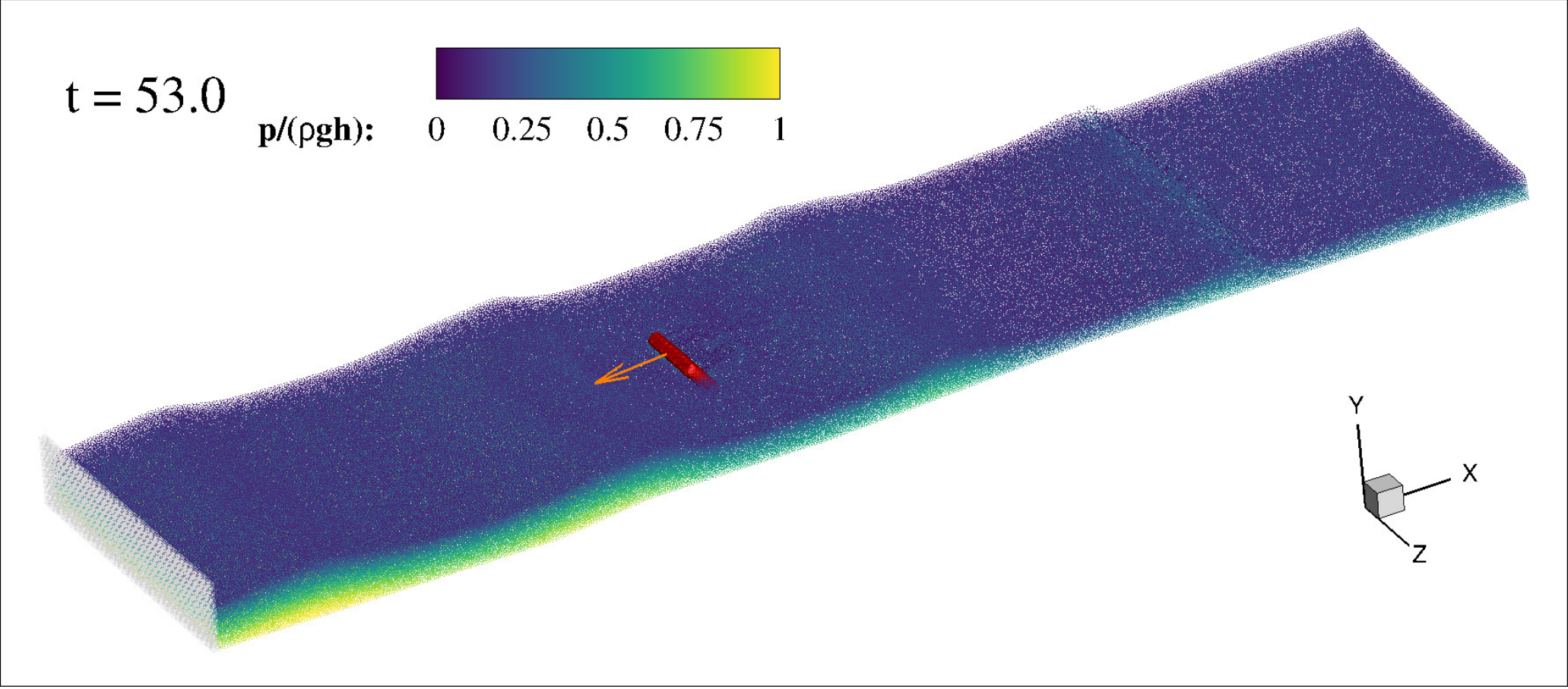}
		\caption{Normalized pressure}
		\label{figs:owsc-surface-p}
	\end{subfigure}
	%\newline 
	\begin{subfigure}[b]{0.49\textwidth}
		\centering
		\includegraphics[trim = 2mm 2mm 2mm  2mm, clip, width=\textwidth]{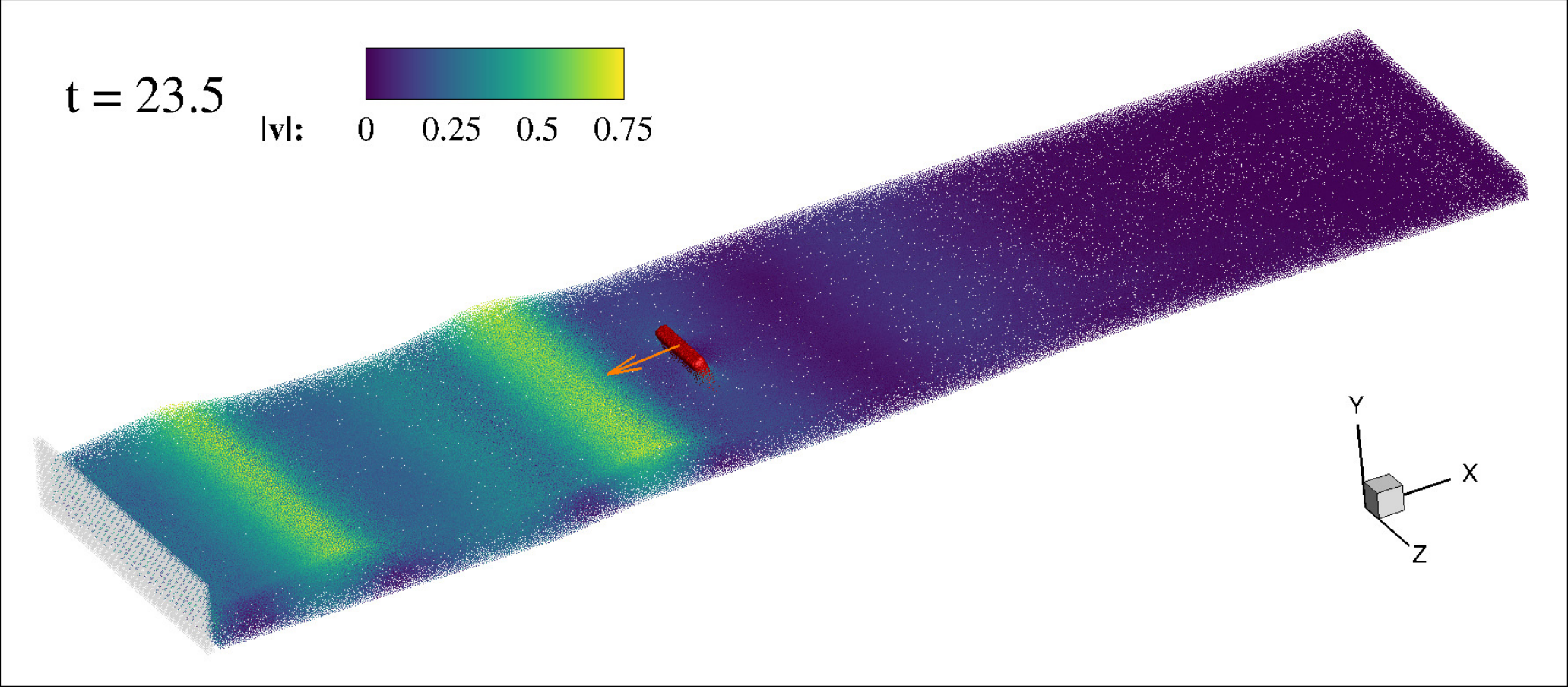}\\
		\includegraphics[trim = 2mm 2mm 2mm  2mm, clip, width=\textwidth]{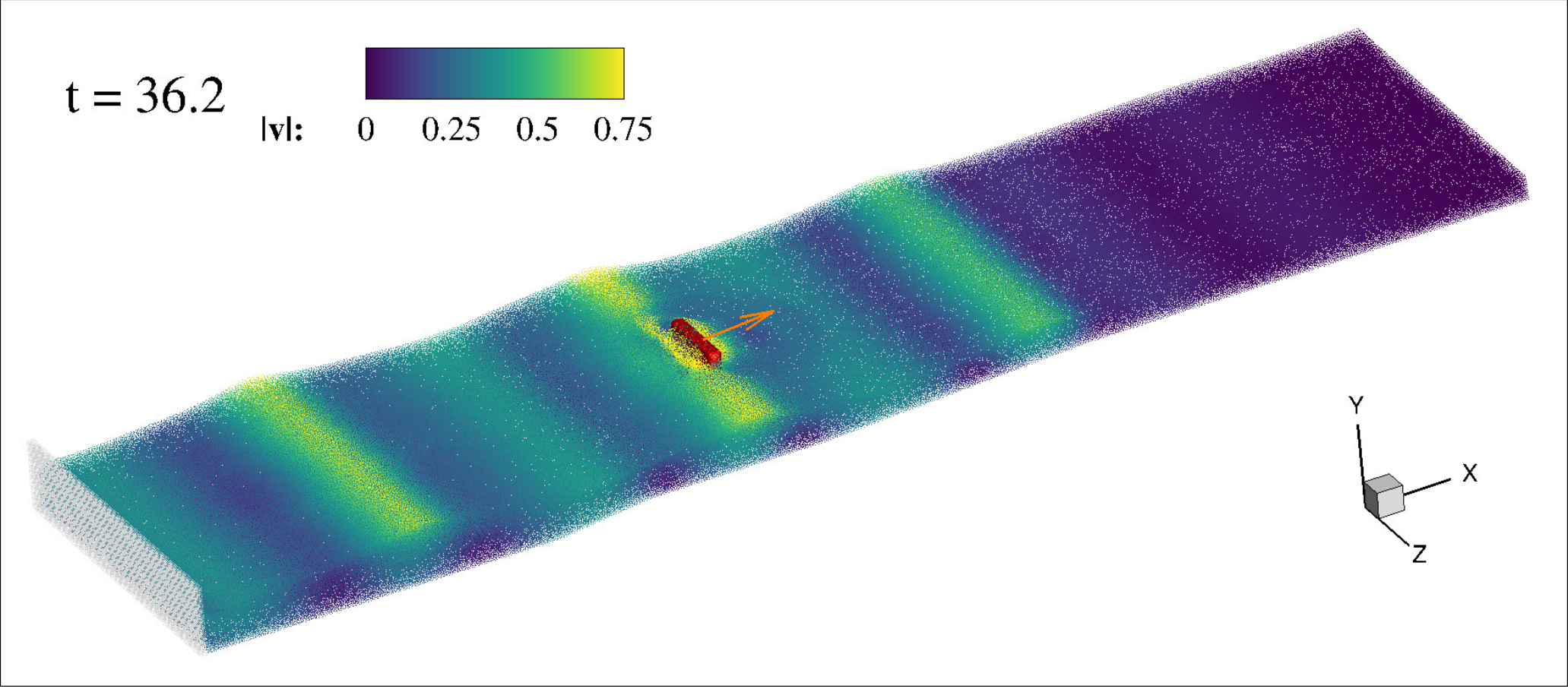}\\
		\includegraphics[trim = 2mm 2mm 2mm  2mm, clip, width=\textwidth]{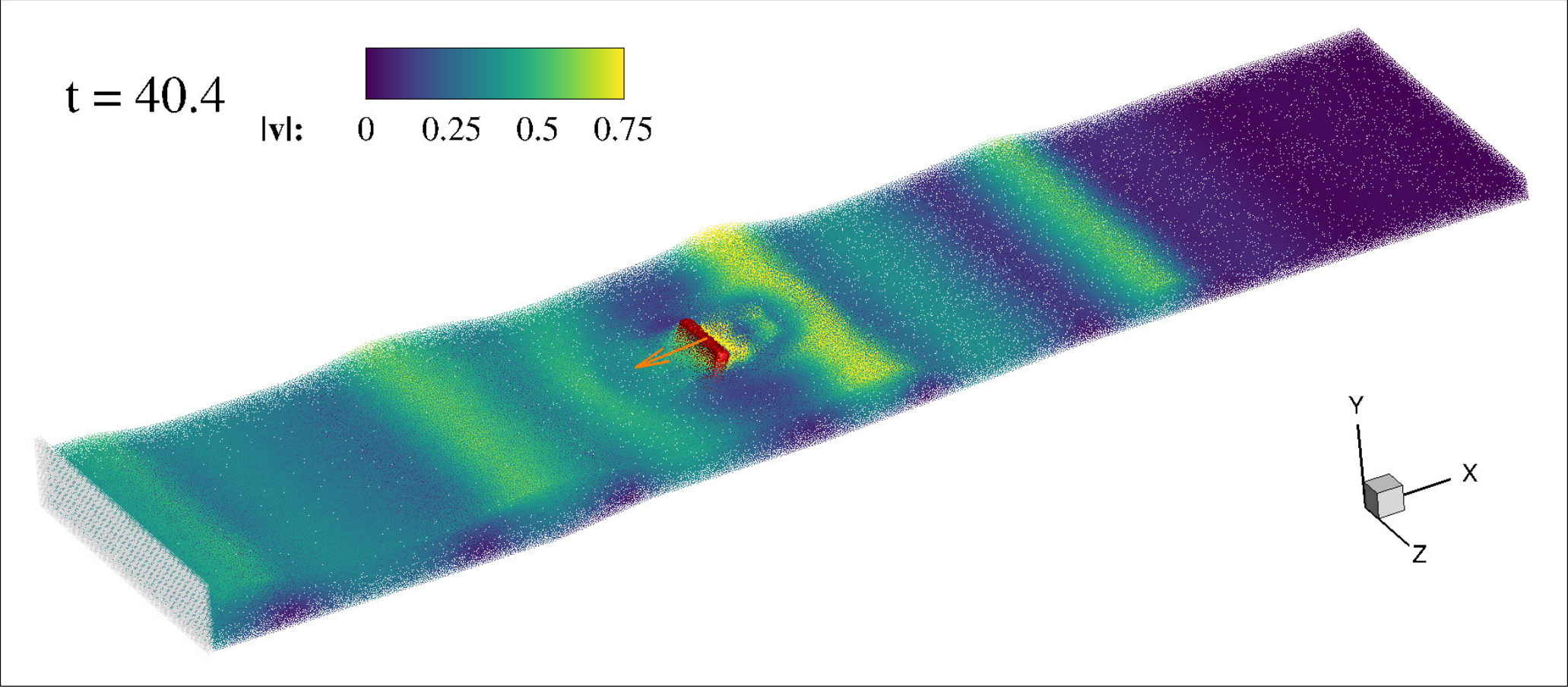}\\
		\includegraphics[trim = 2mm 2mm 2mm  2mm, clip, width=\textwidth]{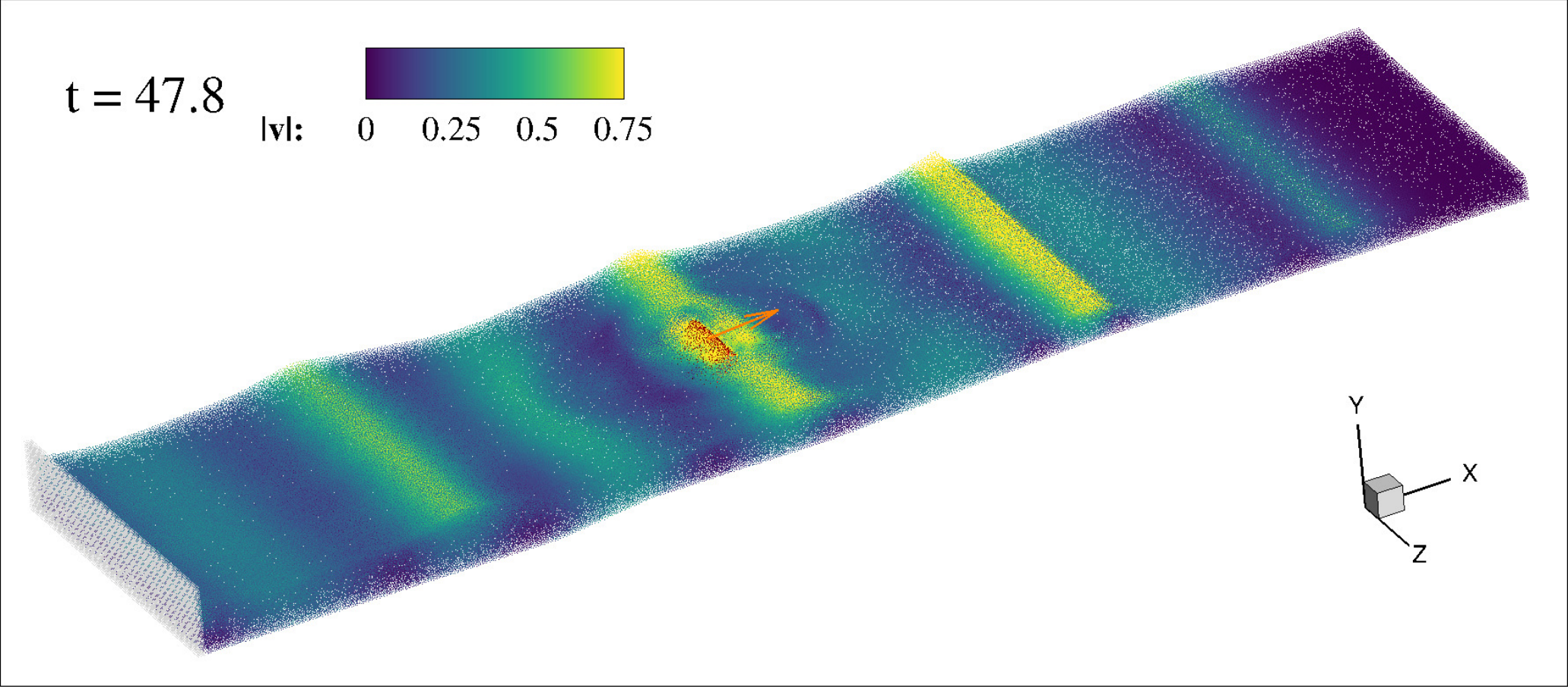}\\
		\includegraphics[trim = 2mm 2mm 2mm  2mm, clip, width=\textwidth]{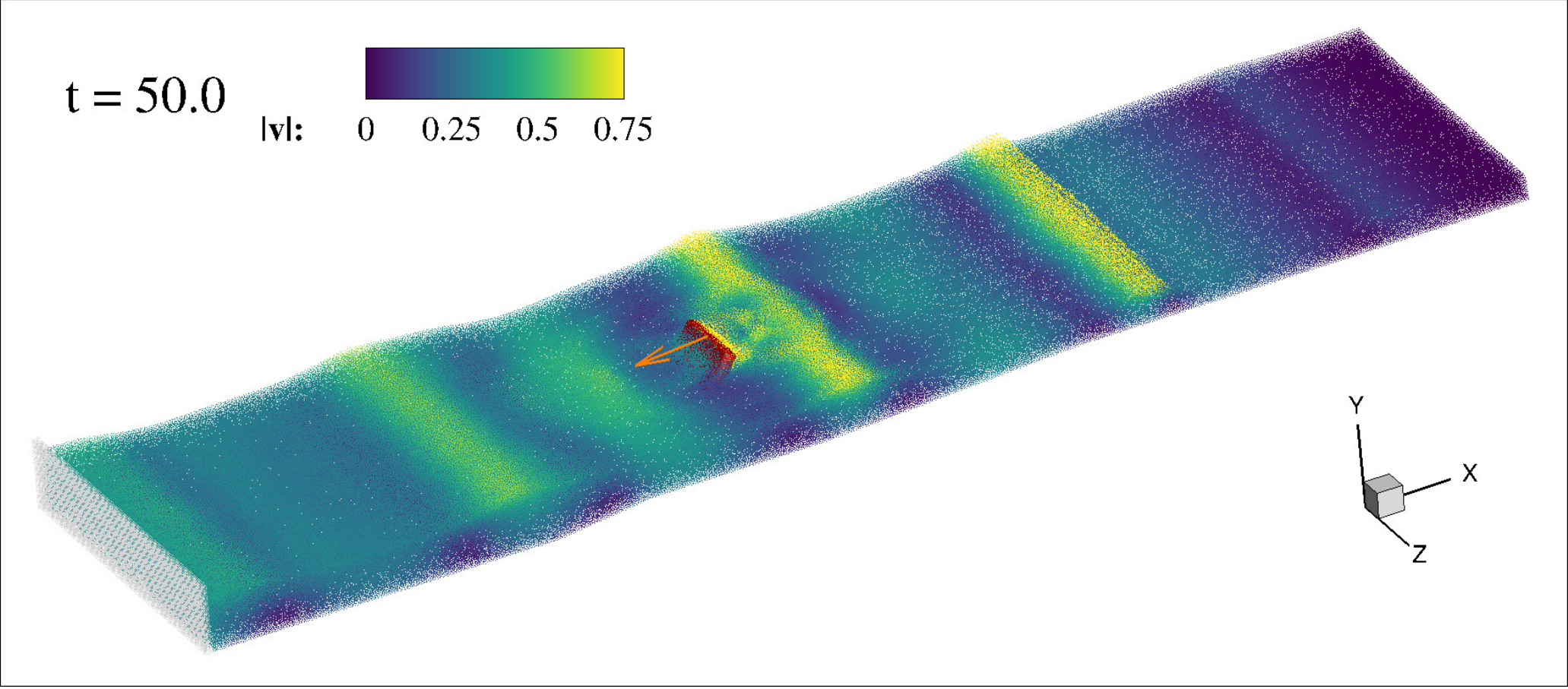}\\
		\includegraphics[trim = 2mm 2mm 2mm  2mm, clip, width=\textwidth]{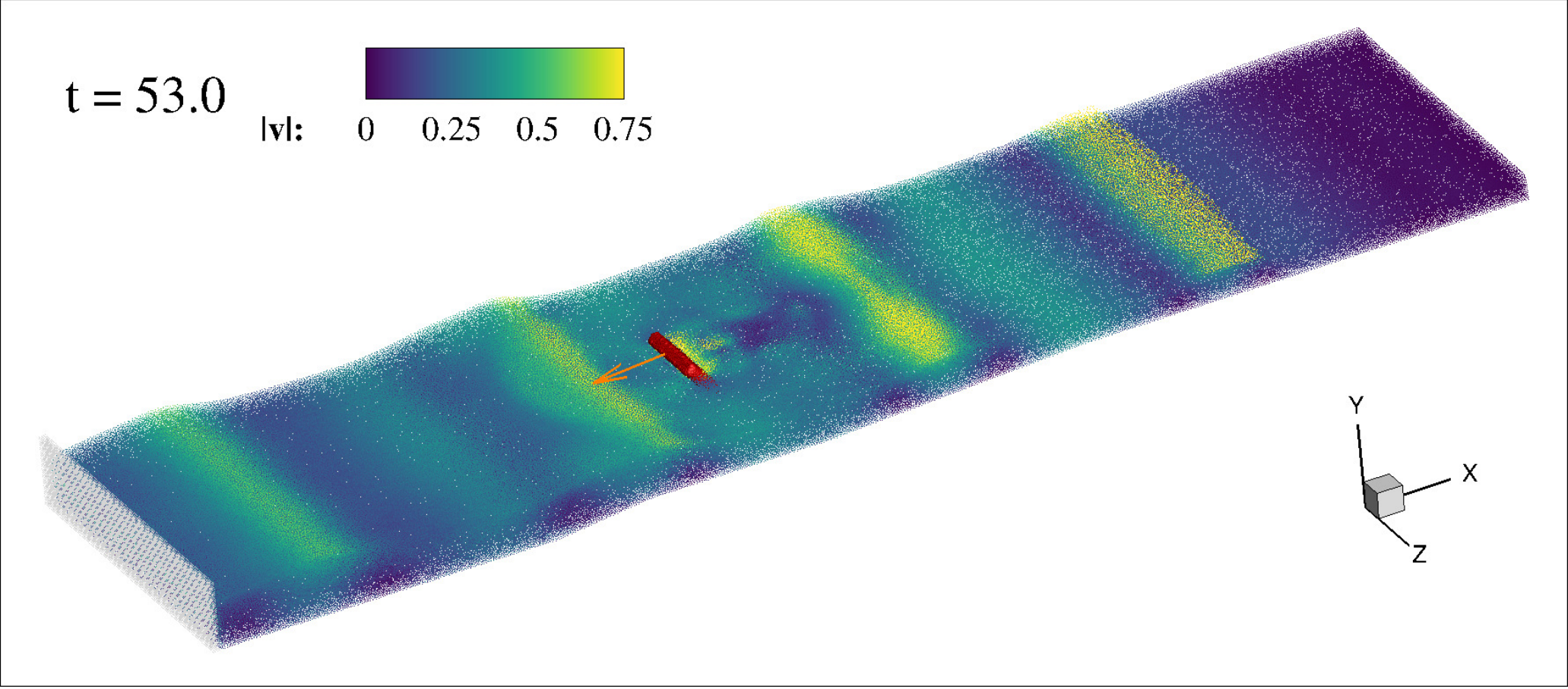}
		\caption{Velocity magnitude}
		\label{figs:owsc-surface-v}
	\end{subfigure}
	\caption{Modeling of OWSC with SPHinXsys: Free surfaces and the flap motion for wave height $H = 5.0 \text{m}$ and wave period $T = 10.0 \text{s}$. 
		(a) Fluid particles are colored by normalized pressure and (b) fluid particles are colored by velocity magnitude (For color interpretation, the reader is referred to the web version of this paper).}
	\label{figs:owsc-surface}
\end{figure*}

Figure \ref{figs:owsc-surface} presents several snapshots showing the free surface colored 
by normalized pressure \ref{figs:owsc-surface-p} and velocity magnitude \ref{figs:owsc-surface-v}, 
and the flap's rotation predicted by the present solver. 
It can be observed that smooth pressure and velocity fields are produced even when complex interactions between the wave and the flap are involved. 
It also worth noting that wave refection and breaking can be seen in the region near the flap during the interaction. 
These phenomena can be clearly observed in the time history of the wave elevation presented in the following. 

\begin{figure*}
	\centering
	%add desired spacing between images, e. g. ~, \quad, \qquad, \hfill etc. 
	\begin{subfigure}[b]{0.95\textwidth}
		\centering
		\includegraphics[trim = 2mm 2mm 2mm 2mm, clip, width=0.5\textwidth]{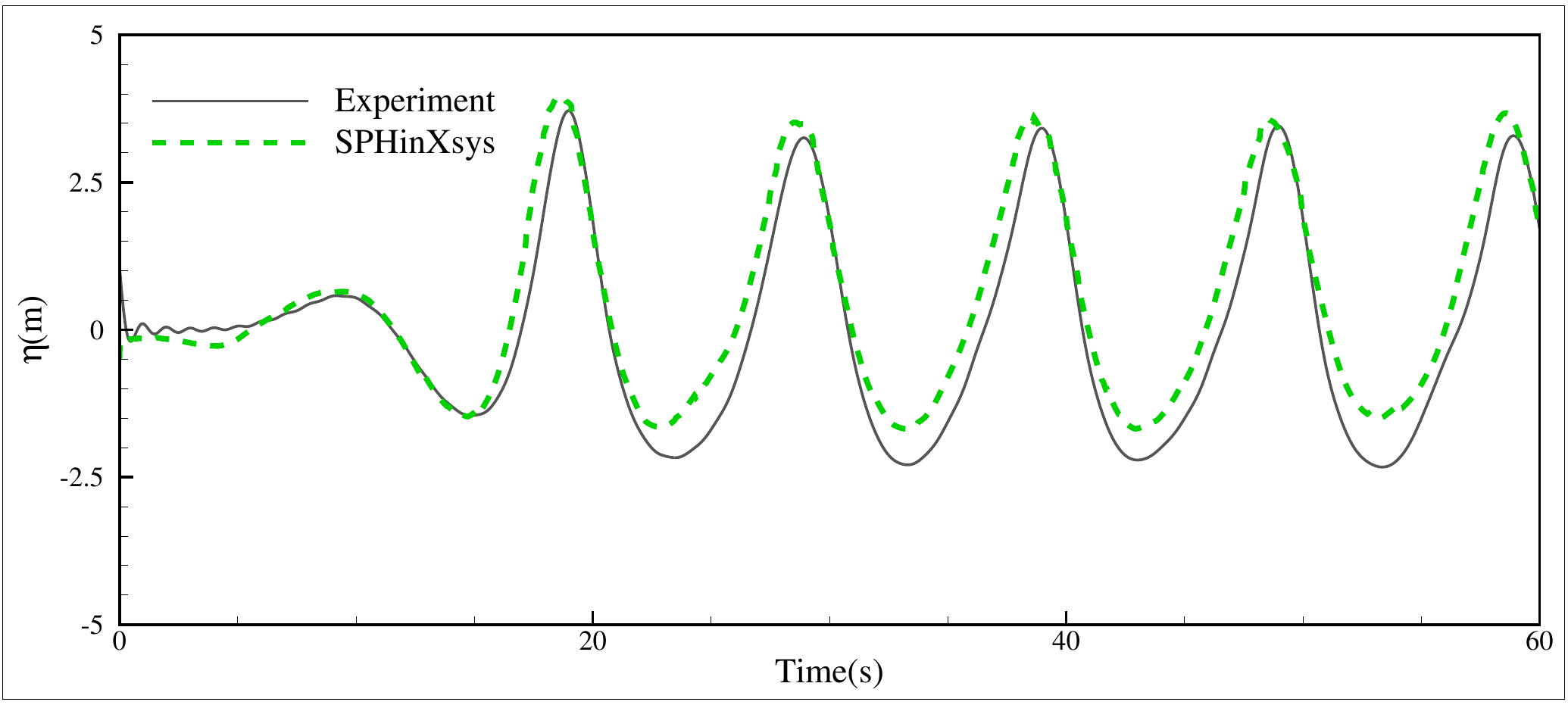}
		\caption{Wave probe $\text{WP}04$}
		\label{figs:owsc-wp-04}
	\end{subfigure}
	\newline 
	\begin{subfigure}[b]{0.49\textwidth}
		\centering
		\includegraphics[trim = 2mm 2mm 2mm  2mm, clip, width=0.95\textwidth]{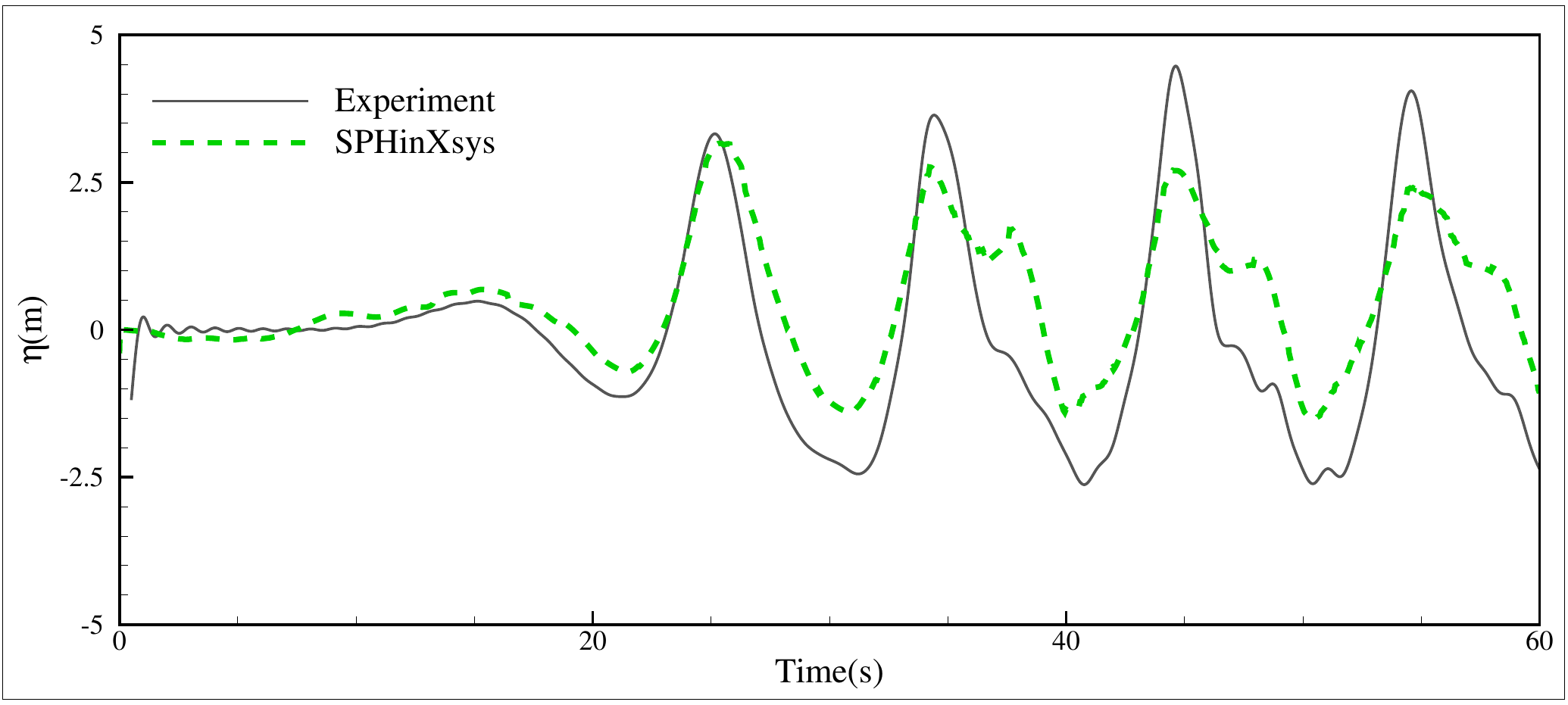}
		\caption{Wave probe $\text{WP}05$}
		\label{figs:owsc-wp-05}
	\end{subfigure}
	\begin{subfigure}[b]{0.49\textwidth}
	\centering
	\includegraphics[trim = 2mm 2mm 2mm  2mm, clip, width=0.95\textwidth]{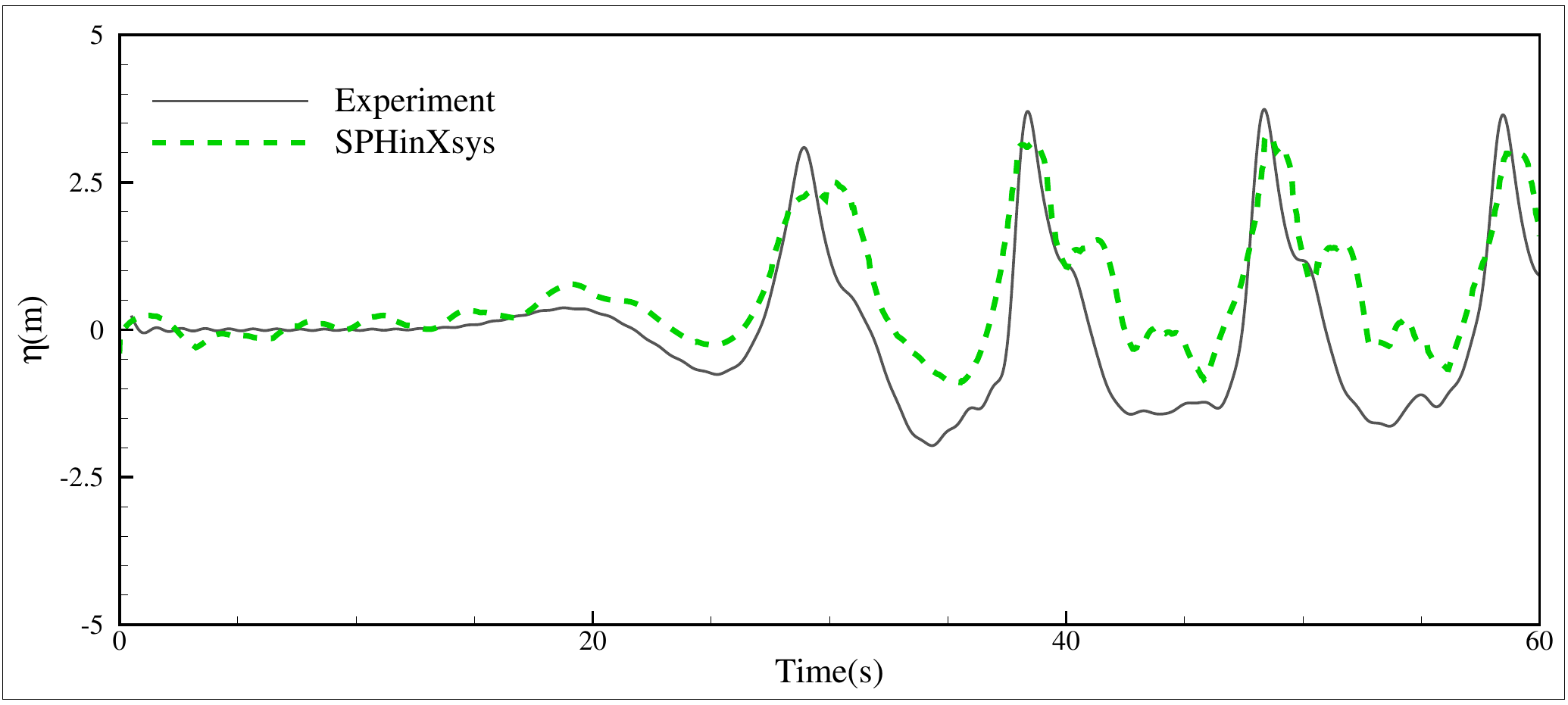}
	\caption{Wave probe $\text{WP}12$}
	\label{figs:owsc-wp-12}
\end{subfigure}
	\caption{Modeling of OWSC with SPHinXsys: Comparison of free surface elevations for wave height $H = 5.0 \text{m}$ and wave period $T = 10.0 \text{s}$. 
	The wave probe number : $\text{WP}04$, $\text{WP}05$ and$\text{WP}12$ and their corresponding locations are given in Table \ref{tab:wp}.}
\label{figs:owsc-wp}
\end{figure*}

Figure \ref{figs:owsc-wp} shows the comparison of the present numerical prediction and the laboratory observation \cite{wei2015wave} 
for the time histories of the water elevation at probes $\text{WP}04$, $\text{WP}05$ and $\text{WP}12$. 
For $\text{WP}04$ which is in the seaward of the flap and $3.93 \text m$ far away from it, 
a good agreement with the experimental data is noted. 
For $\text{WP}05$ and $\text{WP}12$ which measure the wave elevation right before and after wave passing the flap, 
slight discrepancies are noted due to the wave reflection and breaking. 
It is also interesting to compare the present results with those obtained with ANSYS FLUENT in Ref. \cite{wei2015wave}. 
In present work, 
the wave breaking phenomena are more visible compared with those of FLUENT (see Figure 11 (b) in Ref. \cite{wei2015wave}) 
and this is due to the Lagrangian nature of SPH method. 
%
%%%%%%%%%%%%%%%%%%%%%%%%%%%%%%%%%%%%%%%%%%%%%%%%%%%%%%%%%%%%%
% Section
%%%%%%%%%%%%%%%%%%%%%%%%%%%%%%%%%%%%%%%%%%%%%%%%%%%%%%%%%%%%%
\subsubsection{Wave induced rotation of the flap}\label{sec:flaprotation}
\begin{figure*}
	\centering
	%add desired spacing between images, e. g. ~, \quad, \qquad, \hfill etc. 
	\begin{subfigure}[b]{0.95\textwidth}
		\centering
		\includegraphics[trim = 2mm 2mm 2mm 2mm, clip, width=0.85\textwidth]{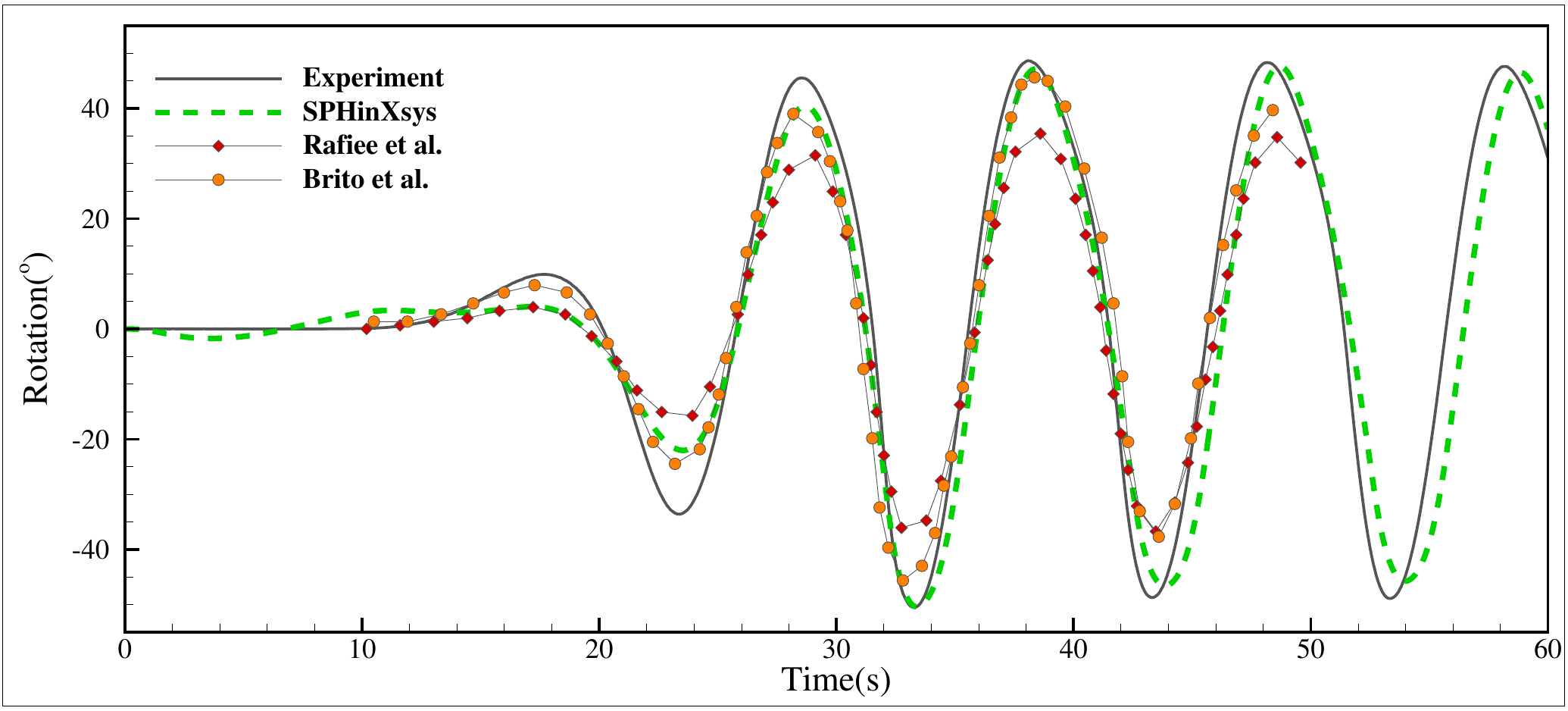}
		\caption{Rotation of the flap}
		\label{figs:owsc-rotation}
	\end{subfigure}
	\newline 
	\begin{subfigure}[b]{\textwidth}
		\centering
		\includegraphics[trim = 2mm 2mm 2mm  2mm, clip, width=0.495\textwidth]{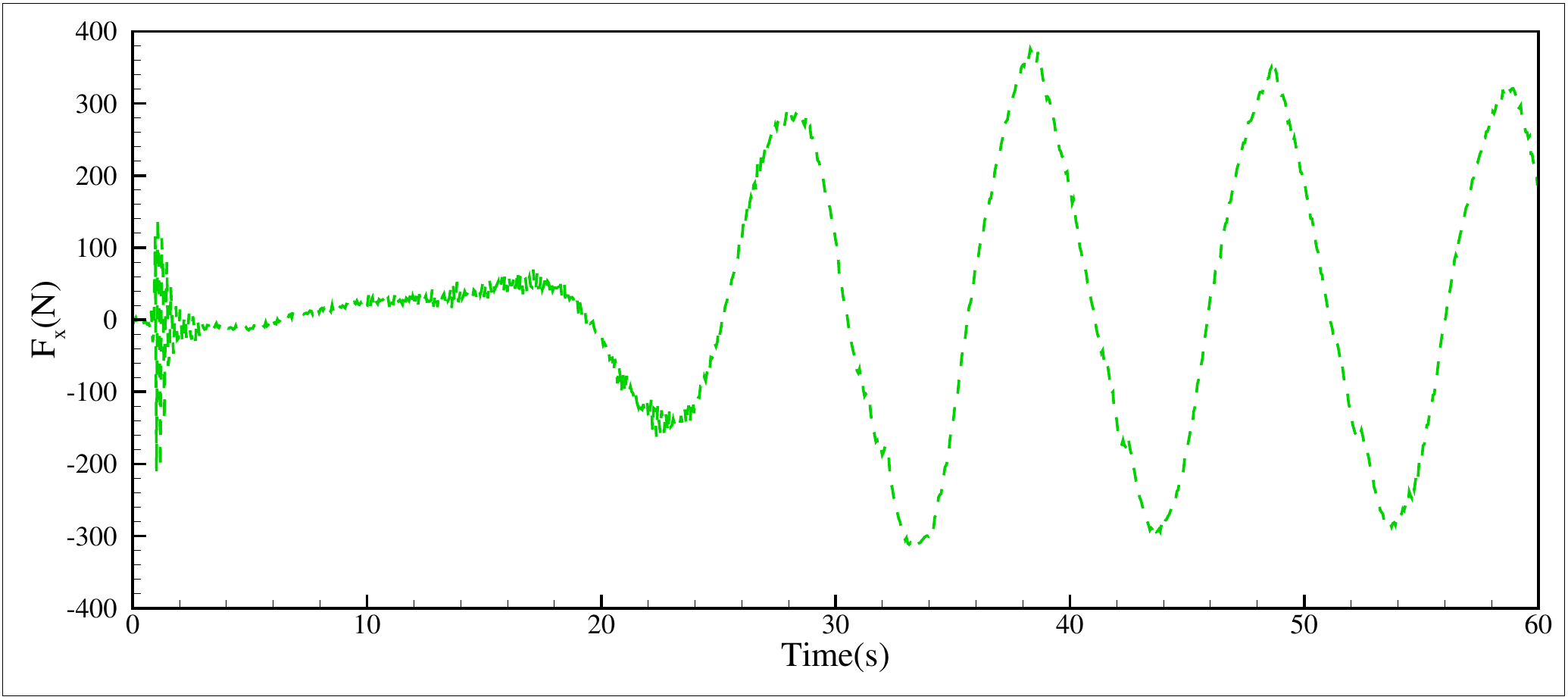}
		\includegraphics[trim = 2mm 2mm 2mm  2mm, clip, width=0.495\textwidth]{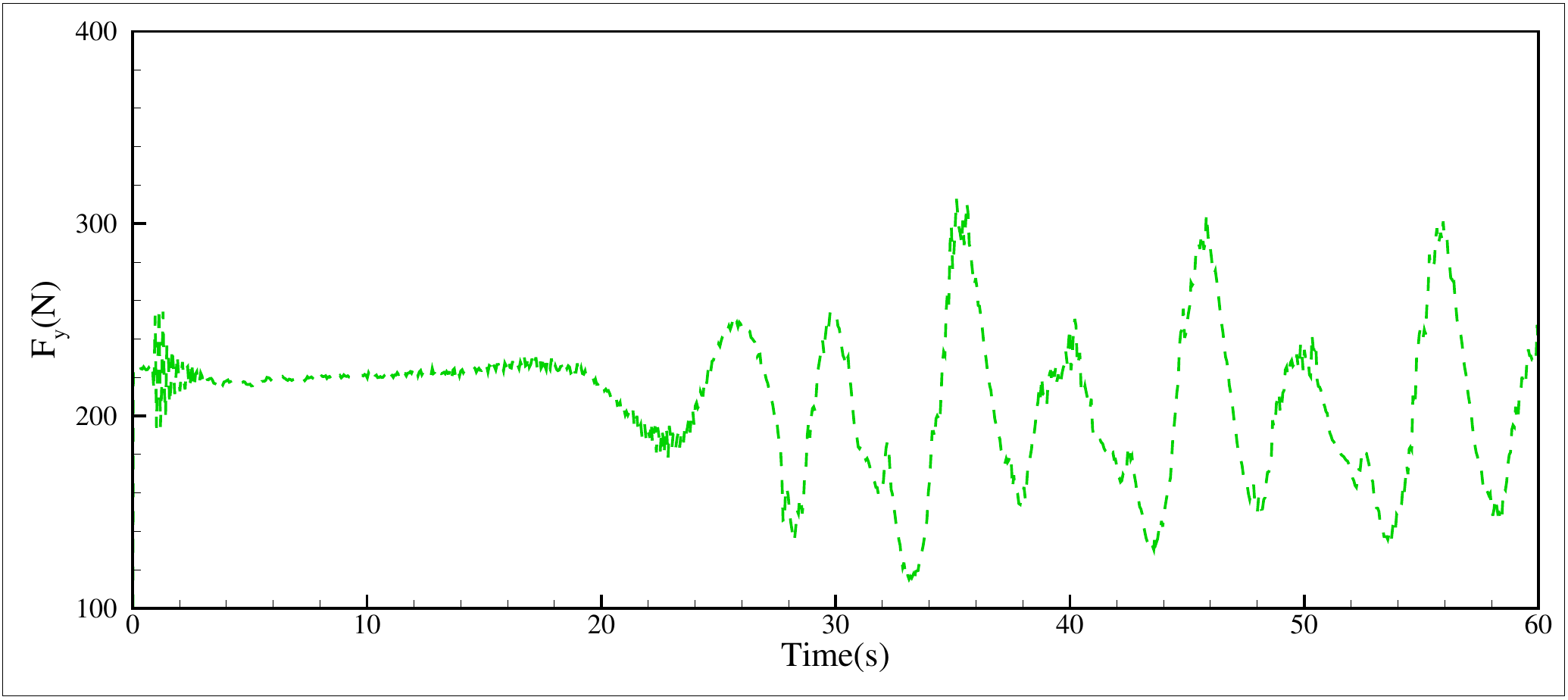}
		\caption{Total force}
		\label{figs:owsc-force}
	\end{subfigure}
	\caption{Modeling of OWSC with SPHinXsys: Comparison of the flap rotation (a)  and 
		the time histories of the total force exerted on flap in $x$-axis (left panel of (b)) and $y$-axis (right panel of (b)). }
	\label{figs:owsc-rotation-force}
\end{figure*}

Figure \ref{figs:owsc-rotation-force} 
shows the comparison of the time history of the flap rotation between the present numerical results and the experimental data 
and also numerical data in literature using SPH-based solver \cite{dias2017analytical, rafiee2013numerical, henry2013characteristics, brito2016coupling}, 
and the corresponding numerical predicted total force on the flap.  
The comparison in Figure \ref{figs:owsc-rotation} shows that the present solver accurately predicted the large amplitude rotation of the flap 
implying a robust and stable feature of the coupling between SPHinXsys and Simbody for WSI applications. 
Compared with results using other SPH-based solver, 
e.g., UCD-SPH \cite{rafiee2013numerical, henry2013characteristics,dias2017analytical} and DualSPHysics \cite{brito2016coupling}, 
the present solver shows improved accuracy in predicting the flap rotation. 
As noted by Wei et al. \cite{wei2015wave}, 
the flap's large amplitude rotation is hard to handle by mesh-based methods which require complex mesh technique for moving interface, 
however, 
the present solver shows its promising capability of  dealing with these difficulties thanks to its very meshless nature. 

Figure \ref{figs:owsc-force} shows the time history of total force on flap in $x$- and $y$-axis. 
As expected, 
the force component in $x$-axis as the main force driving flap rotation shows very similar varying pattern with the rotation profile, 
on other hand, 
the $y$-axis component representing the hydrostatic pressure shows fluctuation due the wave refection and breaking.  
%
%%%%%%%%%%%%%%%%%%%%%%%%%%%%%%%%%%%%%%%%%%%%%%%%%%%%%%%%%%%%%
% Section
%%%%%%%%%%%%%%%%%%%%%%%%%%%%%%%%%%%%%%%%%%%%%%%%%%%%%%%%%%%%%
\subsubsection{Wave loads on the flap}\label{sec:flappressure} 
\begin{figure*}
	\centering
	\includegraphics[trim = 2mm 2mm 2mm  2mm, clip, width=0.495\textwidth]{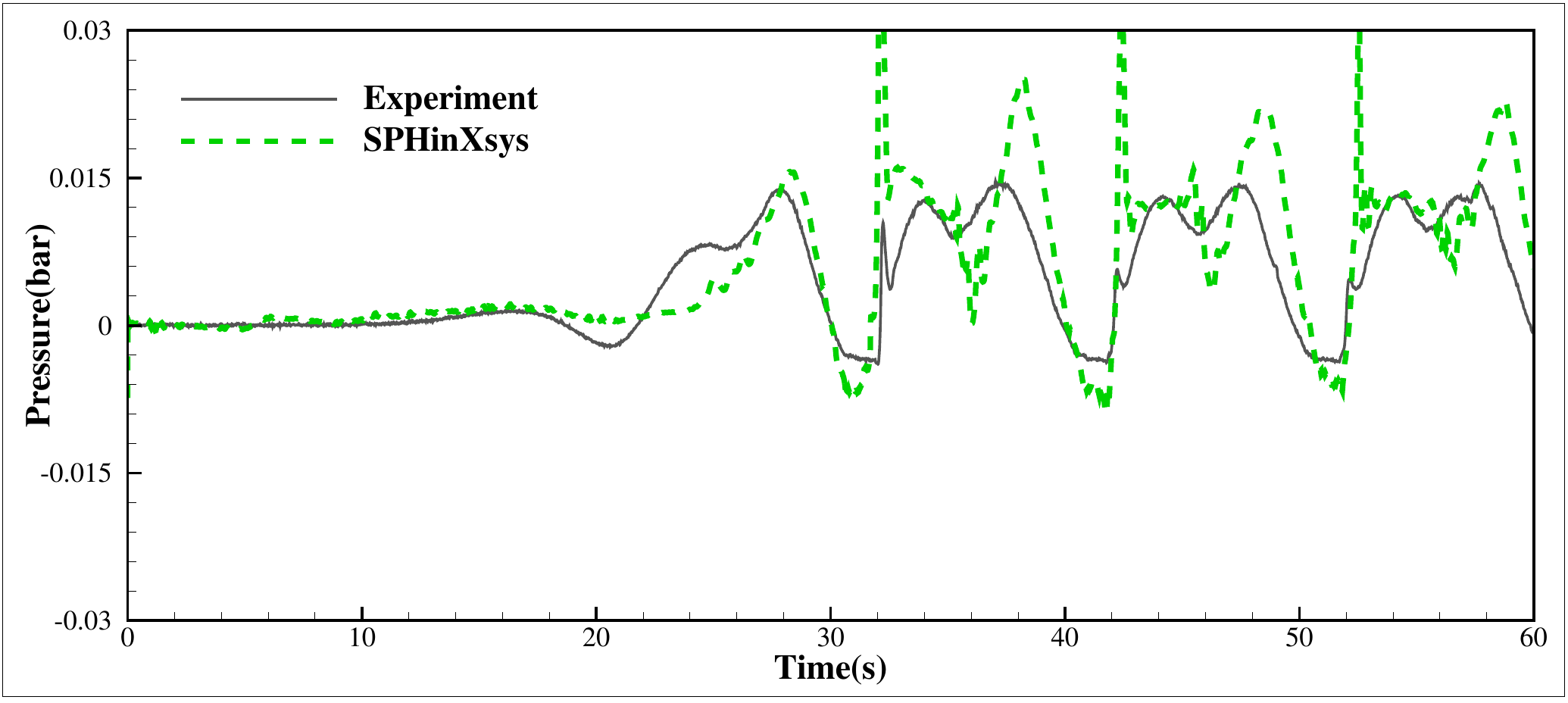}
	\includegraphics[trim = 2mm 2mm 2mm  2mm, clip, width=0.495\textwidth]{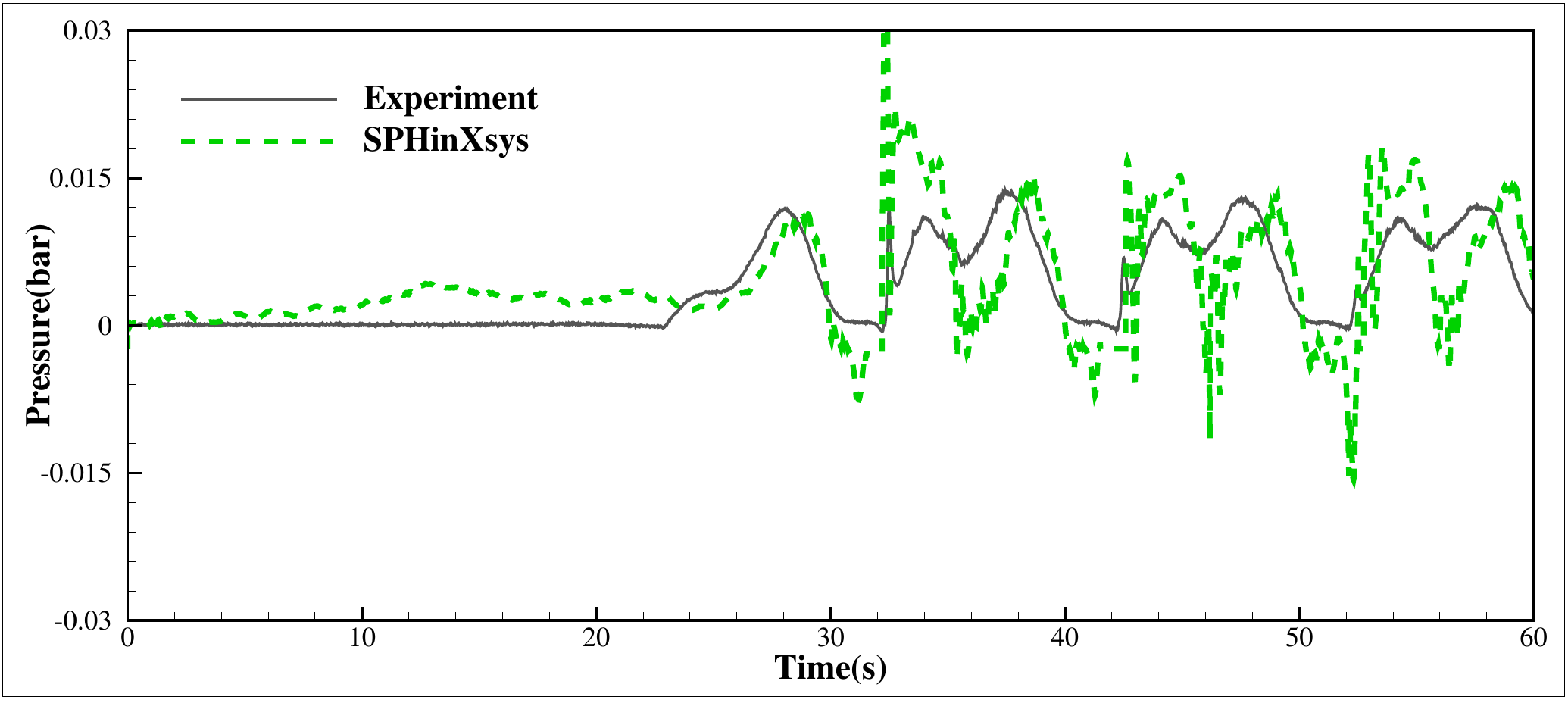}\\
	\includegraphics[trim = 2mm 2mm 2mm  2mm, clip, width=0.495\textwidth]{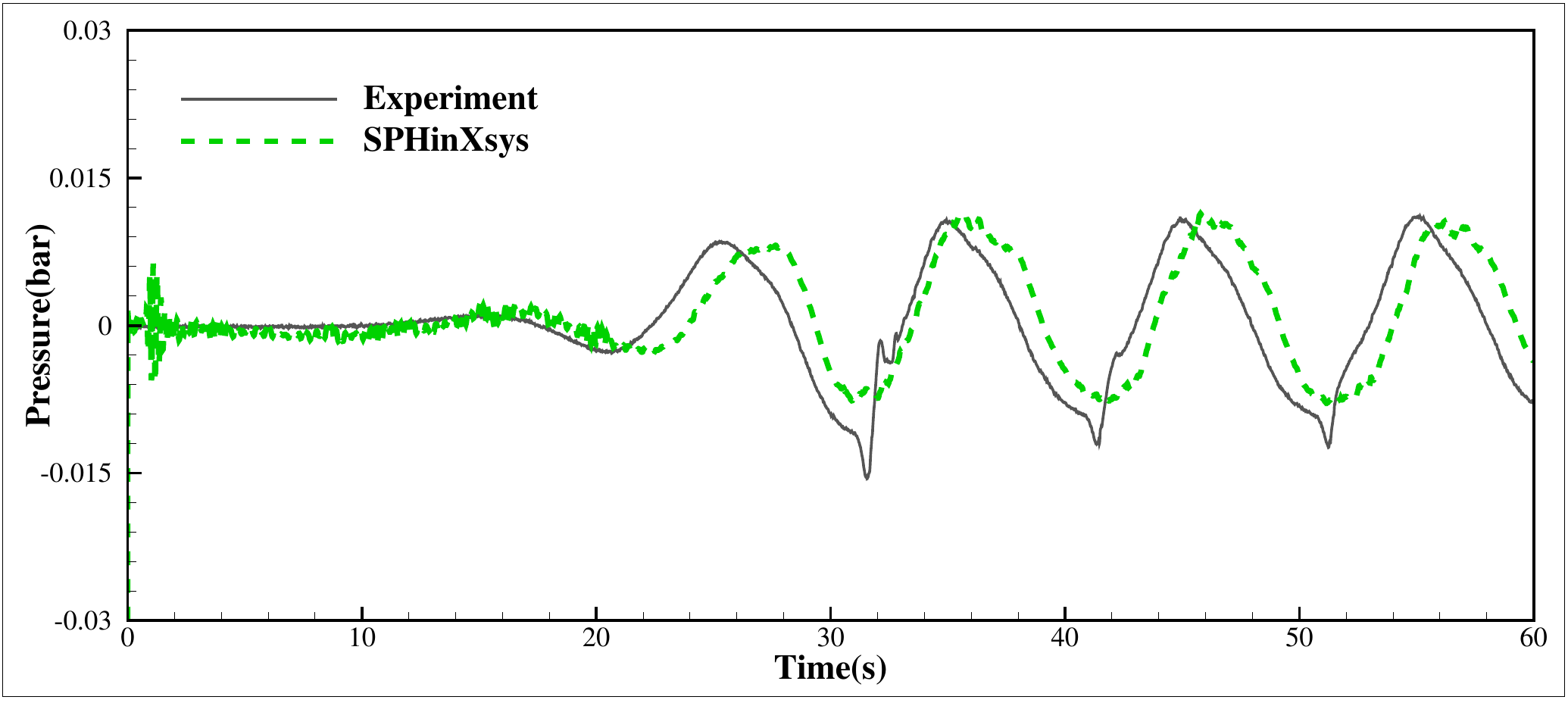}
	\includegraphics[trim = 2mm 2mm 2mm  2mm, clip, width=0.495\textwidth]{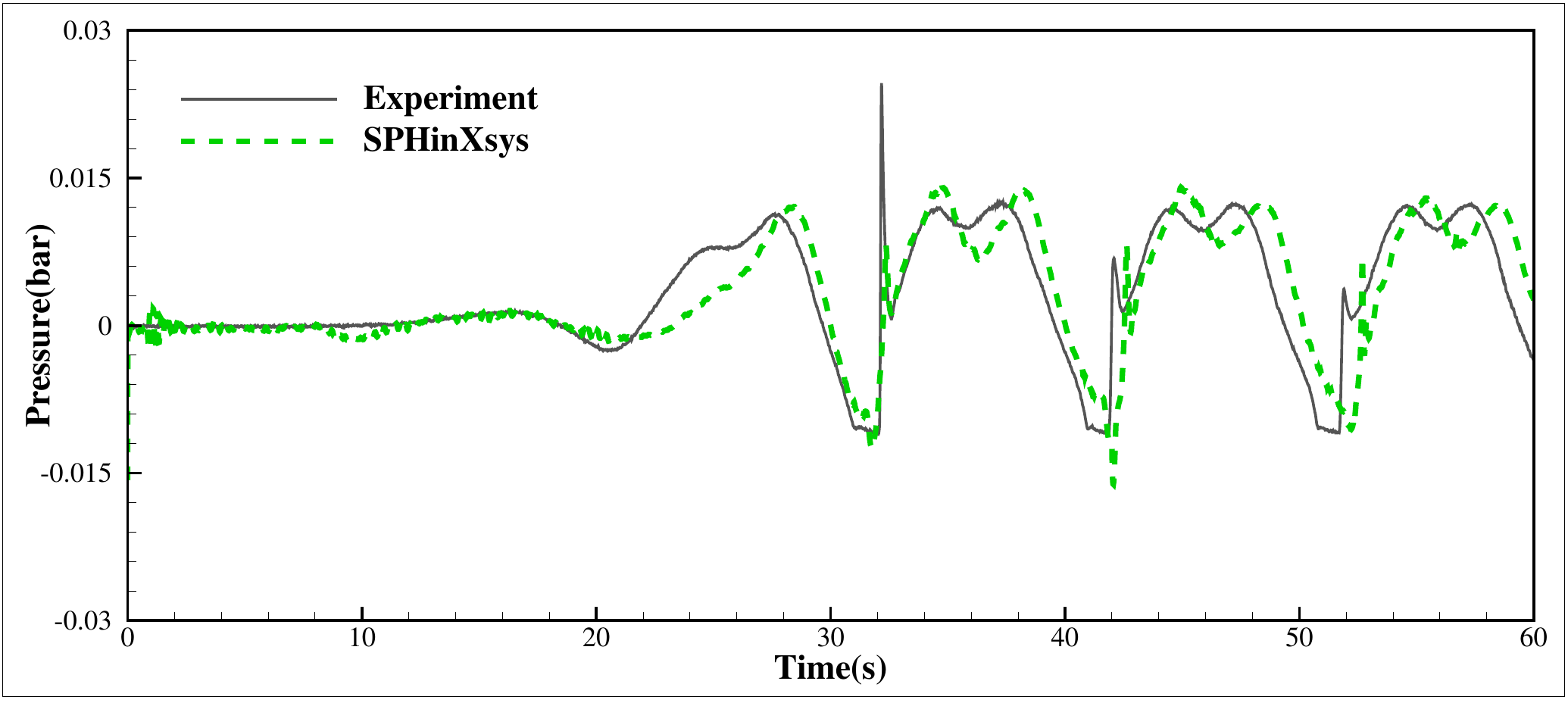}\\
	\includegraphics[trim = 2mm 2mm 2mm  2mm, clip, width=0.495\textwidth]{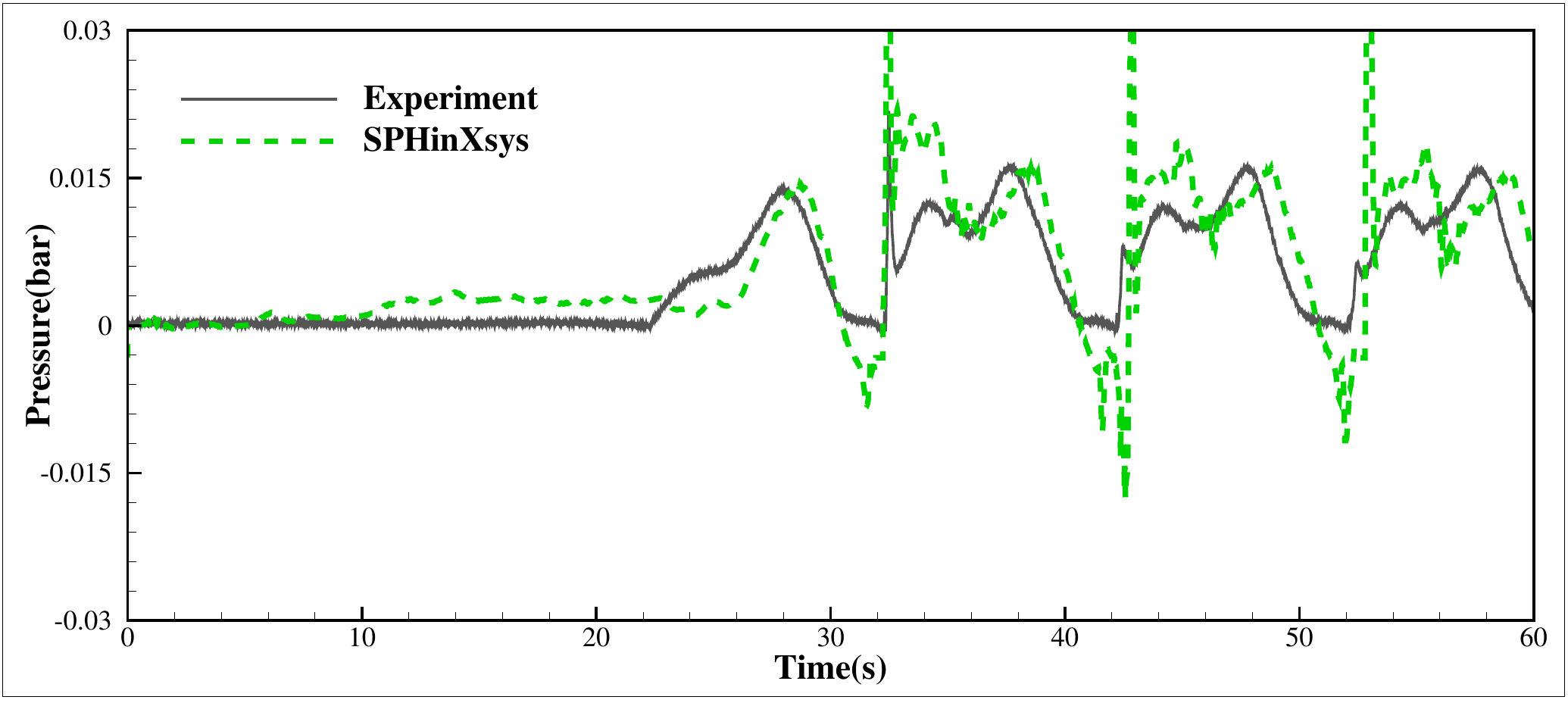}
	\includegraphics[trim = 2mm 2mm 2mm  2mm, clip, width=0.495\textwidth]{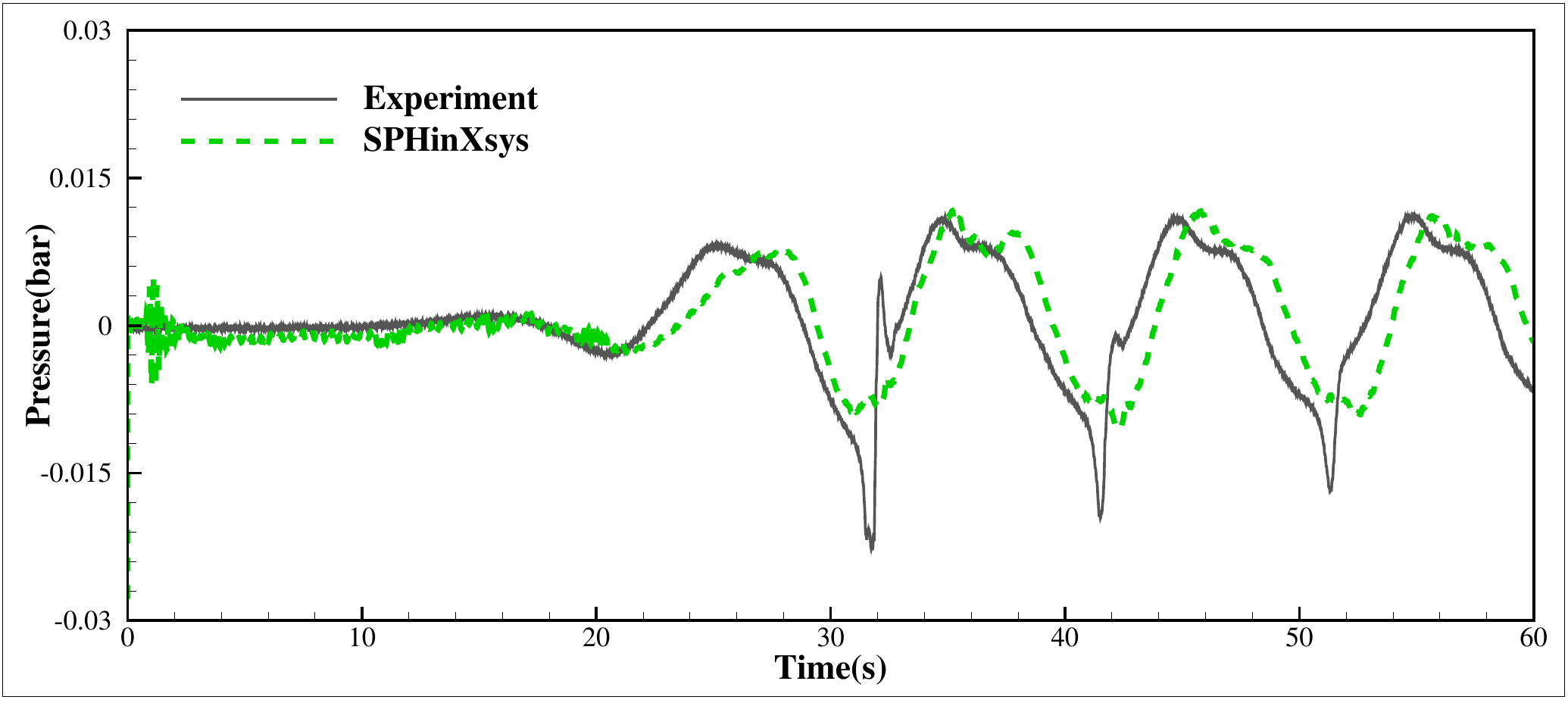}
	\caption{Modeling of OWSC with SPHinXsys: Comparison of the time histories of wave loads on the flap for wave height $H = 5.0 \text{m}$ and wave period $T = 10.0 \text{s}$. 
		The pressure sensor number (from left to right) : $\text{PS}01$ and$\text{PS}03$ (top panel). $\text{PS}05$ and $\text{PS}09$ (middle panel); 
		$\text{PS}11$ and $\text{PS}13$ (bottom panel).}
	\label{figs:owsc-pt}
\end{figure*}

To optimize the structure design of OWSC, 
it is also of significant importance to accurately predict the wave loads on its main operating device, 
e.g., the flap in this study. 
Figure \ref{figs:owsc-pt} shows the time histories of the pressure signals recorded on selected sensors given in Table \ref{tab:ps} 
and its comparison with the experimental data \cite{wei2015wave}. 
Note that the selected sensors are distributed at the top ($\text{PS}03$ and $\text{PS}11$), middle ($\text{PS}05$ and $\text{PS}09$) 
and bottom ($\text{PS}05$ and $\text{PS}13$) of the flap. 
Following Wei et al. \cite{wei2015wave}, 
the initial hydrostatic pressure is subtracted from the recorded pressure signals inducing a  negative drops in some profiles. 
Compared with the experimental data, 
the main plateaus of all the pressure profiles are reasonably well captured by the present solver, 
even if slight discrepancies are noted. 
For sensors $\text{PS}01$, $\text{PS}03$ and $\text{PS}11$, 
large pressure peaks and drops are noted 
since the present SPH model is based on weakly-compressible assumption and air cushion effects are not captured in mono-fluid simulation. 
Similar to the results reported in Ref. \cite{wei2015wave},  
the pressure drops for sensors $\text{PS}05$ and $\text{PS}13$ are underestimated in present results and 
theses discrepancies are related the the wave breaking which results a higher crest elevation compared with experiment as shown in Figure \ref{figs:owsc-surface}. 
Also note that theses discrepancies are also likely to be associated with the stochastic nature of the impact pressures and the lack of exact repeatability of the experiments. 

Compared with the numerical results obtained with ANSYS FLUENT reported in Ref. \cite{wei2015wave}, 
the present pressure peaks show large magnitudes and fluctuations duo to its Lagrangian nature and weakly-compressible assumption. 
It worth noting that these large peak magnitudes are not observed in the results obtained by UCD-SPH code \cite{dias2017analytical, henry2013characteristics, rafiee2013numerical} 
and this may due to the fact that UCD-SPH applied HLLC (Harten Lax and van Leer-Contact) 
Riemann solver for Riemann-based WCSPH \cite{rafiee2012comparative} which induced excessive dissipation \cite{zhang2017weakly}. 
Also, 
it is interesting to conduct SPH simulation with consideration of air cushion to investigate its effects on pressure peak, 
and it will be the main objective of our future work. 
%
%%%%%%%%%%%%%%%%%%%%%%%%%%%%%%%%%%%%%%%%%%%%%%%%%%%%%%%%%%%%%
% Section
%%%%%%%%%%%%%%%%%%%%%%%%%%%%%%%%%%%%%%%%%%%%%%%%%%%%%%%%%%%%%
\subsection{Computational efficiency}\label{sec:computational-efficiency} 
To rigorously assess the computational performance, 
we analyze the total CPU time of the present simulation and those or approximated ones reported in literature 
with different computational models, 
including commercial software package ANSYS FLUENT \cite{wei2015wave}, 
in-house UCD-SPH code \cite{rafiee2013numerical, dias2017analytical}, 
DualSPHysics \cite{crespo2015dualsphysics, brito2020numerical} and GPUSPH\cite{wei2019modeling}. 
For comparison, 
the total CPU time is normalized by total physical time and total cell or particles numbers. 
The UCD-SPH code was developed by Dias and Rafiee \cite{rafiee2013numerical, dias2017analytical} and parallelized based on OpenMP. 
As for DualSPHysics which is implemented with GPU acceleration,
the total CPU time can be approximated by considering the corresponding speedup reported in Ref. \cite{crespo2015dualsphysics}. 

In present work, 
the computations are carried out on an Intel Xeon CPU E5-2620 v3 2.40GHz Desktop computer with 64GB RAM and Scientific Linux system (7.8). 
The ANSYS FLUENT and UCD-SPH code are performed on the Stokes cluster of the Irish Centre for High-End Computing (ICHEC) 
and the Stokes is a SGI Altix ICE 8200EX cluster with 320 compute nodes. 
Each node has two Intel (Westmere) Xeon E5650 hexa-core processors and $24$ GB of RAM 
and the nodes are interconnected via two planes of ConnectX Infiniband (DDR).

Table \ref{tab:computational-time} reports the computation time and the normalized CPU time with the corresponding physical time, 
cell or particle number and system information for different models.  
Although the CPU versions are different,  
there is no doubt that the present solver shows impressive computational performance for three-dimensional large scale simulations. 
Also, 
this performance can be further improved by implementing GPU acceleration which is our ongoing work and will be released soon.  
Note that as an open-source library, 
the code and data-sets accompanying this validation is available on the repository of SPHinXsys in GitHub at \url{https://github.com/Xiangyu-Hu/SPHinXsys}.
\begin{table*}[htb!]
	\renewcommand{\thefootnote}{\roman{footnote}}
	%\tiny
	\small
	\centering
	\caption{Computational efficiency for different models reported in literature. 
				Here, the CPU time denotes the total CPU time for computation of $1\text s$ physical time with $1$million particles or cells. }
	\begin{tabular}{cccccc}
		\hline
		Model   & \begin{tabular}{c} Computing\\ time (hour) \end{tabular} & \begin{tabular}{c} Physical\\ time (s) \end{tabular}   & 
		\begin{tabular}{c} Resolution \\ (million) \end{tabular} &   Device & \begin{tabular}{c} CPU time \\ (h/s/M)  \end{tabular} \\ 
		\hline
		FLUENT \cite{wei2015wave} &$48 \text h$  & $13 \text s$ & \begin{tabular}{c} $1.5$ M  \\hexahedral cells \end{tabular} & \begin{tabular}{c} 24 cores  \\ Intel Xeon E5650 \end{tabular} & $59.08$  \\
		%\hline
		UCD-SPH  \cite{dias2017analytical} &$70 \text h$   & $13 \text s$ &\begin{tabular}{c} $3.24$ M  \\particles \end{tabular} & \begin{tabular}{c} 72 cores \\ Intel Xeon E5650 \end{tabular} & $119.66$  \\ 
		%\hline
		GPUSPH \cite{wei2019modeling} &$2 \text h$  & $2 \text s$ &\begin{tabular}{c} $7$ M  \\particles \end{tabular} & \begin{tabular}{c} 4 GPUs  \\ NVIDIA Tesla K80 \end{tabular} & -  \footnote{CPU information was not provided for this computation \cite{wei2019modeling}, therefore approximated data is not available.}\\ 
		DualSPHysics \cite{brito2020numerical}  &$105 \text h$       & $50 \text s$   & \begin{tabular}{c} $11.4$ M  \\particles \end{tabular}& \begin{tabular}{c} 
			Intel Xeon E5 \footnote{The Intel Xeon E5 family is composed of from 4 up to 22 cores. }
			\\ NVIDIA GTX 2080 \end{tabular}  
		& $\left( 0.74 \sim 4.05\right) \times 25$ \footnote{Note that the total CPU time for DualSPHysics is approximated by multiplying the $25$ speedup with GPU implementation given in Ref. \cite{crespo2015dualsphysics}.} \\ 
		%\hline
		SPHinXsys  &$8.5 \text h$       & $13 \text s$   & \begin{tabular}{c} $2.16$ M  \\particles \end{tabular}& \begin{tabular}{c} 12 Cores \\  Intel Xeon E5-2620 \end{tabular} & $3.63$ \\ 
		\hline
	\end{tabular}
	\label{tab:computational-time}
\end{table*}
%
%%%%%%%%%%%%%%%%%%%%%%%%%%%%%%%%%%%%%%%%%%%%%%%%%%%%%%%%%%%%%
% Section
%%%%%%%%%%%%%%%%%%%%%%%%%%%%%%%%%%%%%%%%%%%%%%%%%%%%%%%%%%%%%
\subsection{Power take-off effects and its energy capture factor}\label{sec:pto-damping}
The Simbody provides a linear damper which acts along or around any mobility coordinate to apply a generalized force there 
which can be applied to imitate the mechanical damping of the PTO operations. 
In present OWSC model, 
this process can be defined by Eq. \ref{eq:newton-euler} where a damping force proportional to angular velocity is introduced. 
Note that the damping coefficient $k_d$ represents the extent of kinematic constraint on the flap. 

Having the PTO, 
the efficiency of wave energy harvesting of OWSC  can be quantitatively described by the capture factor (CF), 
which represents the ratio of the power extracted by OWSC to the incident wave power, 
defined as
\begin{equation}\label{eq:cf}
	CF = \frac{P_w}{E_0} ,
\end{equation}
where $P_w$ denotes the time-averaged power extracted by PTO system and 
$E_0$ the time-averaged energy flux of incident wave. 
Following the work of Senol et al. \cite{senol2019enhancing}, 
the extracted power $P_w$ can be computed as
\begin{equation}\label{eq:pto}
	P_w = \frac{1}{2} \omega^2 k_d \left|\theta^2 \right| ,
\end{equation}
where $\omega$ is the angular frequency of the incident wave, 
$k_d$ the damping coefficient and
$\theta$ the rotation amplitude in the pitch component of the flap. 
Also, 
the mean energy flux $E_0$ is calculated as
\begin{equation}
	E_0 = \frac{\rho_0 \mathbf g \omega H^2}{16k} \left( 1 + \frac{2kh_0}{\sinh\left( 2kh_0 \right) } \right)  B ,
\end{equation}
where $B$ represents the width of the flap. 

\begin{figure*}
	\centering
	\includegraphics[trim = 2mm 2mm 2mm 2mm, clip, width=0.75\textwidth]{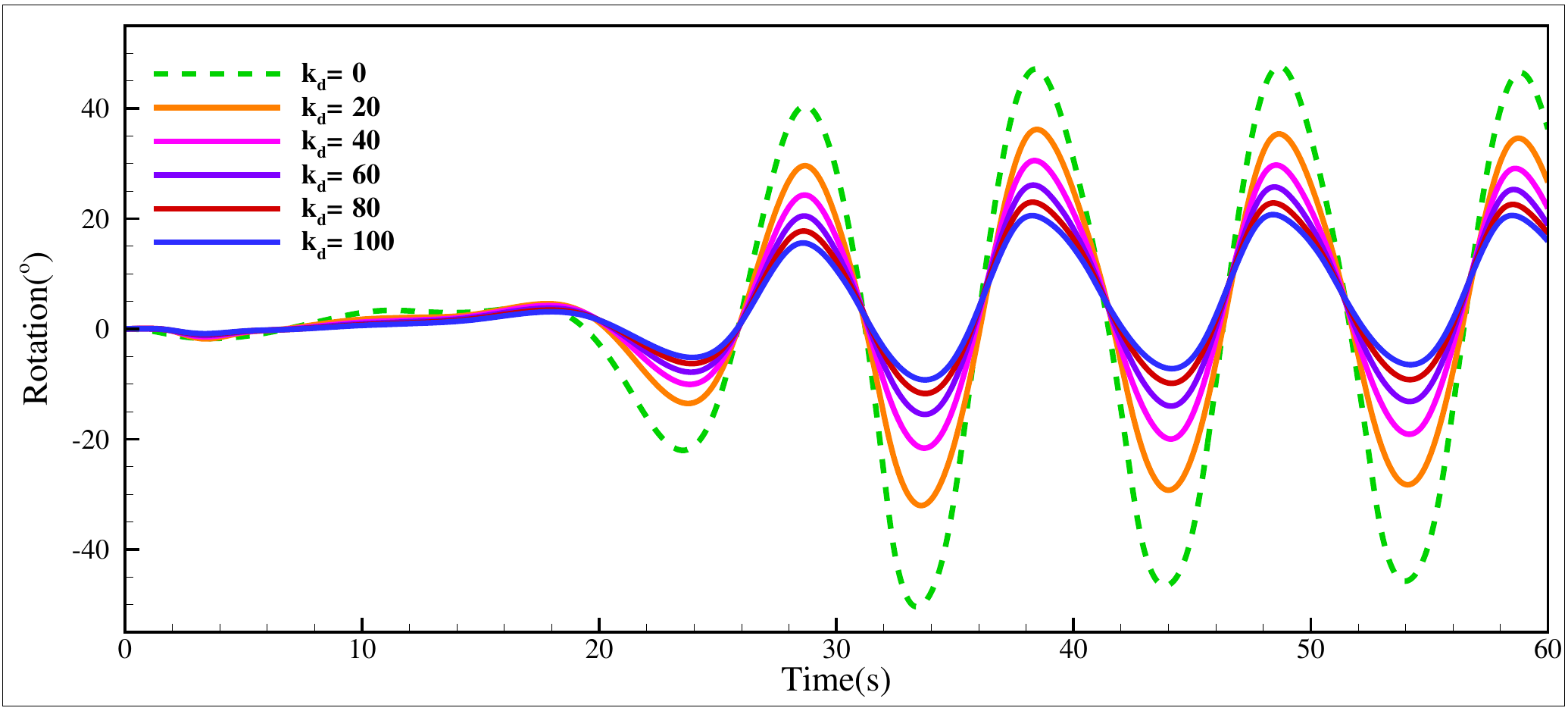}
	\caption{Modeling of OWSC with SPHinXsys: Damping effects on the rotation of the flap for wave height $H = 5.0 \text{m}$ and wave period $T = 10.0 \text{s}$ 
		(For color interpretation, the reader is referred to the web version of this paper). }
	\label{figs:owsc-damping-rotation}
\end{figure*}

\begin{figure*}
	\centering
	\includegraphics[trim = 2mm 2mm 2mm  2mm, clip, width=0.495\textwidth]{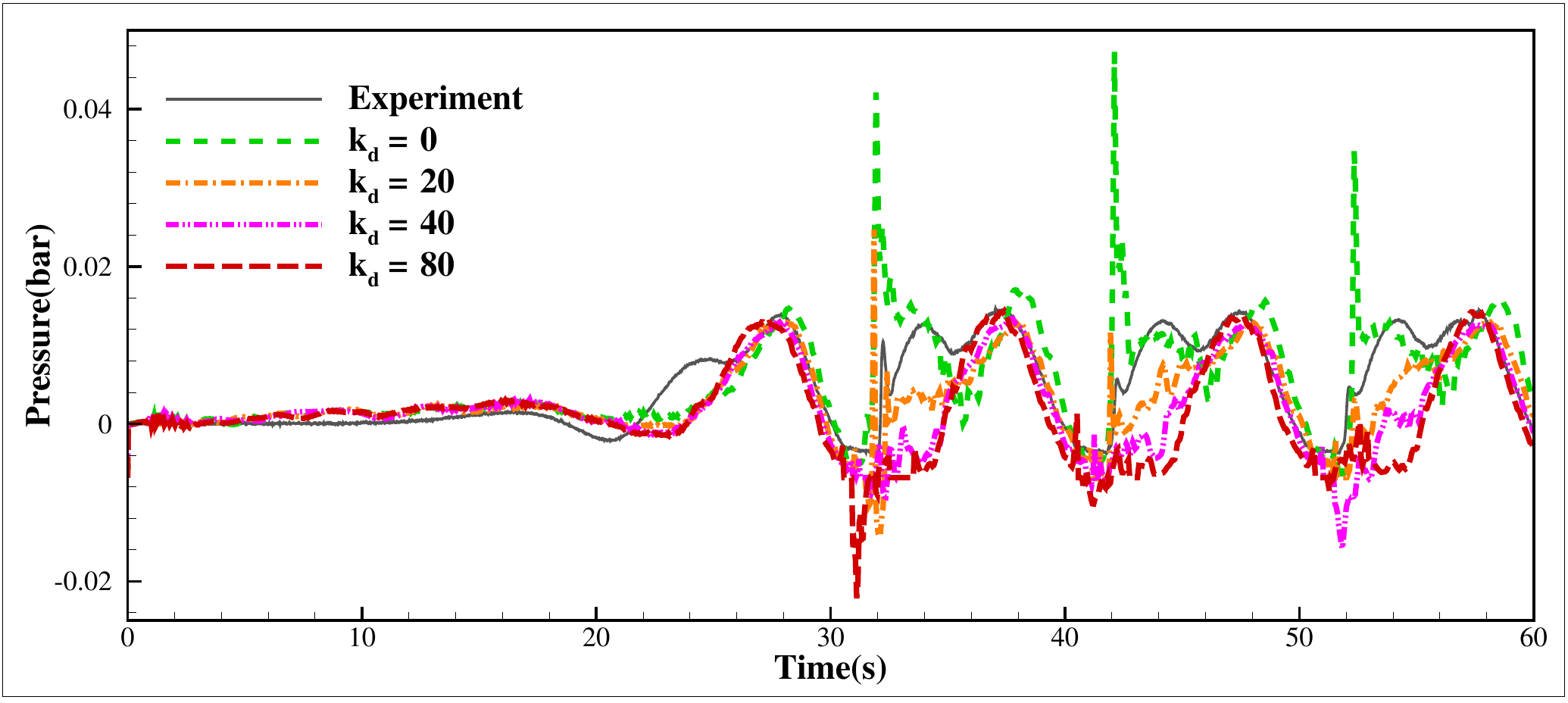}
	\includegraphics[trim = 2mm 2mm 2mm  2mm, clip, width=0.495\textwidth]{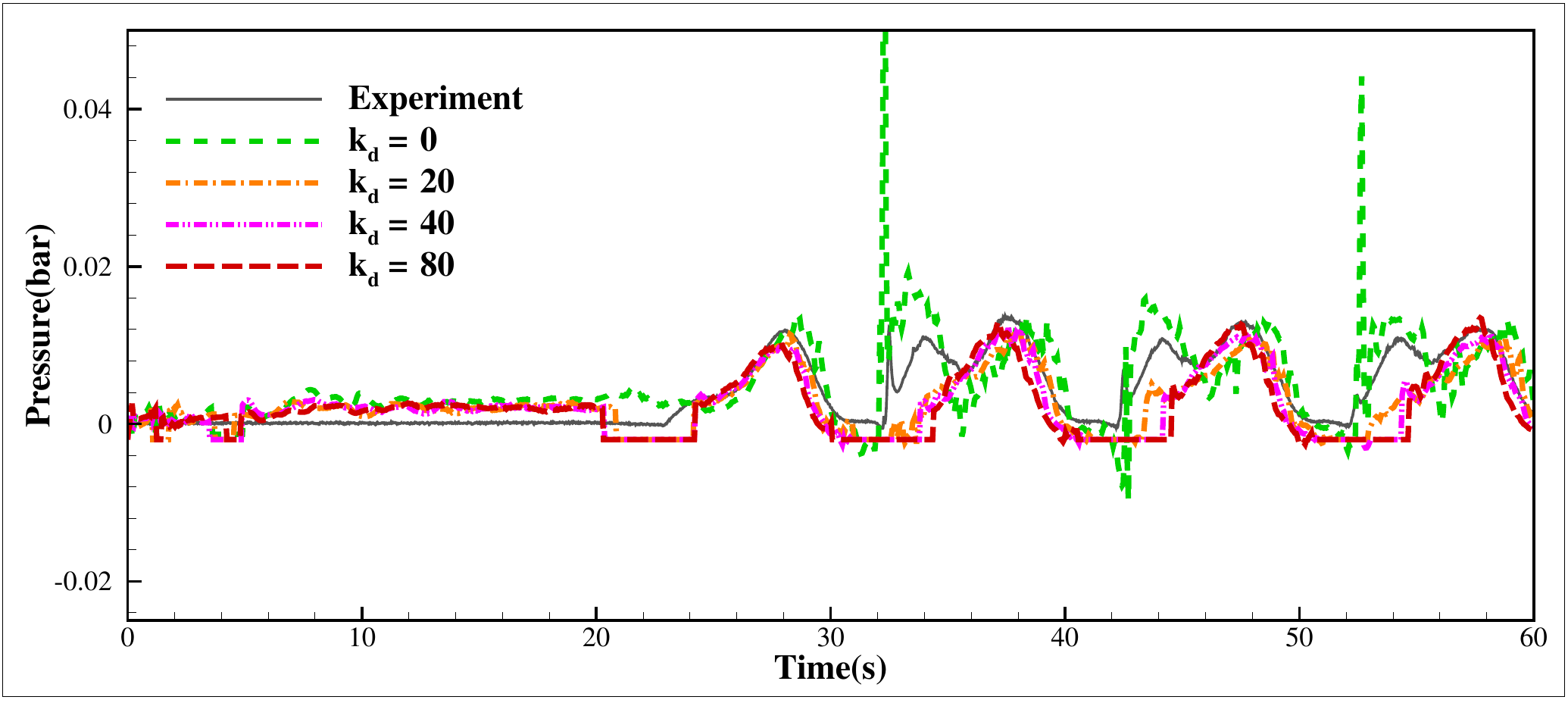}\\
	\includegraphics[trim = 2mm 2mm 2mm  2mm, clip, width=0.495\textwidth]{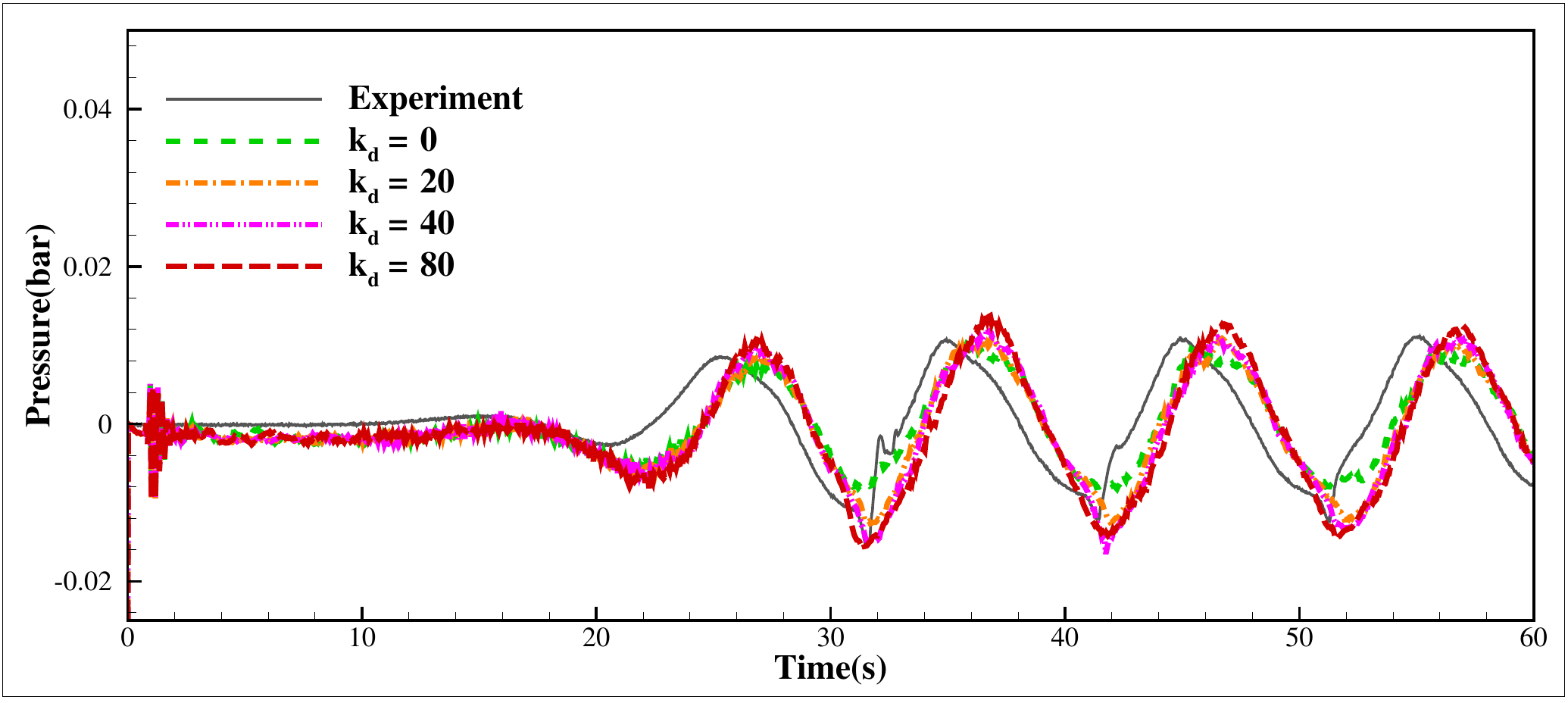}
	\includegraphics[trim = 2mm 2mm 2mm  2mm, clip, width=0.495\textwidth]{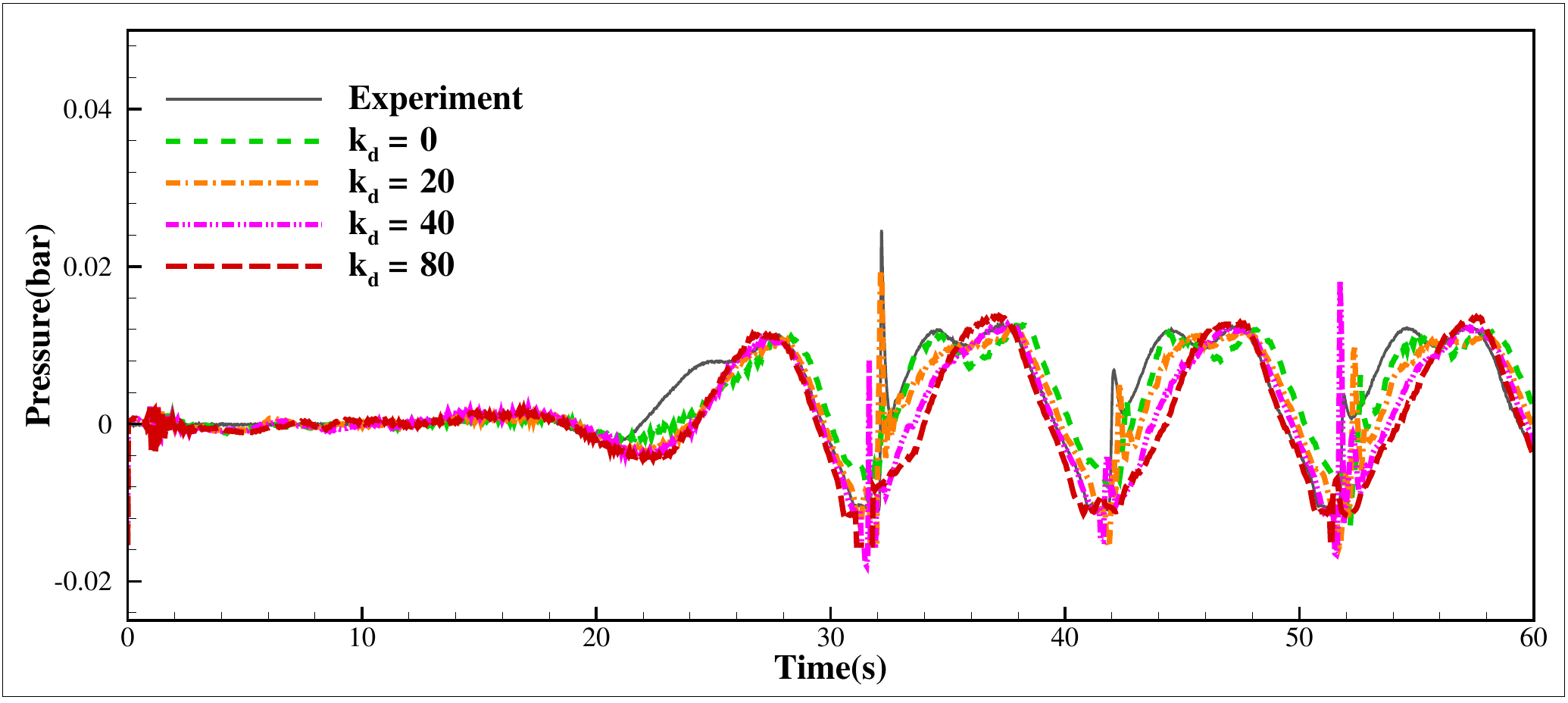}\\
	\includegraphics[trim = 2mm 2mm 2mm  2mm, clip, width=0.495\textwidth]{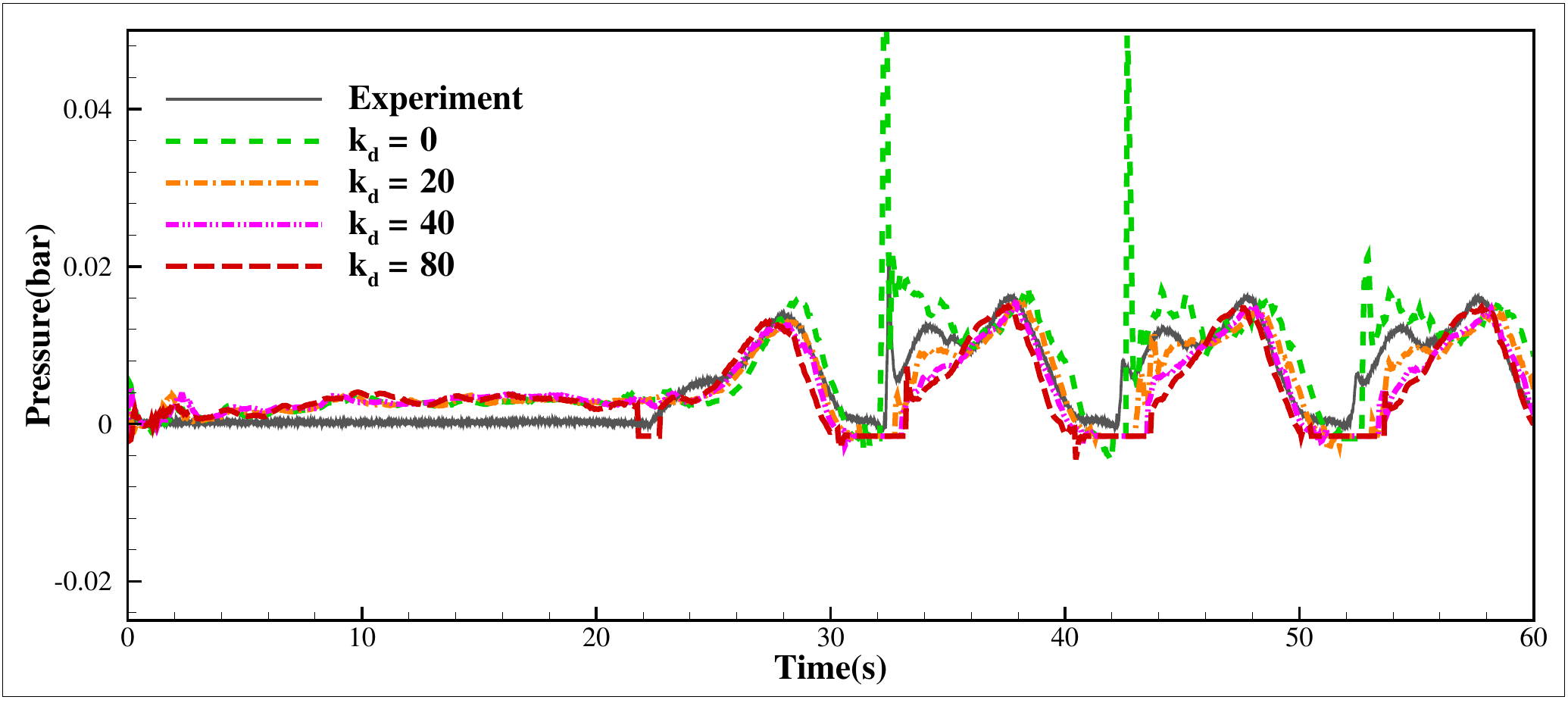}
	\includegraphics[trim = 2mm 2mm 2mm  2mm, clip, width=0.495\textwidth]{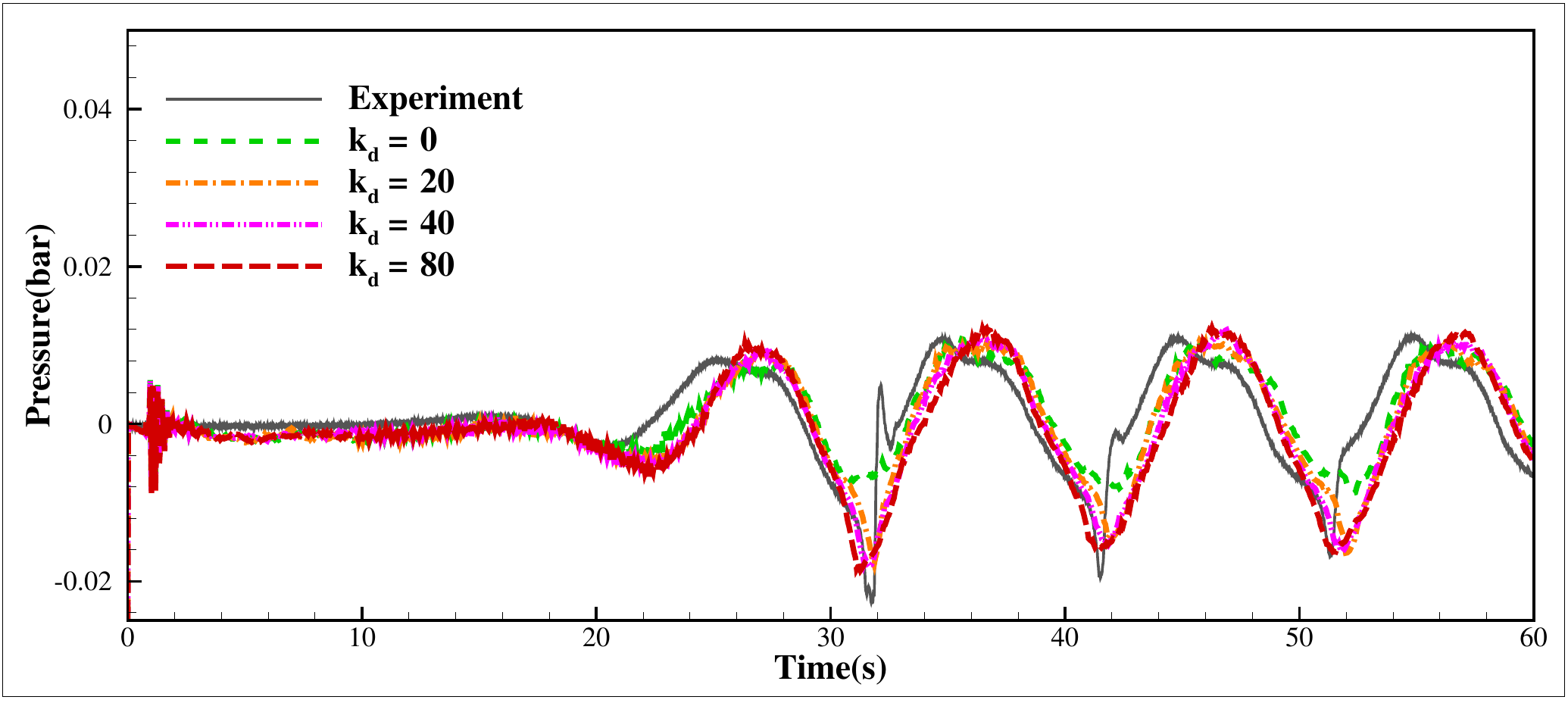}
	\caption{Modeling of OWSC with SPHinXsys: Damping effects on the time history of pressure on the flap for wave height $H = 5.0 \text{m}$ and wave period $T = 10.0 \text{s}$. 
		The pressure sensor number (from left to right) : $\text{PS}01$ and$\text{PS}03$ (top panel). $\text{PS}05$ and $\text{PS}09$ (middle panel); 
		$\text{PS}11$ and $\text{PS}13$ (bottom panel)	(For color interpretation, the reader is referred to the web version of this paper).}
	\label{figs:owsc-damping-pt}
\end{figure*}

Figure \ref{figs:owsc-damping-rotation} shows the time histories of the flap rotation by applying the linear damper 
with different damping coefficients based on the validated model, 
e.g., wave height $H = 5.0 \text{m}$ and wave period $T = 10.0 \text{s}$. 
As expected, 
the rotation amplitude is further reduced as the damping coefficient increase. 
It worth noting that the reduction of the flap rotation does not fit a linear relationship, 
e.g., the rotation reduction ratio from $k_d = 0$ to $k_d = 20$ is not equal to that from $k_d = 20$ to $k_d = 40$, 
indicating that the PTO operation may represents a nonlinear process. 
Figure \ref{figs:owsc-damping-pt} shows the time histories of the pressure recorded by the pressure sensors given in Table \ref{tab:ps}
for damping coefficients $k_d = 20$,  $k_d = 40$ and $k_d = 80$. 
One notable phenomenon is that the pressure peaks are reduced or eliminated 
for pressure sensors $\text{PS}01$,  $\text{PS}03$ and $\text{PS}11$ due to the reduced rotation of the flap. 
This may imply that the strength of the slamming event which leads to the pressure peak \cite{wei2015wave} 
is significantly reduced or even eliminated when damping operation is conducted. 
Also, 
the pressure drops for sensors $\text{PS}05$,  $\text{PS}09$ and $\text{PS}13$ 
show better agreement with experimental data compared with the results where no damping is imposed. 
This indicates that wave reflection and breaking are also reduced during damping operations. 

\begin{figure*} 
	\centering
	\includegraphics[trim = 2mm 2mm 2mm 2mm, clip, width=0.65\textwidth]{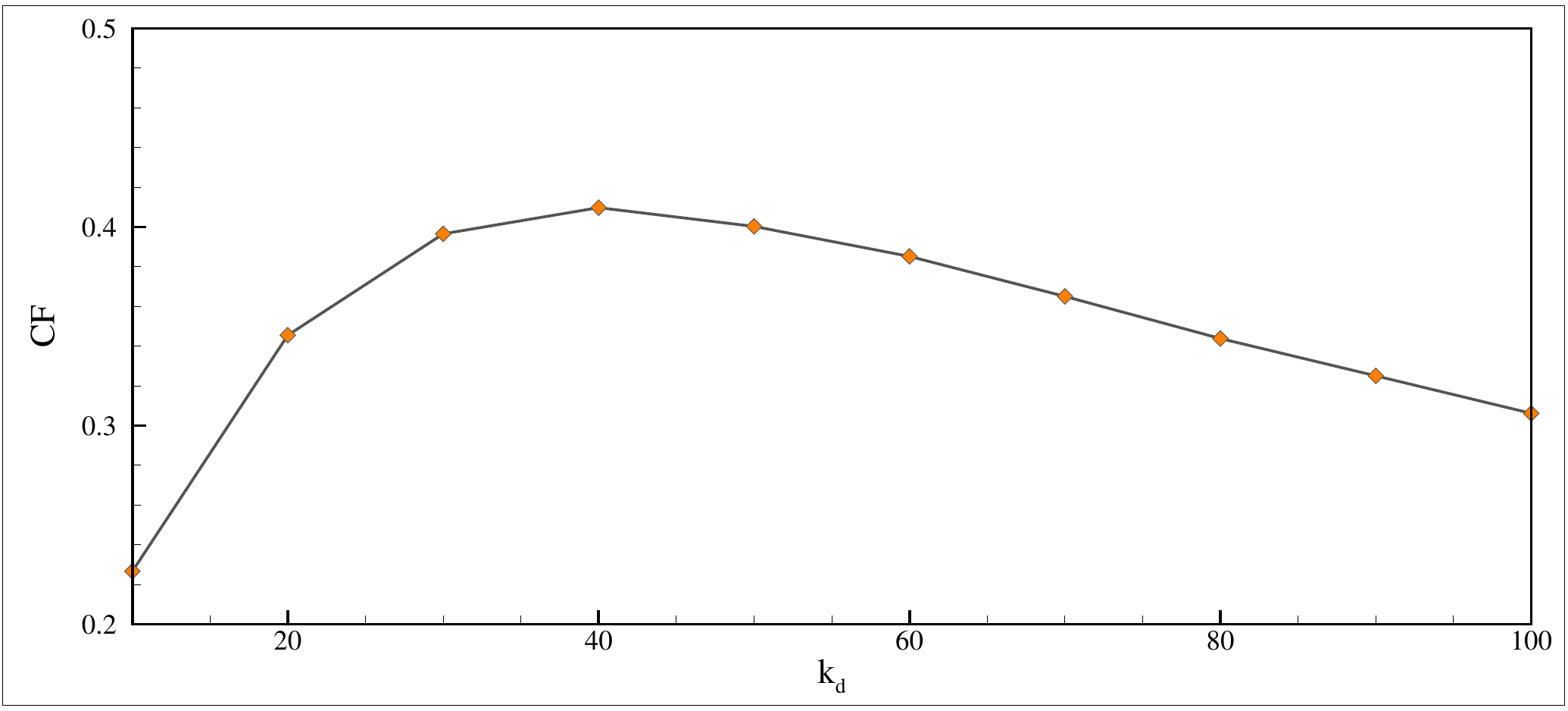}
	\caption{Modeling of OWSC with SPHinXsys: The variations of the CF in terms of damping coefficients for wave height $H = 5.0 \text{m}$ and wave period $T = 10.0 \text{s}$. }
	\label{figs:owsc-cf}
\end{figure*}

Figure \ref{figs:owsc-cf} gives the variation of the CF in terms of damping coefficients $k_d$. 
It can be observed that the CF is enhanced as the damping coefficient increase and 
reaches is maximum about $0.4$ when $k_d = 40$, 
and then shrinks as the damping coefficient further increases, 
implying that its optimized value of the CF may be achieved for moderate damping coefficient. 
%
%%%%%%%%%%%%%%%%%%%%%%%%%%%%%%%%%%%%%%%%%%%%%%%%%%%%%%%%%%%%%
% Section
%%%%%%%%%%%%%%%%%%%%%%%%%%%%%%%%%%%%%%%%%%%%%%%%%%%%%%%%%%%%%
\subsection{Extreme loads}\label{sec:extreme-loads}
Prediction of extreme loads on the device of  OWSC is of significant importance for the structure design, 
however,  
experimental study of the extreme loads is challenging due to its time and economic expensive nature for long time performance of wave tank test. 
In this work, 
we apply the present solver to estimate extreme loads on OWSC by considering extreme wave condition based on focused wave approach. 

In general, 
any wave elevation can be considered as the combination of a certain number of small amplitude waves. 
Therefore, 
the surface elevation at any point in spatio-temporal spaces can be written as
\begin{equation}
	\eta \left( x, t\right)  = \sum_{i = 0}^{N} a_i \cos\left( k_i x + \omega_i t + \phi_i \right) ,
\end{equation}
where $a_i$ is the wave amplitude, $k_i$ the wave number, $\omega$ the angular frequency and $\phi_i$ the phase for $i$th wave component. 
In the focused wave approach, 
the phase of each wave component is defined as
\begin{equation}
	\phi_i = k_i x_f - \omega_i t_f , 
\end{equation}
where $x_f$ and $t_f$ denote the focal position and time, respectively. 
The amplitude of each wave component takes the form of 
\begin{equation}
	a_i = A_f \frac{S(\omega_i) \Delta \omega}{\sum_i S(\omega_i) \Delta \omega} ,
\end{equation}
where $A_f$ is the target amplitude of the focused wave and $S(\omega_i)$ the spectral density. 
In this work, 
we apply the Pierson-Moskowitz spectra for modeling the wave spectra and it is given by
\begin{equation}
	S(\omega_i) = \frac{5}{16} H_s^2 \omega_p^4 \omega_i^{-5} \exp\left[ -\frac{5}{4} \left( \frac{\omega_i}{\omega_p}\right) ^{-4} \right] , 
\end{equation}
where $H_s = 6.0 \text m$ and $\omega_p = 2 \pi$ are the significant wave height and peak spectra wave angular frequency, 
respectively. 

For piston-type wave maker, 
the time history of the stroke is given by \cite{ning2009free} 
\begin{equation}
	S_0\left(t\right)  = \sum_{i = 0}^{N} \frac{a_i}{Tr_i} \sin\left( k_i x + \omega_i t + \phi_i \right) . 
\end{equation}
Here, 
the transfer function $Tr$ represents the relationship between the wave height 
and is given as \cite{gao2016numerical}
\begin{equation}
	Tr_i = \frac{2\left( \cosh\left( 2 k_i h_0\right) - 1\right) }{\sinh\left( 2 k_i h_0\right) + 2 k_i h_0 } .
\end{equation}

\begin{figure*}
	\centering
	\begin{subfigure}[b]{0.495\textwidth}
		\centering
		\includegraphics[trim = 2mm 2mm 2mm  2mm, clip, width=0.95\textwidth]{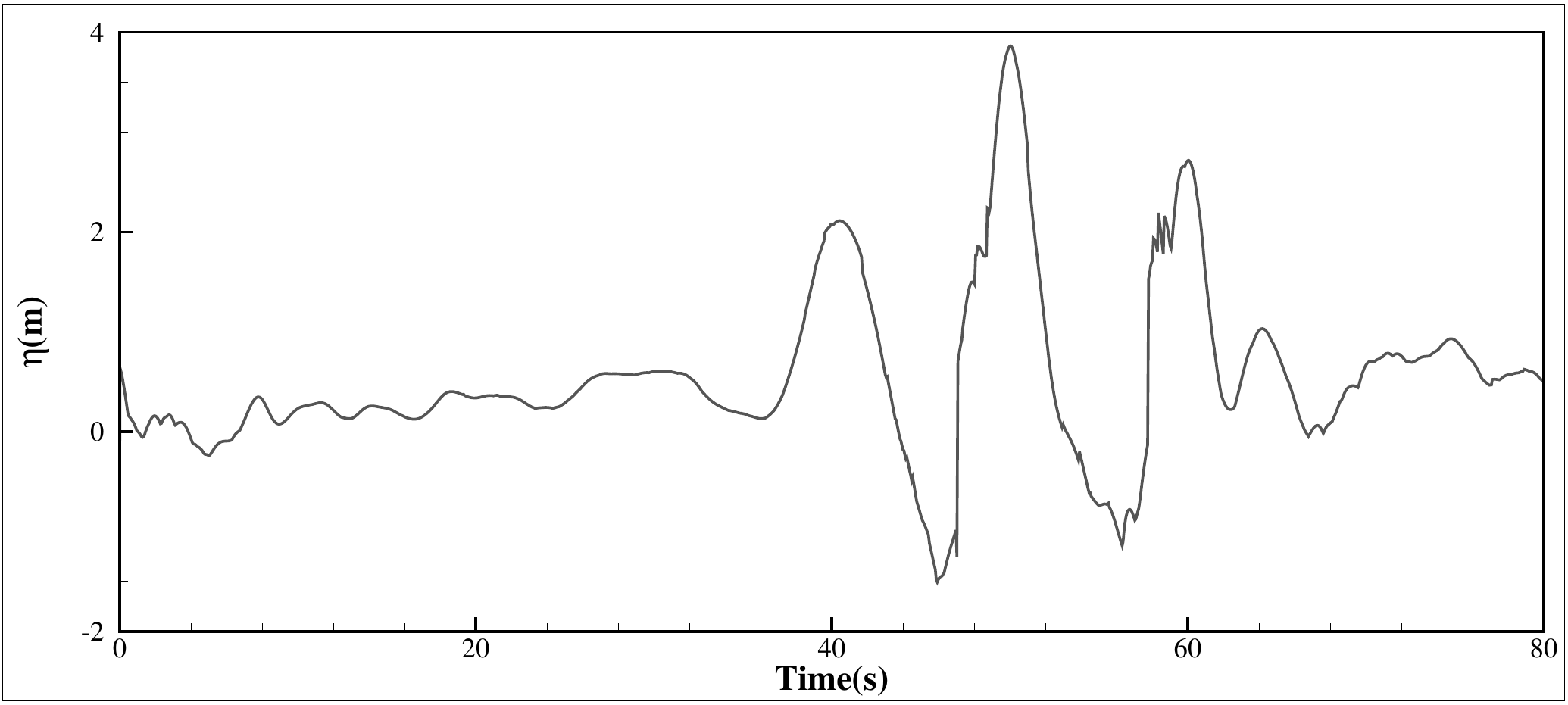}
		\caption{Wave elevation at focused position}
		\label{figs:owsc-focusedwave-surface}
	\end{subfigure}
	%add desired spacing between images, e. g. ~, \quad, \qquad, \hfill etc. 
	\begin{subfigure}[b]{0.495\textwidth}
		\centering
		\includegraphics[trim = 2mm 2mm 2mm 2mm, clip, width=0.95\textwidth]{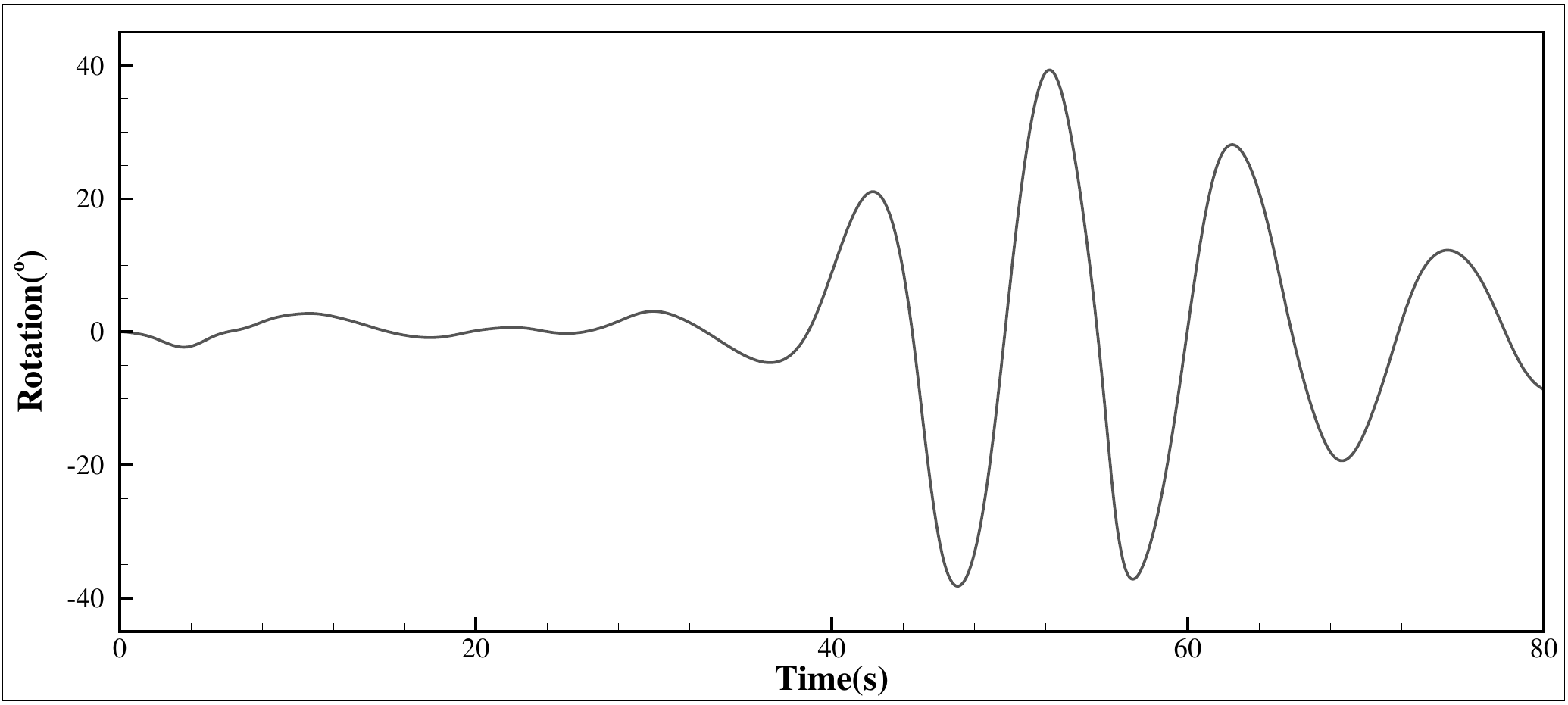}
		\caption{Rotation of the flap}
		\label{figs:owsc-focusedwave-rotation}
	\end{subfigure}
	\caption{Modeling of OWSC with SPHinXsys: The surface elevation and the corresponding  flap rotation under extreme wave condition. }
	\label{figs:owsc-focusedwave-rotation-surface}
\end{figure*}

Figure \ref{figs:owsc-focusedwave-rotation-surface} shows the time history of the wave elevation at the focused location $x_f = 7.6 \text m$ and 
the corresponding flap rotation. 
As shown in Figure \ref{figs:owsc-focusedwave-surface}, 
a sharp wave crest is found at the focused time $t_f = 50 \text s$ and the amplitude of the surface elevation decreases as the energy content of the wave decrease. 
Also, 
wave breaking is noted due to the interaction with the flap. 
Figure \ref{figs:owsc-focusedwave-rotation} gives the time history of the flap rotation under the interaction with the focused wave. 
As expected, 
the sharp wave crest induces a large amplitude rotation of the flap and the amplitude decreases as the wave crest past by. 

\begin{figure*}
	\centering
	\includegraphics[trim = 2mm 2mm 2mm  2mm, clip, width=0.495\textwidth]{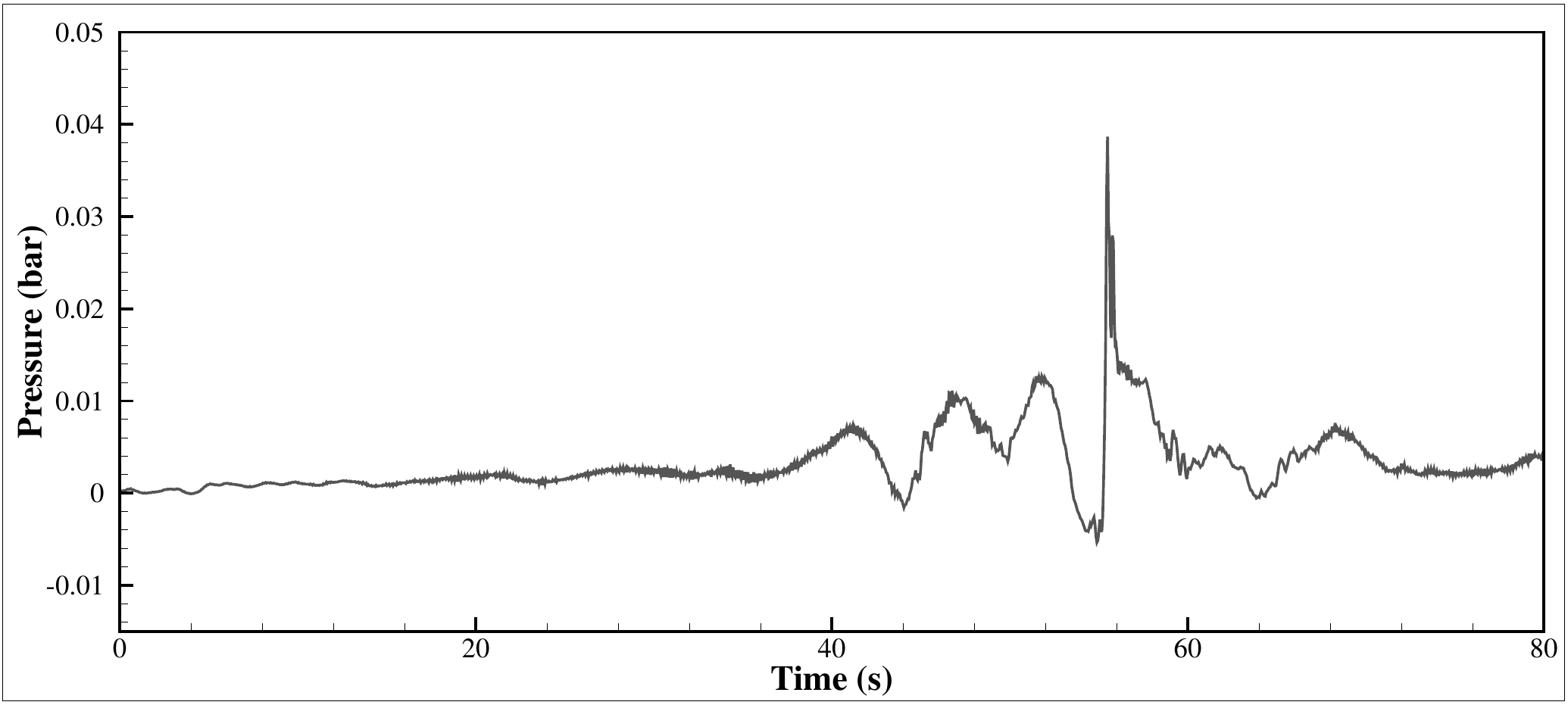}
	\includegraphics[trim = 2mm 2mm 2mm  2mm, clip, width=0.495\textwidth]{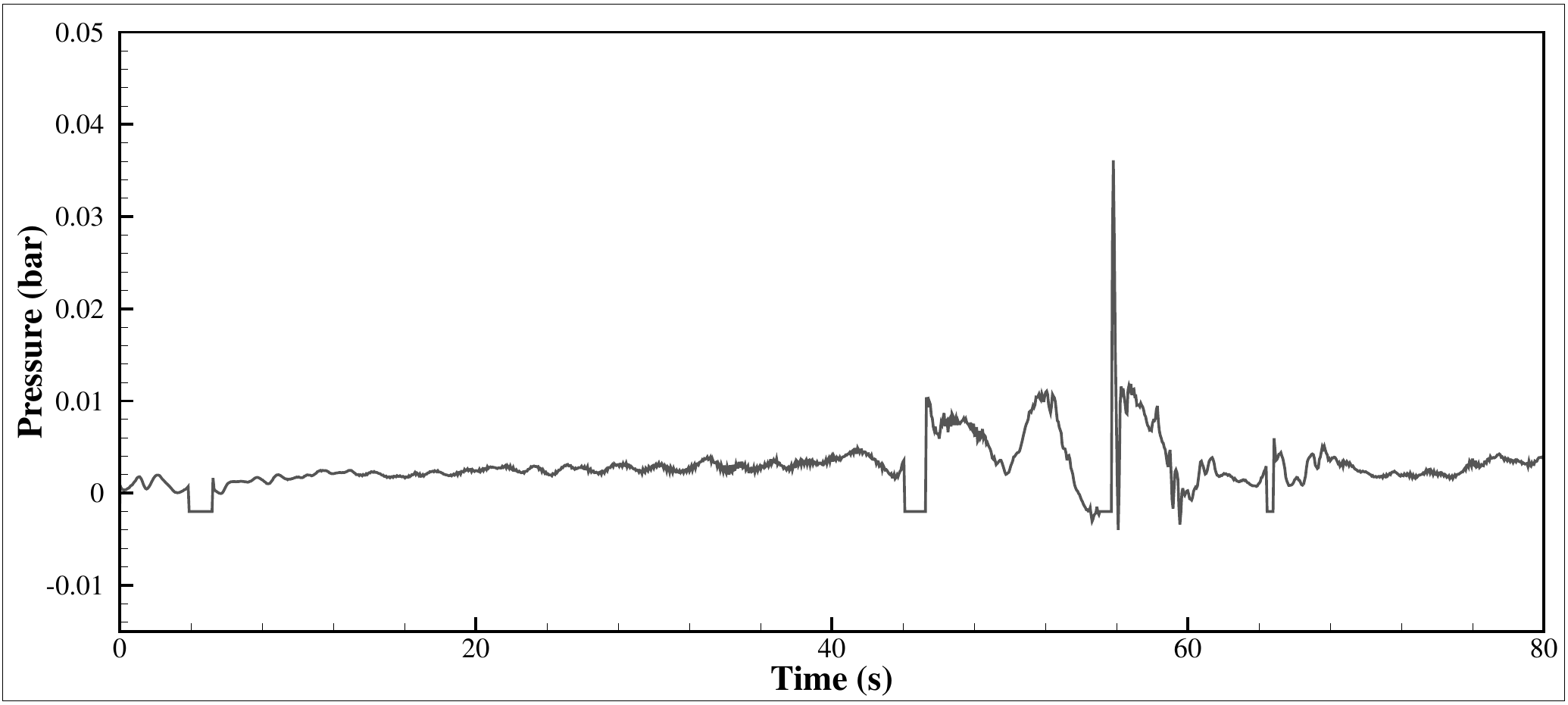}\\
	\includegraphics[trim = 2mm 2mm 2mm  2mm, clip, width=0.495\textwidth]{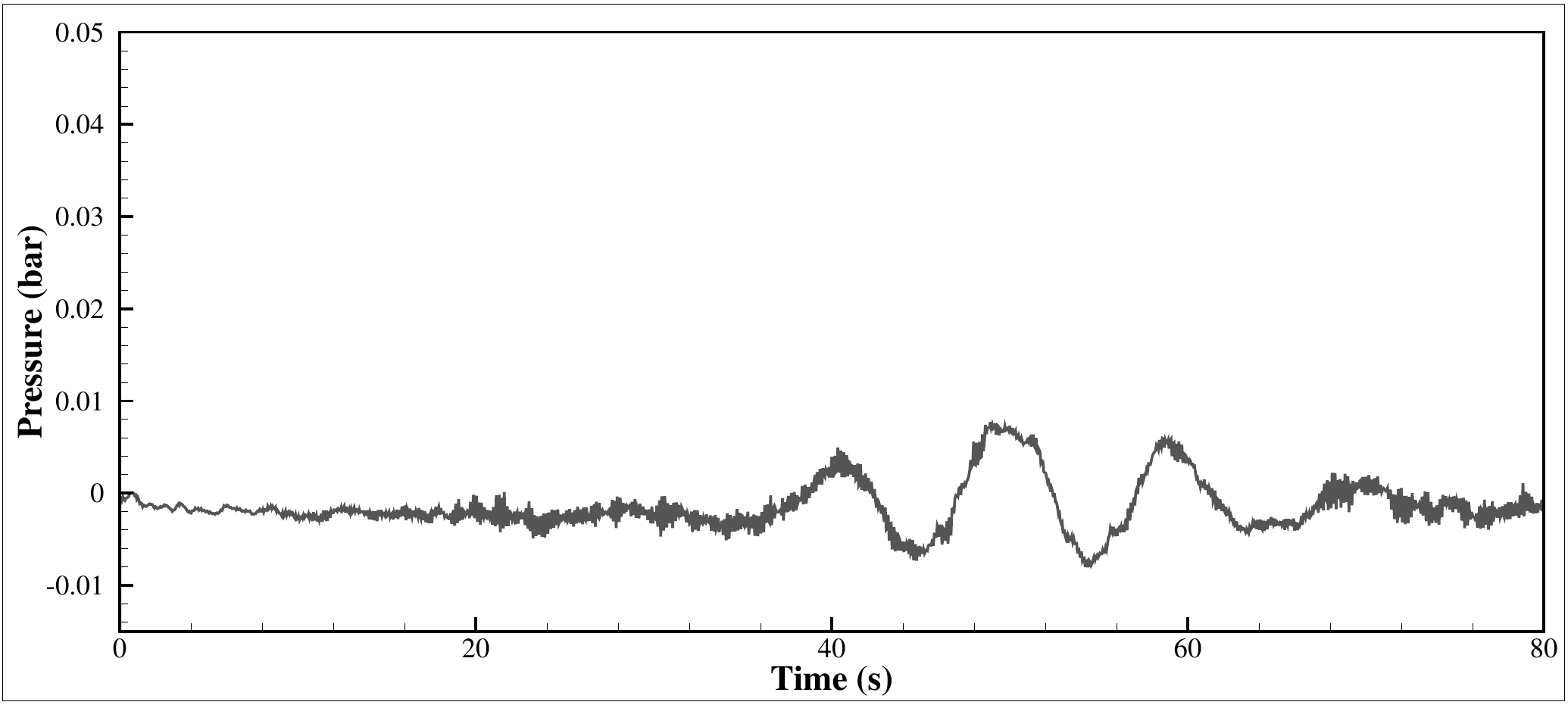}
	\includegraphics[trim = 2mm 2mm 2mm  2mm, clip, width=0.495\textwidth]{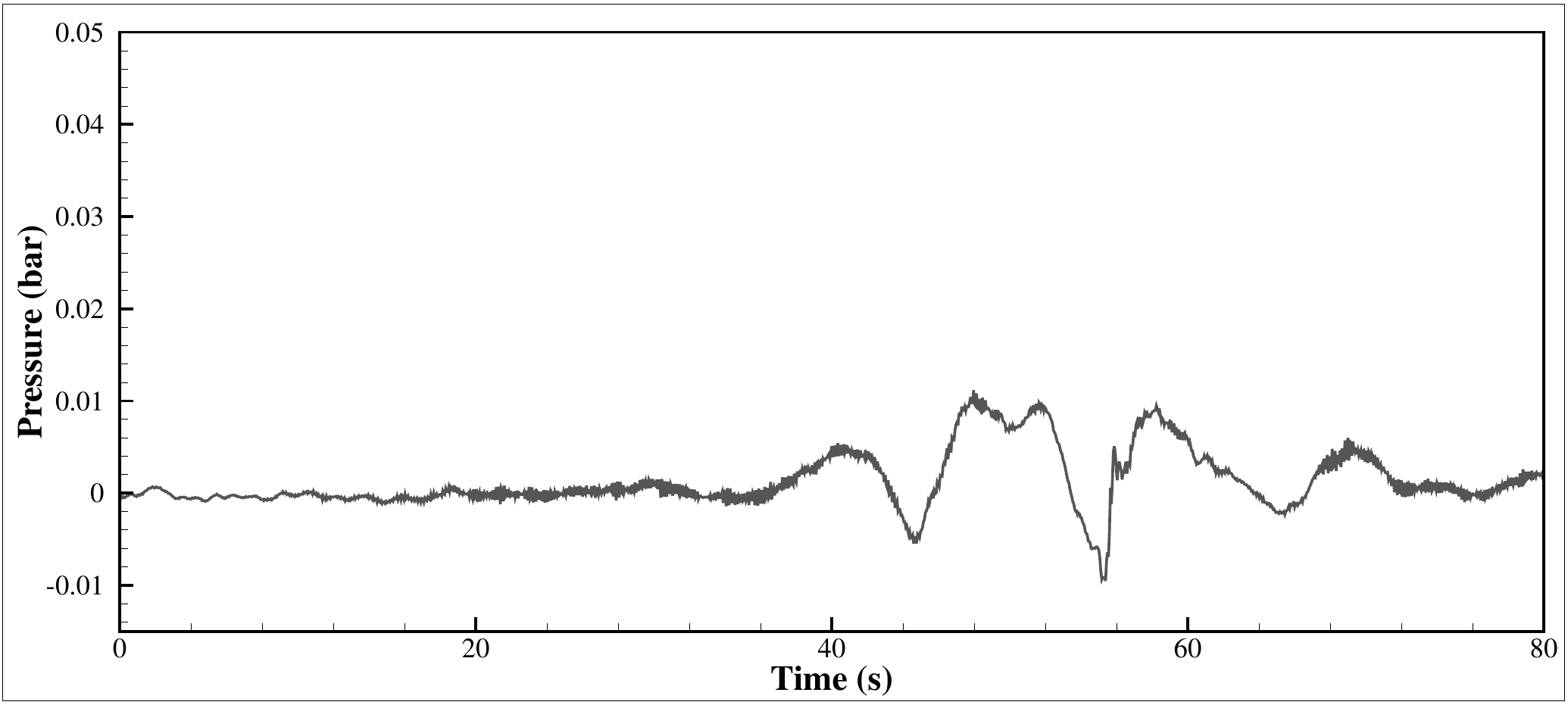}\\
	\includegraphics[trim = 2mm 2mm 2mm  2mm, clip, width=0.495\textwidth]{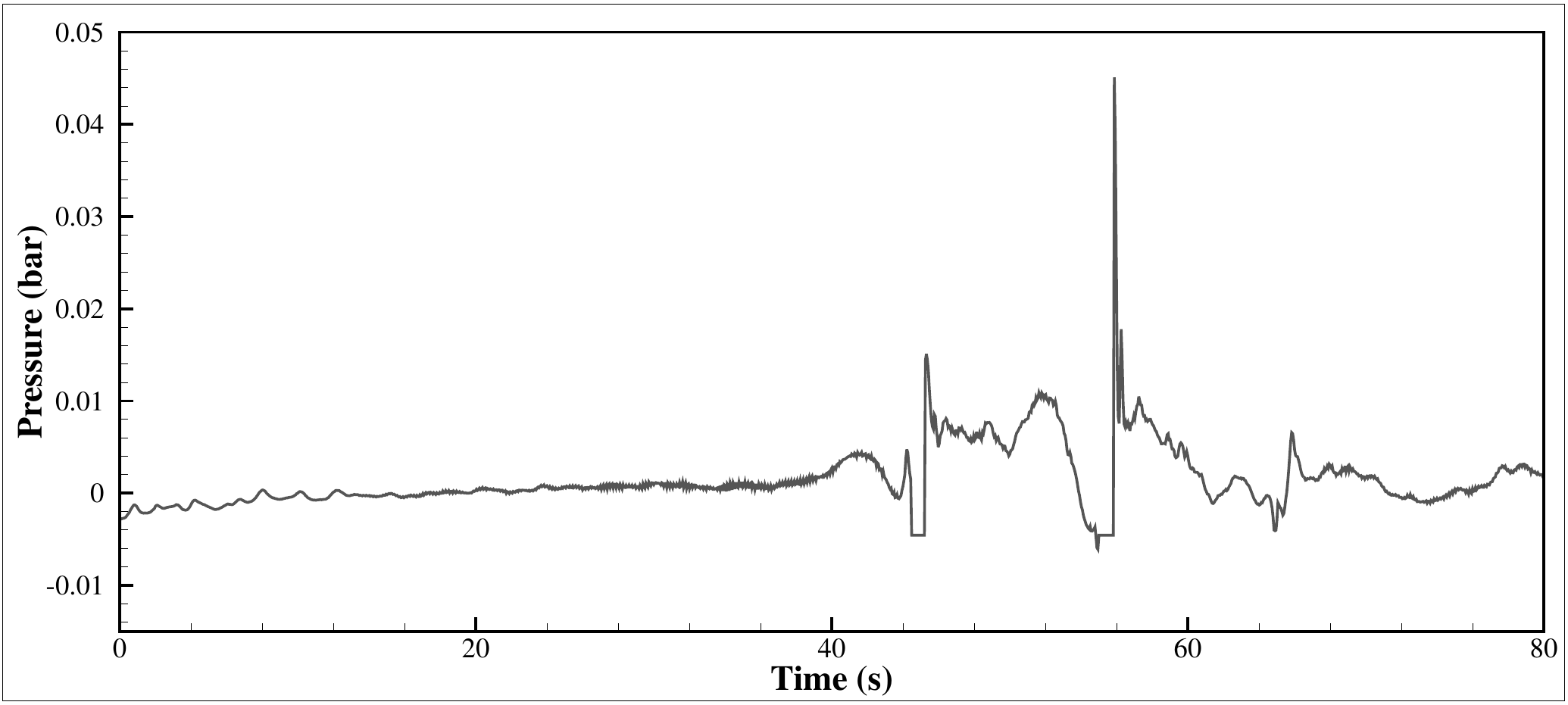}
	\includegraphics[trim = 2mm 2mm 2mm  2mm, clip, width=0.495\textwidth]{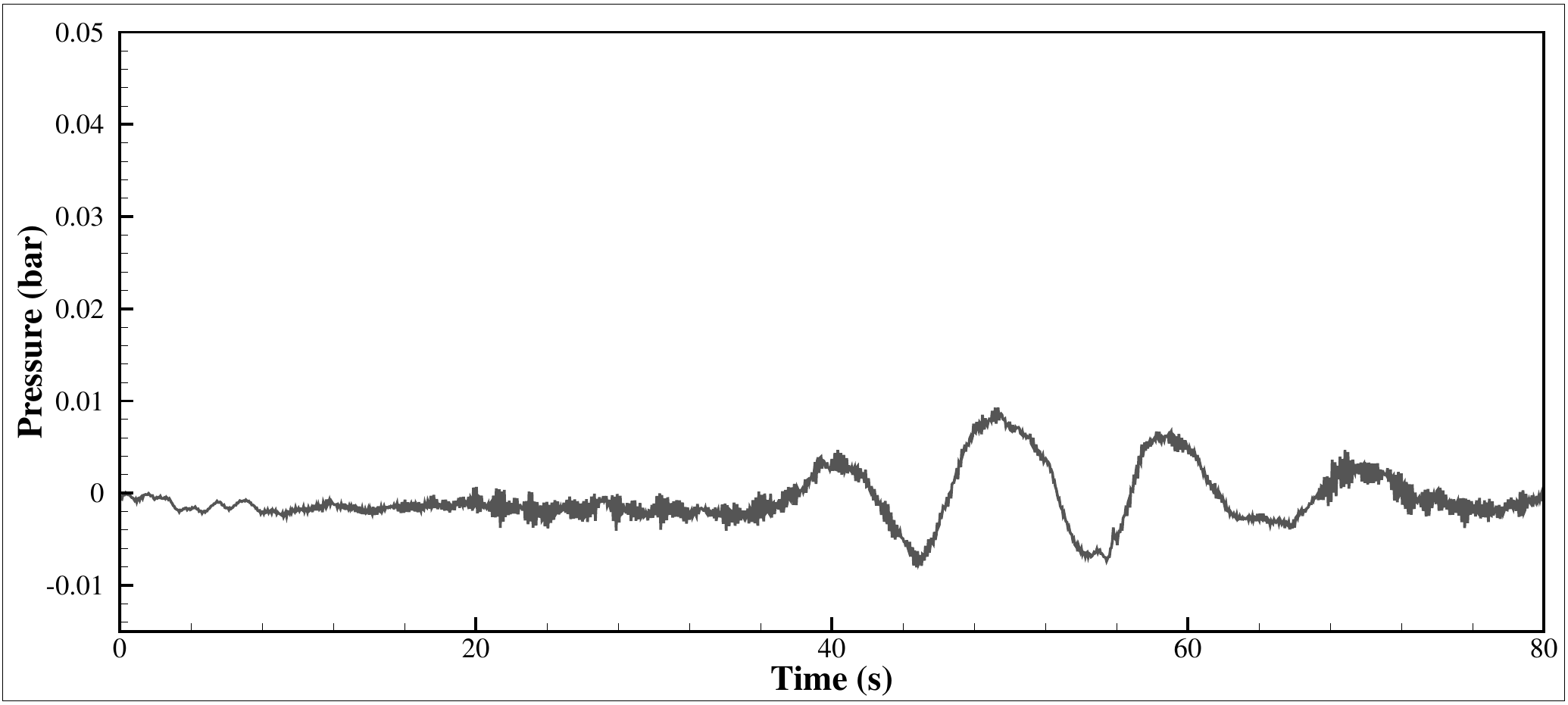}
	\caption{Modeling of OWSC with SPHinXsys: The time histories of wave loads on the flap under extreme wave condition. 
		The pressure sensor number (from left to right) : $\text{PS}01$ and$\text{PS}03$ (top panel). $\text{PS}05$ and $\text{PS}09$ (middle panel); 
		$\text{PS}11$ and $\text{PS}13$ (bottom panel).}
	\label{figs:owsc-focusedwave-pt}
\end{figure*}

Figure \ref{figs:owsc-focusedwave-pt} shows the time histories of the pressure signals recorded on selected sensors given in Table \ref{tab:ps} in extreme wave condition.  
For pressure sensors $\text{PS}01$,  $\text{PS}03$ and $\text{PS}11$, 
which are located close to the surface, 
a very large pressure peak is observed when the wave crest is impacting on the flap, 
indicating that a freak wave may induce extreme loads on the flap. 
%
%%%%%%%%%%%%%%%%%%%%%%%%%%%%%%%%%%%%%%%%%%%%%%%%%%%%%%%%%%%%%
%
% Section
%
%%%%%%%%%%%%%%%%%%%%%%%%%%%%%%%%%%%%%%%%%%%%%%%%%%%%%%%%%%%%%
\section{Concluding remarks}\label{sec:conclusion}
An efficient, robust and accurate numerical solver is proposed by coupling open-source multi-physics SPH-based library, 
SPHinXsys, 
with multi-body physics library, 
Simbody, 
for modeling wave interaction with an bottom hinged OWSC. 

The proposed numerical solver is validated by comparing the predicted wave elevation, 
flap rotation and wave loading on the flap with experimental data and those obtained  in literature with different numerical models.
The comparisons show that the present solver properly predicts the hydrodynamics properties of OWSC. 
More importantly, 
optimized computational performance is achieved in the present solver. 
Then, 
the validated solver is applied to study the PTO effects and efficiency, 
and investigate the extreme loads on the flap in extreme wave conditions. 
These numerical results demonstrate its	promising potential to future practical applications in the design of high-performance WECs. 

In future work, 
GPU implementation  will be conducted to further improve the computational efficiency and 
the slamming effects \cite{wei2016wave, dias2018slamming} on OWSC will also be studied. 
%%%%%%%%%%%%%%%%%%%%%%%%%%%%%%%%%%%%%%%%%%%%%%%%%%%%%%%%%%%%%
%
% Section
%
%%%%%%%%%%%%%%%%%%%%%%%%%%%%%%%%%%%%%%%%%%%%%%%%%%%%%%%%%%%%%
\section*{CRediT authorship contribution statement}
{\bfseries  Chi Zhang:} Investigation, Methodology, Visualization, Validation, Formal analysis, Writing - original draft, Writing - review \& editing; 
{\bfseries  Yanji Wei:} Investigation, Writing - review \& editing;
{\bfseries  Frederic Dias:} Investigation, Writing - review \& editing; 
{\bfseries  Xiangyu Hu:} Investigation, Supervision, Writing - review \& editing.
%%%%%%%%%%%%%%%%%%%%%%%%%%%%%%%%%%%%%%%%%%%%%%%%%%%%%%%%%%%%%
%
% Section
%
%%%%%%%%%%%%%%%%%%%%%%%%%%%%%%%%%%%%%%%%%%%%%%%%%%%%%%%%%%%%%
\section*{Declaration of competing interest }
The authors declare that they have no known competing financial interests 
or personal relationships that could have appeared to influence the work reported in this paper.
%%%%%%%%%%%%%%%%%%%%%%%%%%%%%%%%%%%%%%%%%%%%%%%%%%%%%%%%%%%%%
%
% Section
%
%%%%%%%%%%%%%%%%%%%%%%%%%%%%%%%%%%%%%%%%%%%%%%%%%%%%%%%%%%%%%
\section{Acknowledgement}
C. Zhang and X.Y. Hu would like to express their gratitude to Deutsche Forschungsgemeinschaft (DFG) 
for their sponsorship of this research under grant numbers 
DFG HU1527/10-1 and HU1527/12-1. 
The work of F. Dias has been funded by Science Foundation Ireland (SFI) under Marine Renewable Energy Ireland (MaREI), 
the SFI Center for Marine Renewable Energy (grant 12/RC/2302).
%%%%%%%%%%%%%%%%%%%%%%%%%%%%%%%%%%%%%%%%%%%%%%%%%%%%%%%%%%%%%
%
% Section
%
%%%%%%%%%%%%%%%%%%%%%%%%%%%%%%%%%%%%%%%%%%%%%%%%%%%%%%%%%%%%%
\section*{Reference}
\bibliography{mybibfile}
%%%%%%%%%%%%%%%%%%%%%%%%%%%%%%%%%%%%%%%%%%%%%%%%%%%%%%%%%%%%%
%
% Document end
%
%%%%%%%%%%%%%%%%%%%%%%%%%%%%%%%%%%%%%%%%%%%%%%%%%%%%%%%%%%%%%
\end{document}